\documentclass[12pt]{article}

\usepackage{hyperref}
\hypersetup{
    colorlinks=true,
    linkcolor=blue,
    citecolor=blue,
    filecolor=magenta,      
    urlcolor=green,
}
\usepackage{natbib}
\usepackage{graphicx}
\usepackage{amsmath,amsthm,amssymb}
\usepackage{algpseudocode} 
\usepackage{algorithm}     
\usepackage{multirow}
\usepackage{booktabs}
\let\orintop\intop
\usepackage{mathabx}
\usepackage{mathrsfs}
\usepackage{physics} 
\let\intop\orintop
\usepackage{enumitem}
\usepackage{makecell}
\usepackage{subcaption}

\newcommand{\blind}{1}

\def\diag{{\rm diag}}

\DeclareMathOperator*{\Var}{Var}
\DeclareMathOperator*{\argmin}{arg\,min}
\DeclareMathOperator*{\argmax}{arg\,max}

\newcommand{\single}{\spacingset{1}}

\newtheorem{proposition}{Proposition}[section]
\newtheorem{theorem}{Theorem}[section]
\newtheorem{lemma}{Lemma}[section]
\newtheorem{corollary}{Corollary}[section]

\begin{document}

\def\spacingset#1{\renewcommand{\baselinestretch}%
{#1}\small\normalsize} \spacingset{1}

\if1\blind
{
\title{\bf Sliced Wasserstein Regression}
\author{Han Chen\thanks{The first two authors contributed equally to this work.} , Yidong Zhou$^*$, and Hans-Georg M\"uller\thanks{H.G.M. was partially supported by NSF grants DMS-2014626 and DMS-2310450.}\hspace{.2cm}\\
Department of Statistics, University of California, Davis, USA}
\maketitle
} \fi
\if0\blind
{
  \bigskip
  \bigskip
  \bigskip
  \begin{center}
    {\LARGE\bf Sliced Wasserstein Regression}
\end{center}
  \medskip
} \fi

\bigskip
\begin{abstract}
While statistical modeling of distributional data has gained increased attention, the case of multivariate distributions has been somewhat neglected despite its relevance in various applications. This is because the Wasserstein distance, commonly used in distributional data analysis, poses challenges for multivariate distributions. A promising alternative is the sliced Wasserstein distance, which offers a computationally simpler solution. We propose distributional regression models with multivariate distributions as responses paired with Euclidean vector predictors. The foundation of our methodology is a slicing transform from the multivariate distribution space to the sliced distribution space for which we establish a theoretical framework, with the Radon transform as a prominent example. We introduce and study the asymptotic properties of sample-based estimators for two regression approaches, one based on utilizing the sliced Wasserstein distance directly in the multivariate distribution space, and a second approach based on a new slice-wise distance, employing a univariate distribution regression for each slice. Both global and local Fr\'echet regression methods are deployed for these approaches and illustrated in simulations and through applications. These include the joint distribution of excess winter death rates and winter temperature anomalies in European countries as a function of base winter temperature, and also data from finance.
\end{abstract}

\noindent%
\vfill

\newpage
\newcommand{\double}{\spacingset{1.75}}
\double 

\section{Introduction}
It is increasingly common for statisticians to encounter data that consist of samples of multivariate distributions. Examples include distributions of anthropometric data  \citep{hron:20},  stock price returns for multiple stocks or indices \citep{gueg:18} and systolic and diastolic blood pressure data \citep{fan2021conditional}. Distributional data differ from functional data in that they do not form a vector space. In the emerging field of distributional data analysis, the focus has been on approaches designed for univariate distributions \citep{mata:21,ghos:21,pete:22} while there is a lack of methodology for the case where samples feature multivariate distributions \citep{dai:22}. In this paper, we propose a new regression approach for situations where responses are multivariate distributions and predictors are Euclidean vectors. Our approach is based on a novel general slicing transform framework with the Radon transform as a prominent example and we also introduce a new slice-wise Wasserstein distance. The regression models are developed for both the previously established sliced (in the following referred to as slice-average)  Wasserstein \citep{bonneel2015sliced} and the new slice-wise Wasserstein paradigms. 
  
For the case of univariate distributions, global bijective transformations, including the log quantile density transform and log hazard transform, have been used to map univariate distributions to a Hilbert space $L^2$ \citep{mull:16:1}, where established functional regression methods can then be deployed.  Alternative transformations evolved in the field of compositional data analysis, referred to as Bayes Hilbert space \citep{hron:16,mena:18}. 
When using the Wasserstein metric in the space of one-dimensional distributions, the quasi-Riemannian structure of this space has been exploited by deploying log maps to tangent bundles. One can then develop principal component analysis \citep{bigo:17} and regression models  \citep{chen:20} in tangent spaces,  which are Hilbert spaces so that classical functional data analysis regression models can be applied. This approach comes with some caveats as the inverse map is not well-defined on the entire tangent space \citep{pego:22} and its extension to multivariate distributions has only been considered for the special case of multivariate Gaussian distributions \citep{okano2023distribution}.

Another line of work in distributional regression focuses on learning optimal transports between distributions. \citet{ghodrati2022distribution} directly estimated transport maps from predictor to response distributions and provided a theoretical discussion of potential extensions to the multivariate case, but without implementing or illustrating these extensions in \citet{ghodrati2023transportation}. \citet{zhu2023autoregressive} instead employed rudimentary algebraic operations in the space of optimal transports to construct regression models, demonstrating the approach through autoregressive modeling of time series of one-dimensional distributions. Fr\'{e}chet regression is yet another approach that provides an asymptotically consistent regression method for univariate distributions as responses with vector-valued predictors \citep{mull:19:3}, but this approach lacks theoretical guarantees for the case of multivariate distributions. 
Overall, much less is known about regression models for multivariate distributional data. In addition to \citet{ghodrati2023transportation}, a Bayes Hilbert space approach to model bivariate density functions has been discussed \citep{gueg:18, hron:20}, without any theoretical guarantees.  

Due to the computational and theoretical difficulties when applying optimal transport and Wasserstein distances for multivariate distributions,
the sliced Wasserstein distance \citep{bonneel2015sliced}, a computationally more efficient alternative to the Wasserstein distance,  has gained popularity in statistics and machine learning   \citep{courty2017joint, kolouri2019generalized, rustamov2020intrinsic, tanguy2023reconstructing, quellmalz2023sliced}. To the best of our knowledge, current studies employing  the sliced Wasserstein distance for regression have been limited to the application of kernel methods in machine learning 
\citep{kolouri2016sliced, meunier2022distribution, zhang2022nonlinear} and lack a focus on statistical data analysis and asymptotic convergence. 

These considerations motivate the development of a generalized regression framework utilizing the sliced Wasserstein distance based on a slicing transform from the multivariate distribution space to the slicing space.  We implement this novel framework with multivariate distributions as responses with two regression techniques, the first of which utilizes the sliced Wasserstein distance in the multivariate distribution space, while the second slice-wise approach utilizes a regression step for each slice, followed by an inverse transform from the sliced space to the original distribution space. These approaches come with theoretical guarantees on the convergence of the fitted regressions under suitable regularity conditions. Our results build on the strengths of univariate distribution regression while accounting for the effects of the inverse transform. We provide additional assumptions and details for the case where the response distributions need to be recovered from random samples generated by the underlying distributions in Section \ref{sec:adx:dens} of the Appendix.

The paper is organized as follows. We introduce analytical tools for slicing transforms with emphasis on the Radon transform in Section \ref{sec:transformation} and delineate the slicing space in Section \ref{sec:space}. In Section \ref{sec:regression} we present the two regression models as outlined above, where corresponding estimates are discussed in Section \ref{sec:estimation}. Asymptotic convergence results are presented in Section \ref{sec:convergence} and practical algorithms in Section \ref{sec:numerical}. We report the results of simulation studies in Section \ref{sec:simulation} and illustrate the proposed methods with data on excess winter mortality in European countries and also data on weekly returns of the S\&P500 index and VIX index in the United States in Section \ref{sec:data}, followed by a discussion in Section \ref{sec:disc}. Additional results and proofs are in the Appendix. 

\section{Slicing Transforms for Multivariate Distributions} 
\label{sec:transformation}
\subsection{Preliminaries}

We denote by $\|\cdot\|_2$  the $L^2$ norm and by $\|\cdot \|_\infty$ the sup norm for vectors or functions and throughout use $C_0, C_1, \dots$ to denote various constants and their dependence on relevant quantities $R$ will be indicated by writing $C_0(R), C_1(R), \dots$. All notations are listed in Section \ref{sec:adx:notation} of the Appendix. We assume that the multivariate distributions under consideration possess density functions and share a common support $D$, assumed to be known or diligently chosen, usually from subject-matter considerations, where  
\begin{enumerate}[label=(D\arabic*), leftmargin=1cm]
\item The support $D\subset \mathbb{R}^p$ is compact and convex.
\end{enumerate}
Denote the space of multivariate density functions on $\mathbb{R}^p$ with compact support $D$ by
\begin{align}
\label{formula:multiDensF}
\mathcal{F} = \left\{f \in L^1(\mathbb{R}^p):\,f(z)\geq 0,\, \int_{\mathbb{R}^p} f(z)dz = 1, \, \text{ support }I(f) = D, f \text{ satisfies (F1)} \right\}, 
\end{align}
where 
\begin{enumerate}[label=(F\arabic*), leftmargin=1cm]
\item There exists a constant  $M_0>0$ and an integer $k\geq 2$ such that for all $f\in\mathcal{F}$, \\ $\max\{\|f\|_\infty, \|1/f\|_\infty\}\leq {M_0}$ on $D$ and  $f$ is continuously differentiable of order $k$ on $D$ and has uniformly bounded partial derivatives.
\item $k>p/2$.
\item $k\geq p+1$. 
\end{enumerate}

Smoothness of higher order leads to faster decay rate of Fourier-transformed functions and Conditions (F2)--(F3) will be utilized to obtain rates of convergence. We also require a set
\begin{align*}
    \mathcal{G} = \left\{g \in L^1(\mathbb{R}):\,g(u)\geq 0,\, \int_{\mathbb{R}} g(u)du = 1, g \text{ satisfies (D2) and (G1)}\right\}
\end{align*}
of univariate density functions. Denoting the support of $g\in\mathcal{G}$ by  $I(g)$, we require
\begin{enumerate}[label=(D\arabic*), leftmargin=1cm]
\setcounter{enumi}{1}
\item For all $g\in\mathcal{G}$, the support $I(g)$ is compact and $\bigcup_{g\in\mathcal{G}}I(g)$ is bounded.
\end{enumerate}
\begin{enumerate}[label=(G\arabic*), leftmargin=1cm]
\item For an integer $k$ as in (F1),  there exists a constant $M_1>0$ such that for all $g \in \mathcal{G}$, $\max\{\|g\|_\infty, \|1/g\|_\infty\}\leq {M_1}$
 on $I(g)$ and $g$ is continuously differentiable of order $k$ on $I(g)$ with uniformly bounded  derivatives.
\end{enumerate}

Throughout, we utilize the unit sphere to denote a slicing parameter set in $\mathbb{R}^p$, \newline 
${\Theta = \{z\in\mathbb{R}^p:\, \|z\|_2=1\}.}$ 
Defining the density slicing space $\Lambda_{\Theta}$ as a family of maps from $\Theta$ to $\mathcal{G}$, 
\begin{align*}
    \Lambda_{\Theta} = \left\{\lambda :\, \Theta \rightarrow \mathcal{G}, \int_{\Theta}\int_{\mathbb{R}}[\lambda(\theta)(u)]^2\,dud\theta <\infty \right\}.
\end{align*}
$\Lambda_{\Theta}$ can be endowed with a  $L^2$ metric, 
\begin{align}
\label{formula:LambdaL2}
    d_2(\lambda_1, \lambda_2) = \left(\int_{\Theta}\int_{\mathbb{R}}\left(\lambda_1(\theta)(u) - \lambda_2(\theta)(u)\right)^2\,dud\theta\right)^{1/2},\quad \text{ for all }  \lambda_1, \lambda_2 \in \Lambda_{\Theta},
\end{align}
where we note that the integral is well defined because of the Cauchy-Schwarz inequality. Two maps $\lambda_1, \lambda_2$ will be considered to be identical if they coincide except on a set of measure zero.

\subsection{Radon Transform and Slicing Transforms}
\begin{figure}[tb]
    \single
    \centering
    \includegraphics[width=0.4\linewidth]{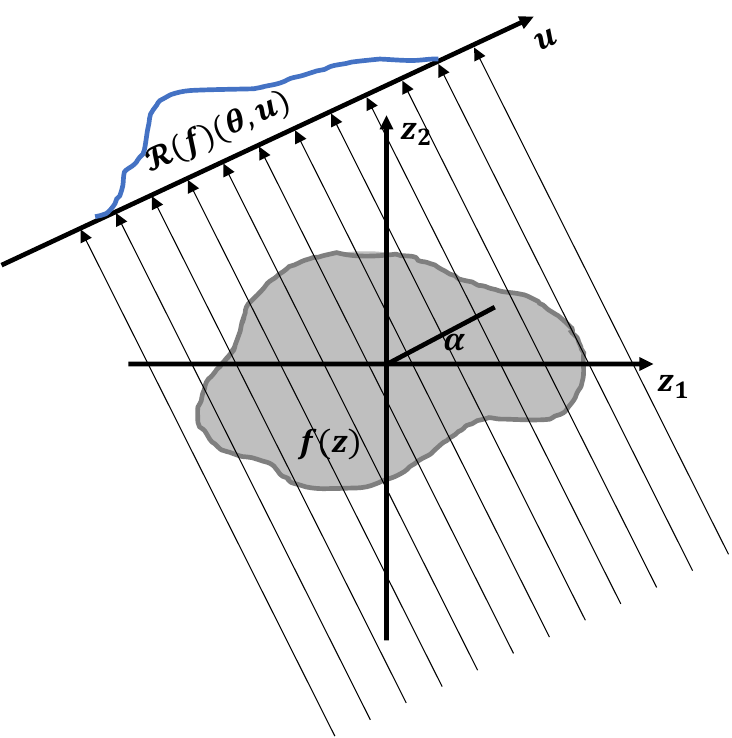}
    \caption{Schematic diagram for the two-dimensional Radon transform. For each unit vector $\theta(\alpha)=({\cos(\alpha), \sin(\alpha))}$, the Radon transform integrates along the line $\langle z,\theta\rangle=u$ for each $u\in\mathbb{R}$.  }
    \label{fig:radon}
\end{figure}

The Radon transform $\mathcal{R}$ \citep{radon20051} is an integral transform, which maps an integrable $p$-dimensional function to the infinite set of its integrals over the hyperplanes of $\mathbb{R}^p$. Following the notation in \citet{epstein2007introduction}, let $\theta$ be a unit vector in $\Theta$, $u$ be an element in $\mathbb{R}$, and  $l_{u,\theta}$  the affine hyperplane represented as 
$l_{u,\theta} = \{z \in \mathbb{R}^p : \,\langle z,\theta \rangle = u\}$.
Employing an  orthonormal basis $\{\theta, e_1,...,e_{p-1} \}$ for $\mathbb{R}^p$ with  
$\langle \theta,e_j \rangle = 0 \text{ and } \langle e_j, e_l \rangle = \delta_{jl}, \text{ for }j,l=1,...,p-1$, the $p$-dimensional Radon transform $\mathcal{R}: \,\mathcal{F} \rightarrow \Lambda_{\Theta}$ is defined through integrals over  $l_{u,\theta}$,  
\begin{align}
\label{formula:radon}
    \mathcal{R}(f)(\theta, u) = \int_{\mathbb{R}^{p-1}}f\left(u\theta + \sum_{j=1}^{p-1}s_je_j\right)ds_1\cdots ds_{p-1},\quad \text{for } \theta\in \Theta \text{ and } u\in \mathbb{R},
\end{align}
where we write  $\mathcal{R}(f)(\theta, u)$ for $\mathcal{R}(f)(\theta)(u)$; see Figure \ref{fig:radon} for a schematic illustration of the two-dimensional Radon transform. Since $l_{u,\theta}$ and $l_{-u,-\theta}$ are the same line, the Radon transform is an even function that satisfies $\mathcal{R}(f)(\theta,u) = \mathcal{R}(f)(-\theta,-u)$.

The inverse Radon transform is related to the Fourier transform. The $p$-dimensional Fourier transform $\mathcal{J}_{p}:\, L^1(\mathbb{R}^p)\rightarrow \{ f : \mathbb{R}^p \to \mathbb{C} \}$ is 
\begin{align*}
    \mathcal{J}_{p}(f)(\iota) = \int_{\mathbb{R}^p}e^{-i\langle z,\iota \rangle}f(z)dz,  \quad \text{ for all }  \iota\in\mathbb{R}^p,
\end{align*}
with the one-dimensional Fourier transform $\mathcal{J}_{1}:\,L^1(\mathbb{R}) \rightarrow \{ f : \mathbb{R} \to \mathbb{C} \}$ given through the formula $\mathcal{J}_{1}(g)(r)= $$\int_\mathbb{R}g(u)e^{-iur}du$ for all $r \in \mathbb{R}$. The connection between the Fourier transform and the Radon transform is as follows.

\begin{proposition}[Central Slicing Theorem \citep{bracewell1956strip}]
    Assume $f \in \mathcal{F}$. For any real number $r$ and a unit vector $\theta\in \Theta$, one has $\mathcal{J}_{1}\left({\mathcal{R}(f)}(\theta)\right)(r) = \mathcal{J}_{p}(f)(r\theta)$.
\end{proposition}
The inverse Radon transform has been well investigated both theoretically and numerically  \citep{abeida2012iterative, herman2009fundamentals, mersereau1974digital}. A common  device  to reconstruct the original multivariate function from its Radon transform is the filtered back-projection, 
\begin{align}
\label{formula:backproj}
    \mathcal{R}^{-1}\left(\lambda\right)(z) = \frac{1}{2(2\pi)^p}\int_{\Theta}\int_\mathbb{R}\mathcal{J}_{1}\left({\lambda}(\theta)\right)(r)e^{ir\langle \theta,z\rangle}|r|^{p-1}drd\theta,
\end{align}
which satisfies ${\mathcal{R}^{-1}\left(\mathcal{R}(f)\right) = f}$ for any $f\in\mathcal{F}$ \citep{natterer2001mathematics, epstein2007introduction}. The inversion formula can be decomposed into two steps. The first step acts as a high-pass filter, suppressing low-frequency components and amplifying high-frequency components. The second step implements an angular integral which can be interpreted as a back-projection of the filtered Radon transform \citep{epstein2007introduction, helgason2010integral}. However, $\mathcal{R}^{-1}$ is not a continuous map, and small errors in the reconstruction of $\mathcal{R}(f)$ are amplified \citep{epstein2007introduction}. Therefore, regularization is usually applied when reconstructing the original function from its Radon transform \citep{horbelt2002discretization, qureshi2005inverse, shepp1982maximum}, often through the use of a ramp filter that has a cut-off in the high-frequency domain \citep{epstein2007introduction, kak2001principles}. 

Defining a regularizing function $\phi_\tau(r)$ with tuning parameter $\tau$ as $\phi_\tau(r) =  1$ for $|r|\leq \tau$ and  $\phi_\tau(r) =  0$ for $|r|> \tau$, the regularized inverse map $\check{\mathcal{R}}^{-1}_\tau: \Lambda_{\Theta}\rightarrow L^1(\mathbb{R}^p)$ is obtained by cutting off the high-frequency components in the filtered back-projection \eqref{formula:backproj} through
\begin{align}
\label{formula:regback0}
    \check{\mathcal{R}}_\tau^{-1}\left(\lambda\right)(z) = \frac{1}{2(2\pi)^p}\int_{\Theta}\int_\mathbb{R}\mathcal{J}_{1}\left({\lambda}(\theta)\right)(r)e^{ir\langle \theta,z\rangle}|r|^{p-1}\phi_\tau(r)drd\theta. 
\end{align}
As this regularized inverse is not guaranteed to be a multivariate density function, normalization is applied to map the resulting $L^1$ function into the multivariate density space $\mathcal{F}$ via 
\begin{equation}
    \mathcal{R}_\tau^{-1}\left(\lambda\right)(z) =
\begin{cases}
\left|\check{\mathcal{R}}_\tau^{-1}(\lambda)(z)\right|/\int_{D}\left|\check{\mathcal{R}}_\tau^{-1}(\lambda)(z)\right|dz & \text{if} \int_{D}\left|\check{\mathcal{R}}_\tau^{-1}(\lambda)(z)\right|dz>0,\\
1 / |D| & \text{otherwise},
\end{cases}
\label{formula:regback}
\end{equation}
where $|D|$ is  the Lebesgue measure of the domain set $D$. Note that $\mathcal{R}_\tau^{-1}\left(\lambda\right)(z)$ satisfies the differentiability assumption in (F1) while the boundedness assumption can be enforced by projecting to the space where (F1) is satisfied. 

If $\mathcal{R}(f)$ is approximated by $\widetilde{\mathcal{R}(f)}$, the reconstruction function $\widetilde{f}_\tau = \mathcal{R}^{-1}_\tau\circ\widetilde{\mathcal{R}(f)}$ will approximate the original function $f$ with the reconstruction error represented as the sum of two error terms
\begin{align*}
\Delta_\tau = f-\widetilde{f}_\tau = \left(f-\mathcal{R}_\tau^{-1}\circ\mathcal{R}(f)\right) + \mathcal{R}_\tau^{-1}\circ\left(\mathcal{R}(f)-\widetilde{\mathcal{R}(f)}\right).
\end{align*}
Let  $\Delta_{\tau,1} = f-\mathcal{R}_\tau^{-1}\circ\mathcal{R}(f)$ and $\Delta_{\tau,2} = \mathcal{R}_\tau^{-1}\circ\left(\mathcal{R}(f)-\widetilde{\mathcal{R}(f)}\right) $.
\begin{theorem}
    \label{thm:reginv}
Assume (D1), (F1) and (F3). If $f\in\mathcal{F}$, as $d_2\left({\mathcal{R}(f)}, \widetilde{\mathcal{R}(f)}\right)\rightarrow 0$ and $\tau\rightarrow\infty$,
\begin{gather*}
    \|\Delta_{\tau,1}\|_\infty = O\left(\tau^{-(k-p)}\right), \quad
    \|\Delta_{\tau,2}\|_\infty = O\left(\tau^{p}d_2\left(\mathcal{R}(f),\widetilde{\mathcal{R}(f)}\right)\right),\\
    \|\Delta_{\tau}\|_\infty = O\left(\tau^{-(k-p)} + \tau^{p}d_2\left(\mathcal{R}(f),\widetilde{\mathcal{R}(f)}\right)\right).
\end{gather*}
\end{theorem}

The first error term $\Delta_{\tau,1}$ arises from the regularized inverse map and  decreases with $\tau$. The convergence of $\Delta_{\tau,1}$ relies on the smoothness assumption for densities, where higher order smoothness corresponds to a faster convergence rate of the Fourier transform in the frequency domain. The second error term $\Delta_{\tau,2}$ results from the approximation of  $\mathcal{R}(f)$ and increases with $\tau$. The value of the tuning parameter $\tau$ is then ideally chosen to minimize the total error $\Delta_{\tau}$. While the focus of this paper is primarily on the Radon transform, there are other transforms that may also be of interest such as the circular Radon transform \citep{kuchment2006generalized} and the generalized Radon transform \citep{beylkin1984inversion, ehrenpreis2003universality}; see Section \ref{sec:adx:slicingTransform} in the Appendix for further details. 
 
Next, we consider a general {\it slicing transform} $\psi: \mathcal{F} \rightarrow \Lambda_{\Theta}$ that maps a multivariate density function on $D$ into a density slicing function indexed by the slicing parameter set $\Theta$ and satisfies
\begin{enumerate}[label=(T\arabic*), leftmargin=1cm]
\setcounter{enumi}{-1}
\item $\psi$ is injective.
\item For a constant ${C_0}$, $d_2\left(\psi (f_1)(\theta), \psi (f_2)(\theta)\right) \leq C_0 d_2(f_1, f_2)$, for all $f_1, f_2 \in \mathcal{F}$ and $\theta \in \Theta$.
\item An inverse transform $\psi^{-1}$ exists such that $\psi^{-1}\circ\psi(f) = f, \,\text{ for all }  f\in \mathcal{F}$.
\item There exists a sequence of approximating inverses $\psi_\tau^{-1}$ and constants $C_1(\tau)$, $C_2(\tau)$ such that 
\[d_\infty\left(\psi_\tau^{-1}\circ\psi(f), f\right) \leq C_1(\tau)\text{ for all }f \in \mathcal{F}\]
and 
\[d_\infty\left(\psi_\tau^{-1}\circ\psi(f), \psi_\tau^{-1}(\lambda)\right) \leq C_2(\tau) d_2\left(\psi(f), \lambda\right)\text{ for all }\lambda \in \Lambda_{\Theta}, f\in\mathcal{F}\text{ as }d_2\left(\psi(f), \lambda\right)\to 0,\] where $C_1(\tau)$ and $C_2(\tau)$ depend only on $\tau$, and $C_1(\tau)$ is decreasing to $0$ while  $C_2(\tau)$ is increasing as $\tau\to\infty$.
\end{enumerate}

Assumption (T1) is concerned with the continuity of the forward transform,  while Assumption (T2) provides the existence of an inverse transform from the image set $\psi(\mathcal{F})$ to $\mathcal{F}$.
Note that $\psi(\mathcal{F})$ is not required to cover the entire space $\Lambda_{\Theta}$ and the inverse transform $\psi^{-1}$ is only defined on  the image space $\psi(\mathcal{F})$ of  $\mathcal{F}$. The sequence of approximating inverses  $\psi_\tau^{-1}$ in Assumption (T3) is required when mapping elements in $\Lambda_{\Theta}$ that are not in  $\psi(\mathcal{F})$ back to $\mathcal{F}$. To be able to make use of a slicing transform for both forward and inverse transformations, we require an additional property and call a slicing transform {\it valid} if in addition to (T1)--(T3) it satisfies 
\begin{enumerate}[label=(T\arabic*), leftmargin=1cm]
\setcounter{enumi}{3}
\item Under (D1) and (F1), it holds that  $\psi(\mathcal{F}) \subset\Lambda_{\Theta}$ and (D2), (G1) are satisfied for 
 all $\lambda(\theta)$ for which $\lambda \in \psi(\mathcal{F}).$ 
\end{enumerate}

\begin{proposition}
\label{prop:radonT0T1}  The Radon transform 
    $\mathcal{R}$ is a valid slicing transform, i.e., it satisfies Assumptions (T0)--(T4) where  $C_1(\tau) = O(\tau^{-(k-p)})$ and $C_2(\tau) = O(\tau^{p})$ as $\tau \rightarrow \infty$.
\end{proposition}

\section{Distribution Slicing and Sliced Distance}
\label{sec:space}
As a formal device, we introduce the map $\varphi$ that assigns to any distribution its associated density function, and its inverse $\varphi^{-1}$, which maps a density function to the corresponding distribution. Recall from \eqref{formula:multiDensF} that $\mathcal{F}$ denotes the space of multivariate density functions on $\mathbb{R}^p$ with compact support $D$, and $\mathcal{G}$ is the space of univariate density functions. We define $\mathscr{F} := \varphi^{-1}(\mathcal{F})$ and $\mathscr{G} := \varphi^{-1}(\mathcal{G})$ as the corresponding spaces of multivariate and univariate probability measures, respectively. Let $G$ denote the map that assigns to each univariate distribution its quantile function, with $G^{-1}$ as its inverse. We then define $\mathcal{Q} := G(\mathscr{G})$ as the space of quantile functions corresponding to univariate distributions in $\mathscr{G}$.

The metric space $(\mathscr{F}, d)$ is obtained by equipping $\mathscr{F}$ with a metric $d$, such as the Fisher–Rao metric \citep{dai:22,mull:23} or the Wasserstein metric \citep{gibbs2002choosing}, the latter being closely related to the optimal transport problem \citep{kant:06:1,vill:03}.  
The $L^2$-Wasserstein metric between two distributions $\mu_1, \mu_2 \in \mathscr{F}$ is defined as
\begin{align} \label{wm} 
    d_W^2(\mu_1, \mu_2) = \inf_{\substack{\pi \in \mathcal{P}(\mu_1, \mu_2) \\ (Z_1, Z_2) \sim \pi}}
    \mathbb{E}\left( \|Z_1 - Z_2\|_2^2 \right),
\end{align}
where $\mathcal{P}(\mu_1,\mu_2)$ is the set of joint probability measures on $D \times D$ with marginals $\mu_1$ and $\mu_2$, and $Z_1, Z_2$ are $\mathbb{R}^p$-valued random variables distributed according to $\mu_1$ and $\mu_2$, respectively.

The $L^2$-Wasserstein metric \eqref{wm} can be equivalently 
expressed in terms of quantile functions,
\begin{align*}
    d_W^2(\nu_1, \nu_2) = \int_0^1\left(G^{-1}(\nu_1)(s) - G^{-1}(\nu_2)(s)\right)^2ds,\quad \text{ for all } \nu_1, \nu_2 \in \mathscr{G}.
\end{align*}
When the dimension $p$ of the random distribution is larger than $1$, i.e., $p\geq 2$, one does not have an analytic form of the Wasserstein distance and the algorithms to obtain it are complex \citep{rabin2011wasserstein, peyre2019computational}. 
The sliced Wasserstein distance \citep{bonneel2015sliced} is a computationally more efficient alternative 
that utilizes the Radon transform. It has become increasingly popular for multivariate distributions due to its attractive topological and statistical properties \citep{nadjahi2019asymptotic, nadjahi2020statistical, kolouri2016sliced}. 

\begin{figure}[tb]
    \single
    \centering
    \includegraphics[width=0.3\linewidth]{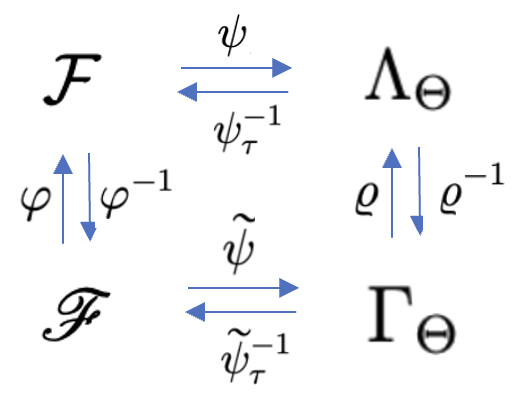}
    \caption{Schematic of maps involved in the slicing transform. Here $\varphi$ maps the multivariate distribution space $\mathscr{F}$ to the multivariate density space  $\mathcal{F}$ and $\varrho$ maps the {quantile slicing space} $\Gamma_{\Theta}$ to the density slicing space $\Lambda_{\Theta}$. 
    A slicing transform $\psi$ and its inverse $\psi_{\tau}^{-1}$ with a tuning parameter $\tau$ between $\mathcal{F}$ and $\Lambda_{\Theta}$ can be naturally extended to a slicing transform  between $\mathscr{F}$ and $\Gamma_{\Theta}$ through the induced maps $\widetilde{\psi}$ and $\widetilde{\psi}_{\tau}^{-1}$.}
    \label{fig:slicingTransform}
\end{figure}

We define the {quantile slicing space} $\Gamma_{\Theta}$ as a family of maps from  $\Theta$ to $\mathcal{Q}$,  
\begin{align*}
    \Gamma_{\Theta} = \left\{ \gamma: \Theta \rightarrow \mathcal{Q}, \,\int_{\Theta}\int_{[0,1]}\gamma(\theta)(s)^2dsd\theta<\infty\right\}
\end{align*}
and introduce a map  $\varrho: \Gamma_{\Theta} \rightarrow \Lambda_{\Theta}$ as $\varrho(\gamma)(\theta) = \varphi\circ G^{-1}(\gamma(\theta)), \gamma \in \Gamma_{\Theta}$, which sends a quantile function $\gamma(\theta)$ to its corresponding density function.  
Similarly, $\varrho^{-1}$ can be defined  through $\varrho^{-1}(\lambda)(\theta) = G\circ\varphi^{-1}(\lambda(\theta)),\, \lambda \in \Lambda_{\Theta}$; see Figure \ref{fig:slicingTransform} for a schematic illustration. A slicing transform $\psi$ between $\mathcal{F}$ and $\Lambda_{\Theta}$ can be naturally extended to a transform $\widetilde{\psi}$ between $\mathscr{F}$ and $\Gamma_{\Theta}$, 
\begin{align}
    \widetilde{\psi}(\mu) = \varrho^{-1}\circ\psi\circ\varphi(\mu), \quad \mu\in \mathscr{F}.
\end{align}
The inverse transform $\widetilde{\psi}^{-1}$ and regularized inverse transform $\widetilde{\psi}_{\tau}^{-1}$ can be extended as 
\begin{align}
    \widetilde{\psi}_{\tau}^{-1}(\gamma) = \varphi^{-1}\circ{\psi}_{\tau}^{-1}\circ\varrho(\gamma),\, \gamma \in \Gamma_{\Theta} \text{ and }\widetilde{\psi}^{-1}(\gamma) = \varphi^{-1}\circ\psi^{-1}\circ\varrho(\gamma),\, \gamma\in \widetilde{\psi}(\mathscr{F}). 
\end{align}

Note that the space of $\Gamma_{\Theta}$ is closed and convex. 
Define the distribution-slicing Wasserstein metric on $\Gamma_{\Theta}$ through the aggregated Wasserstein distance across the slices
\begin{align}
    d_{DW}(\gamma_1, \gamma_2) &= \left(\int_{\Theta} d_W^2\left({G^{-1}}(\gamma_1(\theta)), G^{-1}\left(\gamma_2(\theta)\right)\right)d\theta\right)^{1/2} \nonumber\\
    &= \left(\int_{\Theta} \int_{[0,1]}\left(\gamma_1(\theta)(s)-\gamma_2(\theta)(s)\right)^2dsd\theta\right)^{1/2}, 
    \quad\gamma_1, \gamma_2 \in \Gamma_{\Theta}.
\end{align}
Here the integral is well defined because of the Cauchy-Schwarz inequality. 
We then define the slice-averaged Wasserstein distance as
\begin{align}
     d_{SW}(\mu_1,\mu_2) = d_{DW}(\widetilde{\psi}(\mu_1), \widetilde{\psi}(\mu_2)) 
     ,\quad \text{ for all } \mu_1, \mu_2\in\mathscr{F}.
\end{align}
\begin{proposition}
    The slice-averaged Wasserstein distance is a distance over $\mathscr{F}$ if and only if (T0) is satisfied. 
\label{prop:well-defined}
\end{proposition}


\section{Sliced Wasserstein Regression for Multivariate Distributions}
\label{sec:regression}
A general approach for the regression of metric-valued responses and Euclidean predictors is Fr\'{e}chet regression with both global and local versions \citep{mull:19:3, chen2022uniform}, where its application to distributional data extensively discussed in recent studies \citep{fan2022conditional,zhou2023wasserstein}.
We extend  Fr\'{e}chet regression to the case of  multivariate distributions by
providing two types of sliced regression models. The first model applies Fr\'{e}chet regression  to the multivariate distribution space $\mathscr{F}$ equipped with the slice-averaged Wasserstein distance,  while the second model applies a  Fr\'{e}chet regression step for each slice, followed by the regularized inverse transform. 

Suppose $(X, \mu) \sim F$ is a random pair, where predictors $X$ and responses $\mu$ take values in $\mathbb{R}^q$ and $\mathscr{F}$ and $F$ is their joint distribution. 
Denote the marginal distributions of $X$ and $\mu$ as $F_X$ and $F_\mu$ respectively, and assume that  mean $E(X)$ and  variance $\Var(X)$ exist, where $\Var(X)$ is positive definite. The conditional distributions $F_{\mu|X}$ and $F_{X|\mu}$ are also assumed to exist. Given any metric $d$ on $\mathscr{F}$, 
Fr\'{e}chet mean and Fr\'{e}chet variance of the random distribution $\mu$ are defined as
\begin{align}
    \omega_{\oplus} = \argmin_{\omega\in\mathscr{F}}E\left[
    d^2(\mu,\omega)\right],\quad  V_{\oplus} = E\left[ d^2(\mu, \omega_{\oplus})\right].
\end{align}
The conditional Fr\'{e}chet mean, i.e., the  
regression function of $\mu$ given $X=x$, targets 
\begin{align}
\label{formula:Frechet}
    m(x) = \argmin_{\omega\in\mathscr{F}}M(\omega, x), \quad M(\cdot,x) = E\left[d^2(\mu,\cdot)|X=x\right], 
\end{align}
where $M(\cdot,x)$ is  the conditional Fr\'{e}chet function.
The global Fr\'{e}chet regression given $X=x$ is 
\begin{align}
\label{formula:glob}
     m_G(x) = \argmin_{\omega\in\mathscr{F}}M_G(\omega,x) , \quad M_G(\cdot, x) = E\left[s_G(X,x)d^2(\mu,\cdot) \right],
\end{align}
where the weight $s_G(X,x) = 1+(X-E(X))^\intercal\Var(X)^{-1}(x-E(X))$ is linear in $x$. 

The proposed slice-averaged Wasserstein (SAW) regression employs Fr\'{e}chet regression over the multivariate distribution space $\mathscr{F}$ equipped with the slice-averaged Wasserstein distance 
\begin{align}
    \label{formula:swFrechet}
    m^{SAW}(x) = \argmin_{\omega\in\mathscr{F}}M^{SAW}(\omega, x), \quad M^{SAW}(\cdot,x) = E\left[d_{SW}^2(\mu,\cdot)|X=x\right].
\end{align}
{The global slice-averaged Wasserstein (GSAW) regression} given $x$ is defined as 
\begin{align}
\label{formula:swGlob}
     m_G^{SAW}(x) = \argmin_{\omega\in\mathscr{F}}M_G^{SAW}(\omega,x) , \quad M_G^{SAW}(\cdot, x) = E\left[s_G(X,x)d_{SW}^2(\mu,\cdot) \right].
\end{align}

The second proposed model is the slice-wise Wasserstein (SWW) regression, where  Fr\'{e}chet regression is applied over the quantile slicing space $\Gamma_{\Theta}$, equipped with distribution-slicing Wasserstein metric, followed by the regularized inverse transform,
\begin{gather}
    m^{SWW}_{\tau}(x) = \widetilde{\psi}^{-1}_{\tau}\left[\argmin_{\gamma\in\Gamma_{\Theta}}M^{SWW}(\gamma, x)\right],\quad M^{SWW}(\cdot,x) = E\left[d_{DW}^2(\widetilde{\psi}(\mu),\cdot)|X=x\right]. 
    \label{formula:transformFrechetReg} 
\end{gather}
The global slice-wise Wasserstein (GSWW) regression given $X=x$ is
\begin{gather}
    m_{G,\tau}^{SWW}(x) = \widetilde{\psi}_{\tau}^{-1}\left[ \argmin_{\gamma\in\Gamma_{\Theta}}M_G^{SWW}(\gamma,x) \right], \quad M_G^{SWW}(\cdot, x) = E\left[s_G(X,x)d_{DW}^2(\widetilde{\psi}(\mu),\cdot) \right]. 
    \label{formula:transformGlobReg} 
\end{gather}
We note that for SAW based models the minimization is carried out over the space $\mathscr{F}$ and thus is automatically a multivariate distribution, 
while for SWW based models the minimization is carried out slicewise and as the slice-wise minimizers are not guaranteed to be situated in $\psi(\mathscr{F})$, a regularized inverse is needed since the inverse $\psi^{-1}$ is only defined on $\psi(\mathscr{F})$. 
\begin{proposition}
\label{prop:transformFrechet}
    Let $\gamma_{G,x} = \argmin_{\gamma\in\Gamma_{\Theta}}M_G^{SWW}(\gamma, x)$, see \eqref{formula:transformGlobReg}. It can be characterized as 
    \begin{align*}
        \gamma_{G,x}(\theta) = \argmin_{\nu\in\mathscr{G}}E \left[s_G(X,x)d_W^2\left(G^{-1}\left(\widetilde{\psi}(\mu)(\theta)\right), \nu\right)  \right],\quad {\text{for almost all }} \theta \in \Theta.
    \end{align*}
\end{proposition}

This characterization of SWW regression provides a practical implementation of the method. Proposition~\ref{prop:adx:Connection} in the Appendix shows that the search space of SAW is a subset of that of SWW. In addition to the global Fr\'echet regression described above, we also consider local Fr\'echet regression variants for both SAW and SWW. In these local versions, the global weight function $s_G(X, x)$ is replaced by a localized weight function $s_L(X, x, h)$, which depends on a kernel function $K$ and a bandwidth parameter $h>0$. This modification allows the regression to adapt to local neighborhoods of the predictor space, giving higher weight to observations whose predictors are close to the target predictor. The resulting methods are referred to as LSAW for the local version of SAW and LSWW for the local version of SWW. Further implementation details of these local variants are provided in Section~\ref{sec:adx:local} in the Appendix. Overall, the proposed methodology encompasses four regression approaches: GSWW, GSAW, LSWW, and LSAW.

\section{Estimation}
\label{sec:estimation}
Suppose we have a sample of independent random pairs $\{(X_i, \mu_i)\}_{i=1}^n\sim F$. Define the sample mean and covariance as
\[
\bar{X} = n^{-1}\sum_{i=1}^n X_i, 
\quad 
\hat{\Sigma}=n^{-1}\sum_{i=1}^n (X_i-\bar{X})(X_i-\bar{X})^\intercal.
\]

When the random distributions $\mu_i$ are fully observed on domain $D$, sample-based estimators for GSAW and GSWW are given by
\begin{gather}
     \check{m}_G^{SAW}(x) 
     = \argmin_{\omega\in\mathscr{F}}\check{M}_G^{SAW}(\omega,x), 
     \quad 
     \check{M}_G^{SAW}(\cdot, x) 
     = n^{-1}\sum_{i=1}^n s_{iG}(x)\,d_{SW}^2(\mu_i,\cdot), 
     \label{formula:sample0:swGlob}\\
     \check{m}_{G,\tau}^{SWW}(x) 
     = \widetilde{\psi}_{\tau}^{-1}\!\left[
         \argmin_{\gamma\in\Gamma_{\Theta}}
         \check{M}_G^{SWW}(\gamma,x)
       \right], 
     \quad 
     \check{M}_G^{SWW}(\cdot, x) 
     = n^{-1}\sum_{i=1}^n s_{iG}(x)\,d_{DW}^2(\widetilde{\psi}(\mu_i),\cdot), 
     \label{formula:sample0:transformGlob}
\end{gather}
where $s_{iG}(x) = 1 + (X_i-\bar{X})^\intercal\hat{\Sigma}^{-1}(x-\bar{X})$ 
is the sample version of $s_G(X,x)$.

In many applications, the random distributions $\{\mu_i\}_{i=1}^n$ are not directly observed; instead, only samples generated from them are available. 
In such cases, we first estimate the densities $\hat{\mu}_i$ using kernel density estimation as described in Section~\ref{sec:adx:dens} of the Appendix, which also lists the corresponding assumptions for this setting. 
The estimators in \eqref{formula:sample0:swGlob}–\eqref{formula:sample0:transformGlob} then apply with $\mu_i$ replaced by $\hat{\mu}_i$, that is,
\begin{gather}
\hat{m}_G^{SAW}(x)
= \check{m}_G^{SAW}(x)\big|_{\mu_i=\hat{\mu}_i},
\label{formula:sample:swGlob}\\
\hat{m}_{G,\tau}^{SWW}(x)
= \check{m}_{G,\tau}^{SWW}(x)\big|_{\mu_i=\hat{\mu}_i}.
\label{formula:sample:transformGlobReg}
\end{gather}
Sample-based estimators for the local smoothing models and bandwidth selection procedures are discussed in Section~\ref{subsec:adx:estimation} of the Appendix.

A practical, data-driven method for selecting the tuning parameter $\tau$ in the presence of an i.i.d. sample of random pairs $\{(X_i, \mu_i)\}_{i=1}^n$ is to use leave-one-out cross-validation. Specifically, we aim to minimize the discrepancy between predicted and observed distributions, given by
\begin{align*}
    \hat{\tau} = \argmin_{\tau}\sum_{i=1}^nd_{SW}^2({\mu_i}, \hat{m}_{G,\tau,-i}^{SWW}(X_i)),
\end{align*}
where $\hat{m}_{G,\tau,-i}^{SWW}(X_i)$ is the prediction at $X_i$ from the GSWW regression of the $i$th-left-out sample $\{(X_{i^{'}}, \mu_{i^{'}})\}_{i^{'}\neq i}$. When the sample size $n$ exceeds $30$, we substitute leave-one-out cross-validation with 5-fold cross-validation to strike a balance between computational efficiency and accuracy. 

We define a slice-averaged Wasserstein $R^2$ coefficient to quantify the discrepancy between observed distributions and predicted distributions from the regression model as 
\begin{align*}
    R_{\oplus}^2 = 1 - \frac{E[d_{SW}^2(\mu, m(X))]}{E[d_{SW}^2(\mu, \mu_{\oplus})]},
\end{align*}
where $m(x)$ is a regression function through either SAW or SWW regression and $\mu_{\oplus}$ is the slice-averaged Wasserstein Fr\'{e}chet mean $\mu_{\oplus} = E\left[\argmin_{\omega\in\mathscr{F}}d_{SW}^2(\mu, \omega)\right]$, with empirical estimates
\begin{align}
    \hat{R}_{\oplus}^2 = 1 - \frac{\sum_{i=1}^nd_{SW}^2({\mu}_i, \hat{m}(X_i))}{\sum_{i=1}^nd_{SW}^2({\mu}_i, \hat{\mu}_{\oplus})},
\label{formula:SWFVE}
\end{align}
for a sample $\{(X_i, \mu_i)\}_{i=1}^n$. Here  $\hat{\mu}_{\oplus} = \argmin_{\omega\in\mathscr{F}}n^{-1}\sum_{i=1}^nd_{SW}^2({\mu}_i, \omega)$ is the sample Fr\'{e}chet mean and $\hat{m}(x)$ is the sample estimator of the regression function $m(x)$ for SAW or SWW regression.  

\section{Asymptotic Convergence}
\label{sec:convergence}
To study the convergence of SAW/SWW regression,  we require a curvature condition at the true minimizer, as is commonly assumed for M estimators. For Fr\'{e}chet regression such a curvature condition has been established for the case of 
univariate distributions but not for multivariate distributions. Convexity assumptions have also been invoked in previous work \citep{fan2021conditional, boissard2011distribution,zhou2022network}. Specifically, we require 
\begin{enumerate}[label=(A\arabic*), leftmargin=1cm]
\item $\widetilde{\psi}(\mathscr{F})$ is closed and convex.
\item $\argmin_{\gamma\in\Gamma_{\Theta}}M_G^{SWW}(\gamma,x)$ as per \eqref{formula:transformGlobReg} is in $\widetilde{\psi}(\mathscr{F})$. 
\end{enumerate}
Assumption (A1) guarantees the existence and uniqueness of the minimizer of the SAW regression, while Assumption (A2) ensures the underlying minimizer of the GSWW regression belongs to the image space of the slicing transform. The following results are for GSAW and GSWW for the two scenarios where densities are known or must be estimated from data. In the latter case, we require additional assumptions (P1), (K1), (K2) and ($\text{F1}^\prime$) that are provided in the Appendix Section \ref{sec:adx:dens}, which also contains details about the construction of the density estimates, as well as the results for LSAW and LSWW regression in Theorems \ref{thm:adx:swLoc0} and \ref{thm:adx:transformLoc0} for known densities and in Theorems \ref{thm:adx:swLoc} and \ref{thm:adx:transformLoc} for the case where densities must be estimated.
\begin{figure}[t]
    \single
    \centering
    \includegraphics[width=0.8\linewidth]{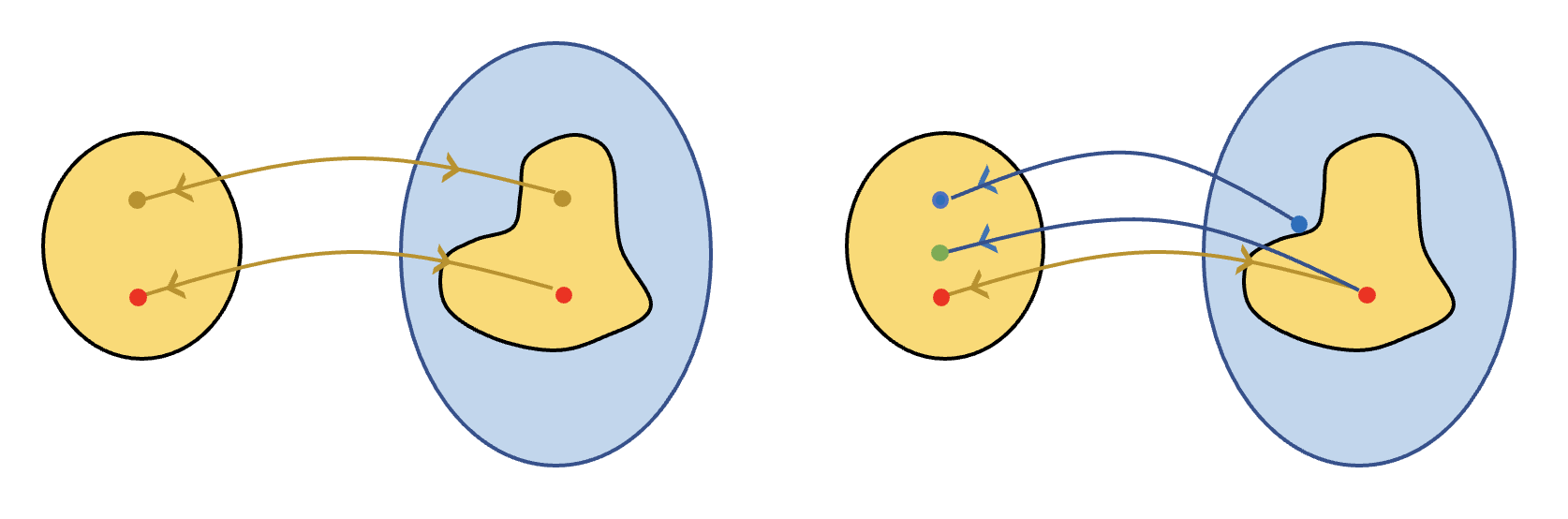}
    \caption{Diagram for SAW (left) and SWW (right) sample estimation. The red dots are population minimizers, the yellow dots are the sample SAW minimizer, the blue dots are the sample SWW minimizers and the green dot is the intermediate state used to facilitate the error analysis of SWW. Bidirectional yellow maps represent $\widetilde{\psi}$ and $\widetilde{\psi}^{-1}$ while directional blue maps  represent the regularized inverse $\widetilde{\psi}_{\tau}^{-1}$. }
    \label{fig:minimizer}
\end{figure}

\begin{theorem}[Convergence of GSAW regression]
\label{thm:swGlob}
    Assume (D1), (F1)--(F2), (A1) and (T0)--(T2). For a fixed $x\in \mathbb{R}^q$, it holds for $m_G^{SAW}(x)$, $\check{m}_G^{SAW}(x)$ as per \eqref{formula:swGlob}, \eqref{formula:sample0:swGlob} that 
    \begin{align*}
        d_{SW}\left(m_G^{SAW}(x), \check{m}_G^{SAW}(x)\right) = O_p(n^{-1/2}), \\  
        \sup_{\|x\|_E\leq B}d_{SW}\left(m_G^{SAW}(x), \check{m}_G^{SAW}(x)\right) = O_p(n^{-1/(2+\epsilon)}),    
    \end{align*}
    for a given $B>0$ and any $\epsilon>0$. When densities are estimated, the same rates hold for $\hat m_G^{SAW}(x)$ as per \eqref{formula:sample:swGlob} under (P1), $(\mathrm{F1}')$, (K1)–(K2) in Section \ref{sec:adx:dens} of the Appendix.
\end{theorem}

The GSAW thus yields a parametric convergence rate that does not depend on the dimension of the distribution space. Since the regularized inverse transform $\widetilde{\psi}_{\tau}^{-1}$ is continuous due to (T3), it is possible to obtain sup norm convergence for SWW. Under Assumption (A2), the population-level target $m_{G}^{SWW}$ is 
\begin{gather}
    m_{G}^{SWW} = \widetilde{\psi}^{-1}\left[\argmin_{\gamma\in\Gamma_{\Theta}}M_G^{SWW}(\gamma, x)\right], \label{formula:transformGlob}
\end{gather}
where $M_{G}^{SWW}$ is as in \eqref{formula:transformGlobReg}.

\begin{theorem}[Convergence of GSWW regression]
\label{thm:transformGlob}
Assume (D1), (F1), (F3), (A2), (T0)--(T4). For a fixed $x\in\mathbb{R}^q$,  for $m_G^{SWW}(x)$, $\check{m}_{G,\tau}^{SWW}(x)$ as per \eqref{formula:transformGlob}, \eqref{formula:sample0:transformGlob}, $C_1(\tau)$ and $C_2(\tau)$ from (T3),  
\begin{align*}
        d_{\infty}\left( m_{G}^{SWW}(x), \check{m}_{G,\tau}^{SWW}(x)\right) = O_p\left({C_1(\tau) + C_2(\tau)}{n^{-2/7}} \right).
\end{align*}
Furthermore, for a given $B>0$ and any $\epsilon>0$,
\begin{align*}
    \sup_{\|x\|_E\leq B}d_{\infty}\left( m_{G}^{SWW}(x), \check{m}_{G,\tau}^{SWW}(x)\right) = O_p\left(C_1(\tau) + C_2(\tau)n^{-2/(7+\epsilon)} \right).
\end{align*}
For estimated densities  one obtains the same rates for $\hat{m}_{G,\tau}^{SWW}(x)$ as per \eqref{formula:sample:transformGlobReg} under (P1), ($\text{F1}^\prime$), (K1)--(K2) in Section \ref{sec:adx:dens} of the Appendix.
\end{theorem}
The reconstruction error has two parts. The first term $C_1(\tau)$ is linked to the approximation bias in the reconstruction, while the second term $C_2(\tau)$ ensures that the approximation in the transformed space $\Gamma_{\Theta}$ is not excessively amplified (see Figure \ref{fig:minimizer}). If the slicing transform is the Radon transform, Corollary \ref{cor:RadonGlob} shows that the curse of dimensionality is manifested in both $C_1(\tau)$ and $C_2(\tau)$, as the required order of smoothness increases with the dimension of the distribution to achieve the same convergence rate. We only provide the more intricate scenario when densities are not fully observed; analogous results hold for the case of known densities. 

\begin{corollary}
\label{cor:RadonGlob}
    When taking the Radon transform $\mathcal{R}$ and the corresponding regularized inverse $\mathcal{R}^{-1}_{\tau}$ as per \eqref{formula:radon} and \eqref{formula:regback}, under  the assumptions of Theorem \ref{thm:transformGlob},
    \begin{gather*}
        d_{\infty}\left( m_{G}^{SWW}(x), \hat{m}_{G,\tau}^{SWW}(x)\right) = O_p\left(\tau^{-(k-p)} + \tau^p{n^{-2/7}} \right), \\
        \sup_{\|x\|_E\leq B}d_{\infty}\left( m_{G}^{SWW}(x), \hat{m}_{G,\tau}^{SWW}(x)\right) = O_p\left(\tau^{-(k-p)} + \tau^pn^{-2/(7+\epsilon)} \right),
    \end{gather*}
    and with $\tau \sim n^{2/7k}$,
    \begin{gather*}
        d_{\infty}\left( m_{G}^{SWW}(x), \hat{m}_{G,\tau}^{SWW}(x)\right) = O_p\left(n^{-2\frac{k-p}{7k}}\right), \\
        \sup_{\|x\|_E\leq B}d_{\infty}\left( m_{G}^{SWW}(x), \hat{m}_{G,\tau}^{SWW}(x)\right) = O_p\left(n^{-2\frac{k-p}{7k+\epsilon}}\right).
    \end{gather*}
\end{corollary}



\section{Numerical Algorithm}
\label{sec:numerical}
\subsection{SAW Regression}
To solve problem \eqref{formula:sample:swGlob} in practice,  we use a numerical optimization process proposed in \citet{bonneel2015sliced} that involves parametrizing a probability measure with equal weights. Specifically, we use random observations  $\mathbf{W}=\{\mathbf{W}_j\}_{j=1}^N\in\mathbb{R}^{p\times N}$ where each observation ${\mathbf{W}}_j \in \mathbb{R}^p$ follows a distribution characterized by a density function $f_{\mathbf{W}}$ in $\mathcal{F}$.  
We introduce the Radon slicing operation $\langle \cdot, \theta \rangle: \mathbb{R}^p\rightarrow \mathbb{R}$ for each random observation, resulting in a univariate distribution with the density function $\mathcal{R}\circ f_{\mathbf{W}}(\theta)$. The multivariate distribution $\mu$ corresponding to the density $f_{\mathbf{W}}$ can be represented as a discrete input measure $\mu_{\mathbf{W}}= \frac{1}{N}\sum_{j=1}^N\delta_{\mathbf{W}_{j}}$. Similarly, we represent the $i$-th distribution $\mu_i$ as a discrete input measure $\mu_{\mathbf{W}^{(i)}}=\frac{1}{N}\sum_{j=1}^N\delta_{\mathbf{W}_{ij}}$ where $\mathbf{W}^{(i)} = \{\mathbf{W}_{ij}\}_{j=1}^{n_i}\in\mathbb{R}^{p\times N}$ and $\mathbf{W}_{ij}\sim\mu_i$. 
The GSAW regression of \eqref{formula:sample:swGlob} given $X=x$ can be represented as
\begin{align}
\label{formula:SAW_Glob_Numeric}
    \argmin_{\mathbf{W}\in\mathbb{R}^{p\times N}}\mathcal{M}_G(\mathbf{W},x) = n^{-1}\sum_{i=1}^n\left[ s_{iG}(x)d_{SW}^2(\mu_{\mathbf{W}^{(i)}}, \mu_{\mathbf{W}})\right]. 
\end{align}

It is known that the target function is smooth for the Radon transform, a result that we state below for reference. We use the notation  $\mathbf{W}(\theta) = (\langle \mathbf{W}_j, \theta\rangle)_{j=1}^N\in \mathbb{R}^N$ and similarly $\mathbf{W}^{(i)}(\theta) = (\langle \mathbf{W}_{ij}, \theta\rangle)_{j=1}^N\in \mathbb{R}^N$. For any $A=\{A_j\}_{j=1}^N\in\mathbb{R}^N$ where $A_j\in\mathbb{R}$, $\Pi_{A}$ is a permutation operator on $A$, such that $\Pi_A(A) = (A_{\Pi(1)}, A_{\Pi(2)},...,A_{\Pi(N)})^\intercal$ with $A_{\Pi(1)} \leq A_{\Pi(2)} \leq \cdots \leq A_{\Pi(N)}$. 

\begin{proposition}[Theorem 1 \citep{bonneel2015sliced}]
\label{prop:alg}
    For each fixed $x$ and  $N_i\equiv N$, we have
    $\mathcal{M}_G(\mathbf{W},x):\mathbb{R}^{p\times N} \rightarrow \mathbb{R}$ is a $L^1$ function with a uniformly $\rho_G$-Lipschitz gradient for some $\rho_G > 0$ given by
    \begin{align*}
        \nabla \mathcal{M}_G(\mathbf{W}, x) = n^{-1}\sum_{i=1}^n\left[ s_{iG}(x)  
        \int_{\Theta}\theta\left(\mathbf{W}(\theta) - \Pi_{\mathbf{W}(\theta)}^{-1}\circ \Pi_{\mathbf{W}^{(i)}(\theta)}\circ \mathbf{W}^{(i)}(\theta)\right)^\intercal d\theta 
        \right].
    \end{align*}
\end{proposition}
The operation $\Pi_{\mathbf{W}(\theta)}^{-1}\circ \Pi_{\mathbf{W}^{(i)}(\theta)}\circ \mathbf{W}^{(i)}(\theta)$ aligns $\mathbf{W}^{(i)}(\theta)$ with $\mathbf{W}(\theta)$ in the sense of non-decreasing order for calculating the empirical Wasserstein distance.  In practice, an equidistant grid $(\theta_1,\theta_2,...,\theta_L)$ along the angular coordinate $\theta$ 
is used to approximate the integration over $\Theta$. Note that \eqref{formula:SAW_Glob_Numeric} is non-convex and we use gradient descent to find a stationary point; see Algorithm \ref{alg:SAW}. 

Since the algorithm implied by Proposition \ref{prop:alg} requires that the sample sizes $N_i$ at which each distribution is sampled are identical,  we set $N = \min_{i=1,...,n}N_i$ in step 2 of the following algorithm.  In instances where $N_i>N$, we choose a randomly selected subsample of size $N$, referred to as downsampling in the following. For an additional discussion about the local modeling approach for SAW, we refer to Section \ref{subsec:adx:alg} in the Appendix. 

\begin{algorithm}[htbp]
\single
\caption{GSAW Algorithm when using the Radon Transform} 
\begin{algorithmic}[1]
    \State Initialize a grid $(\theta_1,\theta_2,...,\theta_L)$ along $\Theta$
    \State Set $N = \min_{i=1,...,n}N_i$, convergence threshold $\varepsilon$ and learning rate $\eta$
    \State For each $\mu_{\mathbf{W}^{(i)}}$, downsample $\mathbf{W}^{(i)}$ such that  all $\mathbf{W}^{(i)} \in \mathbb{R}^{p\times N}$
    \State Initialize $\mathbf{W}^{[0]}\in\mathbb{R}^{p\times N}$ arbitrarily and fix the output predictor $X=x$
    \Repeat 
        \State Calculate $\nabla 
        \mathcal{M}_G(\mathbf{W}^{[k]}, x)$
        \begin{align*}
            \nabla 
        \mathcal{M}_G(\mathbf{W}^{[k]}, x) =  (nL)^{-1}\sum_{i=1}^n 
        \sum_{l=1}^L
        s_{iG}(x)  \theta_l\left(\mathbf{W}(\theta_l) - \Pi_{\mathbf{W}(\theta_l)}^{-1}\circ \Pi_{\mathbf{W}^{(i)}(\theta_l)}\circ \mathbf{W}^{(i)}(\theta_l)\right)^\intercal
        \end{align*}
        \State {$\mathbf{W}^{[k+1]} = \mathbf{W}^{[k]} - \eta \nabla 
        \mathcal{M}_G(\mathbf{W}^{[k]}, x) $} 
    \Until Algorithm converges with $\left\|\mathbf{W}^{[k+1]} - \mathbf{W}^{[k]} \right\|_2 / \left\| \mathbf{W}^{[k]}\right\|_2 < \varepsilon$
    to $\mathbf{W}^{[\infty]}$
    \State  
    Consider each column of $\mathbf{W}^{[\infty]}$ as a sample from $\hat{m}_G^{SAW}(x)$ and 
    apply the kernel density estimator \eqref{formula:adx:kerndens} in the Appendix to derive the density estimator $\hat{f}$ 
	\end{algorithmic}
\label{alg:SAW}
\end{algorithm}

\subsection{SWW Regression}
Since SWW regression is split into completely separate slicewise regression steps, 
one can leverage parallel computing to enhance computing efficiency; the permutation operator is not needed. We note that the local model for the SWW approach can be implemented analogously simply by using local instead of global Fr\'{e}chet regression in the algorithm. 

\begin{algorithm}[htbp]
\single
\caption{GSWW Algorithm for the Radon Transform} 
	\begin{algorithmic}[1]
    \State Initialize a grid $(\theta_1,\theta_2,...,\theta_L)$ along $\Theta$ and fix the output predictor $X=x$
    \State Perform a Radon slicing operation over random observations, i.e. $\langle \mathbf{W}^{(i)}, \theta_l\rangle$, $i=1,...,n, l=1,...,L$
    \For{$l=1,...,L$} in parallel
        \State Apply the Fr\'{e}chet regression for sliced random pairs, $\left(X_i, \langle\mathbf{W}^{(i)}, \theta_l\rangle\right)$, $i=1,...,n$ \citep{mull:19:3, yaqing2013frechet}
        \State Calculate the fitted density on the output $X=x$ as $\hat{\lambda}(\theta_l)$
    \EndFor 
    \State Apply the regularized Radon reconstruction $\hat{f} = \mathcal{R}_{\tau}^{-1}(\hat{\lambda})$ through \eqref{formula:regback}
	\end{algorithmic}
\label{alg:SWW}
\end{algorithm}

\section{Simulation Study}
\label{sec:simulation}
To evaluate the finite-sample performance of the proposed methods, we design a generative framework that produces random $p$-dimensional probability distributions $\mu$, together with a scalar predictor $X$ from a uniform distribution on $[-0.5, 0.5]$.
We begin with multivariate Gaussian responses, where both the mean vector and covariance matrix depend on $X$. Specifically, given $X=x$, the mean vector is generated as $\mathcal{N}(\alpha(x), I_p)$, with $\alpha(x)\in \mathbb{R}^p$, and the covariance matrix is sampled from a Wishart distribution $\mathcal{W}_p(\mathscr{D}(x), p+1)$, with degrees of freedom $p+1$ and scale matrix $\mathscr{D}(x) \in \mathbb{R}^{p\times p}$.  

To generate more complex non-Gaussian responses, we apply a random transport map to the Gaussian distributions constructed above. Specifically, each coordinate of the multivariate distribution is pushed forward by a map $T$, chosen uniformly at random from $T_k(z)=z-\sin(kz)/|k|$ for $k\in\{-2, -1, 1, 2\}$ \citep{pana:16}. The map $T$ is non-decreasing and has expectation equal to the identity. This transformation preserves the true regression function while significantly complicating the distributional structure, yielding non-Gaussian multivariate distributional responses.  

We consider both Gaussian and non-Gaussian distributional settings, each with two different dimensions ($p=2$ and $p=5$), and further distinguish between global and local regression models. This results in a total of eight simulation settings. Table~\ref{tab:simsetting} summarizes all cases: Settings~I--IV correspond to Gaussian responses with parameterizations depending on $X=x$, while Settings~V--VIII represent their non-Gaussian counterparts, obtained by applying a random transport map $T$ to the Gaussian distributions.

\begin{table}[tb]
    \single
    \small
    \centering
    \caption{Simulation settings. Settings I--IV correspond to multivariate Gaussian distributional responses with distributional parameters depending on $X=x$ as indicated. 
    Settings V--VIII are identical to I--IV, but with an additional random transport map $T$, sampled from $\{T_k(z)=z-\sin(kz)/k : k\in\{-2,-1,1,2\}\}$, applied to generate non-Gaussian multivariate distributional responses.}
    \label{tab:simsetting}
    \begin{tabular}{lllll}
        \toprule
         & Distribution & Dimension & Regression & Setting \\\midrule
        I & \multirow{4}{*}{Gaussian} & \multirow{2}{*}{$p=2$} & GSWW, GSAW & \makecell[cl]{$\alpha(x)=(x, x)^\intercal$,\\ $\mathscr{D}(x)=\diag(x+1, x+1)$}\\\cmidrule(lr){4-5}
        II & & & LSWW, LSAW & \makecell[cl]{$\alpha(x)=\frac{1}{2}(\sin(\frac{\pi}{2}x), \frac{1}{2}\sin(\frac{\pi}{2}x))^\intercal$,\\ $\mathscr{D}(x)=\diag(\cos(\frac{\pi}{2}x), \cos(\frac{\pi}{2}x))$} \\\cmidrule(lr){3-5}
        III & & \multirow{2}{*}{$p=5$} & GSWW, GSAW & \makecell[cl]{$\alpha(x)=(x, \ldots, x)^\intercal$,\\ $\mathscr{D}(x)=\diag(x+1, \ldots, x+1)$} \\\cmidrule(lr){4-5}
        IV & & & LSWW, LSAW & \makecell[cl]{$\alpha(x)=(\frac{1}{2}\sin(\frac{\pi}{2}x), \ldots, \frac{1}{2}\sin(\frac{\pi}{2}x))^\intercal$,\\ $\mathscr{D}(x)=\diag(\cos(\frac{\pi}{2}x), \ldots, \cos(\frac{\pi}{2}x))$} \\\cmidrule(lr){2-5}
        V--VIII & non-Gaussian & $p=2, 5$ & \makecell[cl]{GSWW, GSAW; \\ LSWW, LSAW} & \makecell[cl]{Same as Settings I--IV but with \\ a random transport map $T$} \\\bottomrule
    \end{tabular}
\end{table}

The proposed global (GSWW, GSAW) and local (LSWW, LSAW) regression methods were evaluated separately under each of the eight simulation settings. For comparison, we also included a comparison to a previous method \citep{fan2021conditional}, denoted as FM, which is based on the Sinkhorn distance \citep{cutu:13}. The approach in \citet{fan2021conditional} focuses on interpolation and extrapolation of univariate and bivariate distributions. Accordingly, comparisons with FM were performed only for the bivariate case ($p=2$), as its grid-based discretization suffers from the curse of dimensionality and becomes computationally infeasible for higher dimensions (e.g., a $50^5$ grid for $p=5$ would require excessive memory and runtime).

Simulations were conducted for sample sizes $n \in \{50, 100, 200\}$. For each distribution, $N=200$ observations were generated per distribution for $p=2$ and $N=500$ for $p=5$. Each configuration was replicated 100 times, and estimation accuracy for the $k$th Monte Carlo run was quantified by the integrated squared error (ISE):
\begin{align}
\label{formula:simulationMetric}
    \text{ISE}_k = \int_{-0.5}^{0.5} d_{SW}^2\left(\hat{m}_k(x),m(x)\right)dx,
\end{align}
where $m(x)$ is the true regression function and $\hat{m}_k(x)$ the fitted regression function from the $k$th run. For the local methods, bandwidths were set to $0.25n^{-1/5}$.

\begin{table}[tb]
\single
\footnotesize
\centering
\caption{Simulation results for three sample sizes across eight settings using the proposed regression methods (SWW and SAW) and the comparison method (FM) \citep{fan2021conditional}. The FM method becomes computationally infeasible for $p=5$ and is therefore omitted in that case. Each cell reports the mean (standard deviation) of the integrated squared error over 100 replications. Settings~I--IV correspond to Gaussian responses, while Settings~V--VIII correspond to non-Gaussian responses generated via a random transport map~$T$.}
\label{tab:simres}
\begin{tabular}{llllllll}
\toprule
 & Regression & Distribution & Dimension & $n$ & SWW & SAW & FM\\
\midrule
\multirow{3}{*}{I} 
 & \multirow{12}{*}{global} & \multirow{6}{*}{Gaussian} & \multirow{3}{*}{$p=2$} 
 & 50  & 0.080 (0.035) & 0.109 (0.032) & 0.460 (0.070)\\
 &  &  &  & 100 & 0.053 (0.021) & 0.086 (0.019) & 0.438 (0.059)\\
 &  &  &  & 200 & 0.037 (0.012) & 0.073 (0.011) & 0.416 (0.031)\\
\cmidrule(lr){4-8}
\multirow{3}{*}{III} 
 &  &  & \multirow{3}{*}{$p=5$} & 50  & 0.076 (0.020) & 0.115 (0.023) & ---\\
 &  &  &  & 100 & 0.041 (0.010) & 0.088 (0.016) & ---\\
 &  &  &  & 200 & 0.027 (0.006) & 0.076 (0.015) & ---\\
\cmidrule(lr){3-8}
\multirow{3}{*}{V} 
 &  & \multirow{6}{*}{non-Gaussian} & \multirow{3}{*}{$p=2$} 
 & 50  & 0.081 (0.037) & 0.109 (0.034) & 0.515 (0.126)\\
 &  &  &  & 100 & 0.047 (0.021) & 0.081 (0.019) & 0.481 (0.101) \\
 &  &  &  & 200 & 0.031 (0.012) & 0.067 (0.010) & 0.436 (0.049)\\
\cmidrule(lr){4-8}
\multirow{3}{*}{VII} 
 &  &  & \multirow{3}{*}{$p=5$} & 50  & 0.068 (0.019) & 0.110 (0.020) & ---\\
 &  &  &  & 100 & 0.033 (0.009) & 0.081 (0.010) & ---\\
 &  &  &  & 200 & 0.019 (0.005) & 0.071 (0.010) & ---\\
\cmidrule(lr){2-8}
\multirow{3}{*}{II} 
 & \multirow{12}{*}{local} & \multirow{6}{*}{Gaussian} & \multirow{3}{*}{$p=2$} 
 & 50  & 0.150 (0.054) & 0.126 (0.035) & 0.448 (0.063)\\
 &  &  &  & 100 & 0.086 (0.024) & 0.094 (0.019) & 0.414 (0.045)\\
 &  &  &  & 200 & 0.056 (0.014) & 0.076 (0.011) & 0.393 (0.026)\\
\cmidrule(lr){4-8}
\multirow{3}{*}{IV} 
 &  &  & \multirow{3}{*}{$p=5$} & 50  & 0.144 (0.027) & 0.131 (0.023) & ---\\
 &  &  &  & 100 & 0.074 (0.016) & 0.092 (0.016) & ---\\
 &  &  &  & 200 & 0.046 (0.007) & 0.077 (0.014) & ---\\
\cmidrule(lr){3-8}
\multirow{3}{*}{VI} 
 &  & \multirow{6}{*}{non-Gaussian} & \multirow{3}{*}{$p=2$} 
 & 50  & 0.163 (0.062) & 0.131 (0.038) & 0.505 (0.107)\\
 &  &  &  & 100 & 0.086 (0.026) & 0.092 (0.020) & 0.467 (0.093)\\
 &  &  &  & 200 & 0.055 (0.015) & 0.073 (0.011) & 0.422 (0.049)\\
\cmidrule(lr){4-8}
\multirow{3}{*}{VIII} 
 &  &  & \multirow{3}{*}{$p=5$} & 50  & 0.135 (0.024) & 0.125 (0.019) & ---\\
 &  &  &  & 100 & 0.068 (0.015) & 0.086 (0.009) & ---\\
 &  &  &  & 200 & 0.039 (0.006) & 0.074 (0.009) & ---\\
\bottomrule
\end{tabular}
\end{table}

The results for the proposed SWW and SAW approaches and the comparison method FM are shown in Table \ref{tab:simres}. Across all settings, the proposed SWW and SAW approaches substantially outperform FM, yielding considerably smaller mean ISE values. This improvement reflects the efficiency of slicing in capturing directional variations through one-dimensional projections. Furthermore, SWW and SAW demonstrate faster convergence as the sample size $n$ increases, whereas FM suffers from approximation errors due to full-grid discretization and entropic regularization. Between the two proposed methods, SWW shows superior performance compared to SAW, particularly for larger sample sizes. This advantage arises because SWW minimizes a convex objective function independently within each slice, ensuring a unique minimizer, whereas SAW relies on gradient descent to optimize a global non-convex objective over the multivariate distribution space. A visual comparison of the true and fitted bivariate densities obtained by SWW and SAW is provided in Figure~\ref{fig:simcom} in Section~\ref{sec:adx:figures} of the Appendix, where SWW is seen to produce a density estimate that is visually more accurate and better aligned with the true distribution.

\begin{table}[tb]
\single
\centering
\caption{Runtime (in seconds) for the proposed regression methods (SWW and SAW) and the comparison method (FM) \citep{fan2021conditional} across different dimensions and sample sizes. The FM method becomes computationally infeasible for $p=5$ and is therefore omitted.}
\label{tab:runtime}
\begin{tabular}{llllll}
\toprule
 Regression & Dimension & $n$ & SWW & SAW & FM\\
\midrule
\multirow{6}{*}{global} & \multirow{3}{*}{$p=2$} 
 & 50  & 0.44 & 8.22 & 59.38\\
 &  & 100 & 0.47 & 17.74 & 109.10\\
 &  & 200 & 1.42 & 35.00 & 285.62\\
\cmidrule(lr){3-6}
 & \multirow{3}{*}{$p=5$} & 50  & 2.08 & 33.50 & ---\\
 &  & 100 & 2.24 & 66.47 & ---\\
 &  & 200 & 4.83 & 130.45 & ---\\
\cmidrule(lr){2-6}
\multirow{6}{*}{local} & \multirow{3}{*}{$p=2$} 
 & 50  & 0.57 & 8.80 & 55.13\\
 &  & 100 & 0.66 & 18.38 & 107.47\\
 &  & 200 & 1.30 & 33.43 & 266.73\\
\cmidrule(lr){3-6}
 & \multirow{3}{*}{$p=5$} & 50  & 3.39 & 33.52 & ---\\
 &  & 100 & 2.34 & 66.44 & ---\\
 &  & 200 & 4.66 & 132.11 & ---\\
\bottomrule
\end{tabular}
\end{table}

Runtime comparisons for SWW, SAW, and FM are reported in Table~\ref{tab:runtime}. All experiments were run on a 16-inch MacBook Pro with an Apple M3~Max chip (14~CPU cores, 36~GB unified memory) running macOS~Tahoe~26.0.1. Parallelization was configured to use 10~CPU cores. We only report FM runtimes for the bivariate case, as the FM method becomes infeasible for $p=5$. As expected, both SWW and SAW are considerably faster than FM, which involves computationally expensive Sinkhorn iterations. Moreover, SWW is markedly faster than SAW because SAW requires gradient descent to compute the weighted barycenter based on the sliced Wasserstein distance. A related discussion of computational complexity for bivariate Wasserstein barycenters using Sinkhorn and sliced Wasserstein distances can be found in Section~5.3 of \citet{bonneel2015sliced}.

\section{Data Analysis}
\label{sec:data}
\subsection{Excess Winter Deaths in Europe}
\label{subsec:data:winter}

\begin{figure}[tb]
    \single
    \centering
    \includegraphics[width=\linewidth]{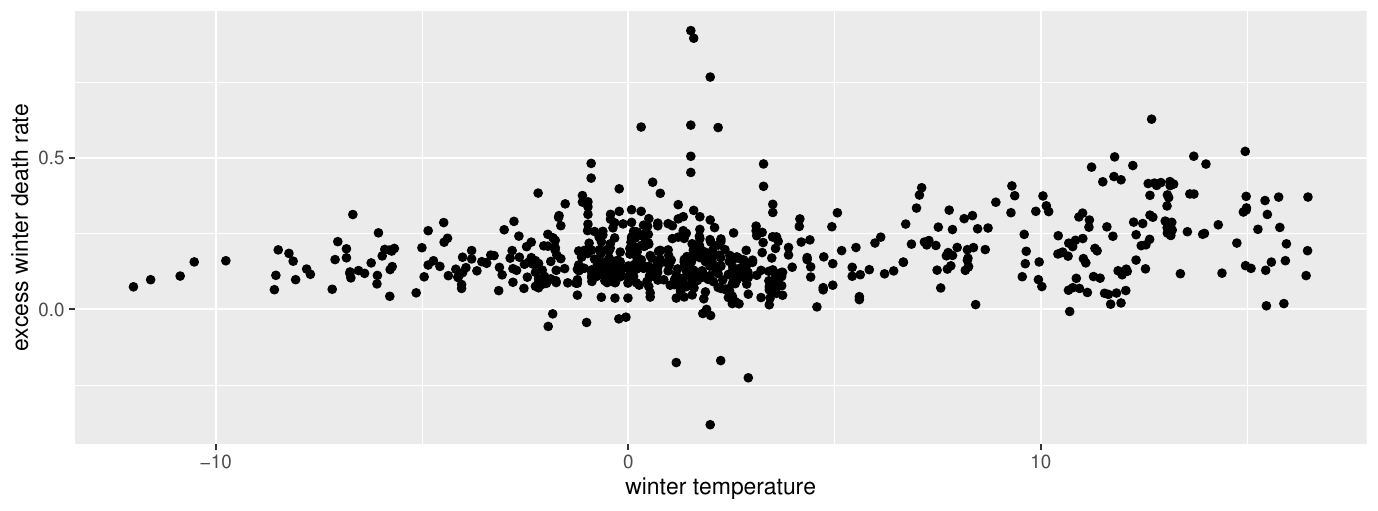}
    \caption{Scatterplot of excess winter death rates ($y$-axis) against  winter temperature ($x$-axis)}
    \label{fig:EUDeathScatter}
\end{figure}

Excess winter deaths are a social and health challenge. For  European countries, they have become an acute problem due to rapidly rising heating costs. It is known that in general Northern European countries have lower excess winter mortality rates compared to Southern European countries \citep{healy2003excess, fowler2015excess}. Our goal in this analysis was to follow up on this by modeling conditional distributions directly, rather than characterizing mortality effects through summary statistics such as sample mean or standard deviation,  as was done in previous studies. 
Simply applying the conventional regression model to the scatter plot between excess winter mortality rates and the absolute winter temperature (see Figure \ref{fig:EUDeathScatter}) makes the analysis susceptible to Simpson's paradox, creating a misleading perception of an association between increased excess winter death rates and higher winter temperature. A better approach is to separate the observations by country.

\begin{figure}[]
    \single
  \centering
  \begin{subfigure}{\textwidth}
    \centering
    \includegraphics[width=\linewidth]{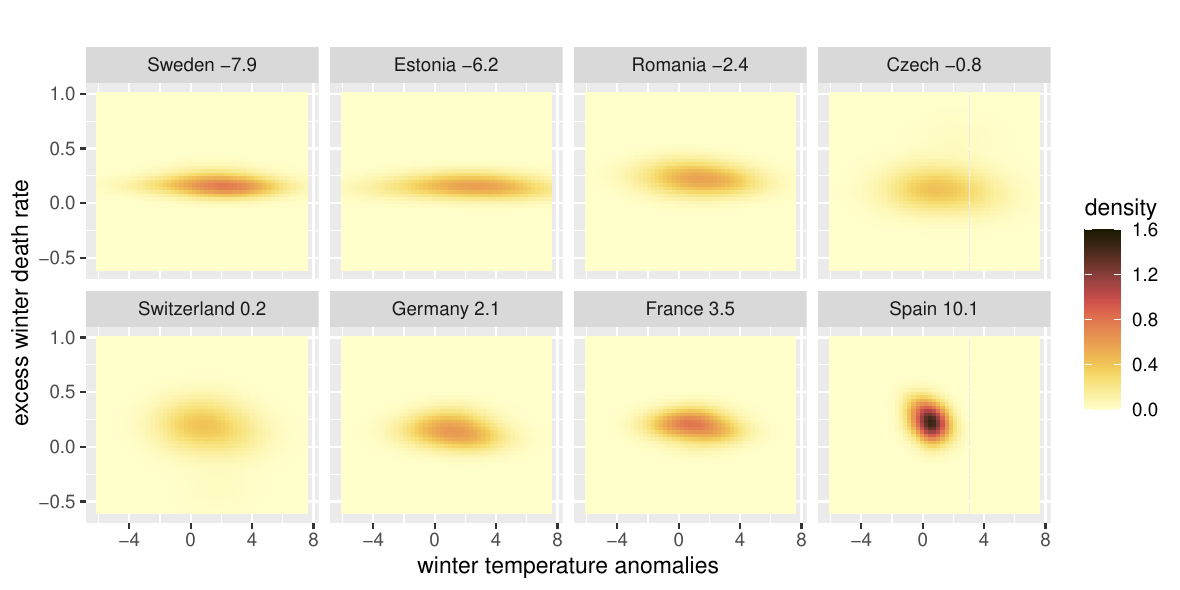}
    \caption{}
  \end{subfigure}\hfill
  \begin{subfigure}{\textwidth}
    \centering
    \includegraphics[width=\linewidth]{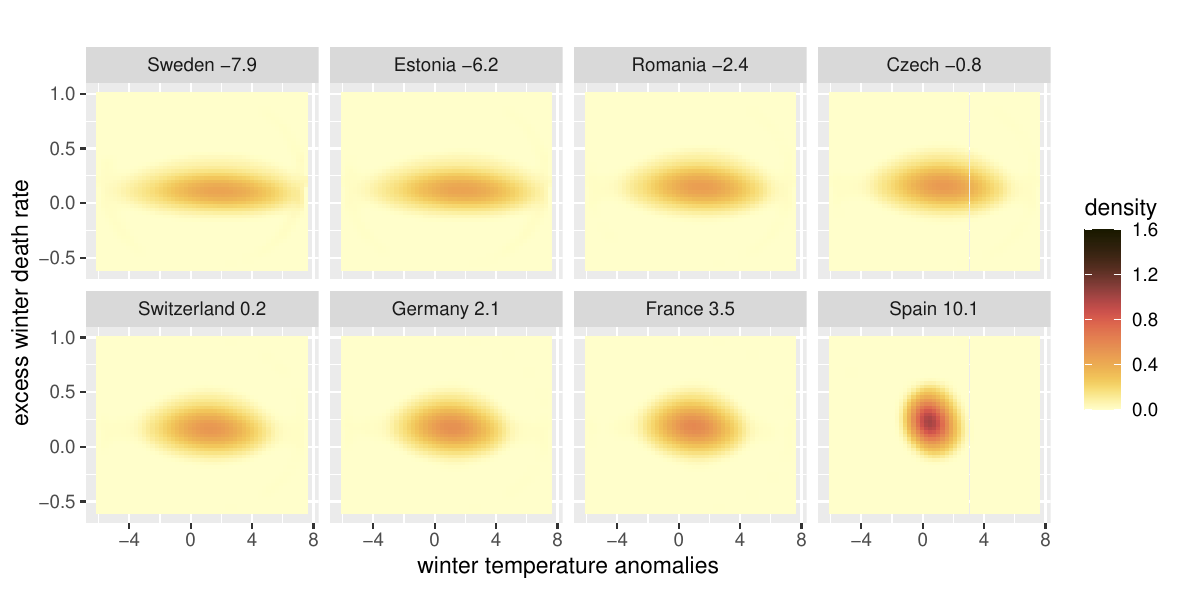}
    \caption{}
  \end{subfigure}
  \caption{(a) Observed smoothed and (b) fitted densities from global slice-wise Wasserstein regression (GSWW) of the joint distributions of excess winter death rates ($y$-axis) and winter temperature anomalies ($x$-axis) for selected European countries, ordered by their base winter temperature. Each panel is labeled with the country name, and the corresponding average base winter temperature (in $^\circ$C) for 1961--1990 is shown. The sliced Wasserstein fraction of variance explained is 0.53.}
  \label{fig:EUDeath}
\end{figure}

We focused on $n=31$ countries with available data from 1999 to 2021, for which we adopted the proposed GSWW and GSAW models with the country-level base winter temperature averaged from 1961-90 as predictor and the country-level bivariate distribution between the excess rate of winter mortality relative to the previous summer and winter temperature anomaly relative to the base winter temperature from 1961-90 as the response. We chose standardized winter temperature anomalies to ensure better overlap of shared subdomains within country-specific bivariate distributions. The absolute number of deaths in each of the 31 European countries from 1999 to 2021 was obtained from the Eurostat database \url{https://ec.europa.eu/eurostat/web/main/data/database} and the winter temperature anomalies and base winter temperatures from 1961-90 from  \url{https://www.uea.ac.uk/groups-and-centres/climatic-research-unit/data}. The observed smoothed bivariate distributions are in Figure \ref{fig:EUDeath} (a) for a few selected countries.

The fitted densities obtained from the GSWW model are shown in Figure \ref{fig:EUDeath} (b), where the corresponding sliced Wasserstein fraction of variance explained is 0.53; the fitted densities when fitting  GSAW for the same countries can be found in Section \ref{sec:adx:figures} in the  Appendix.  
Countries with warmer climates typically experience a higher rate of excess deaths during the winter season compared to colder countries for the same temperature anomaly. For instance, in Spain, a country with a warm climate, there is a roughly 30\% increase in the number of deaths during the winter months compared to the preceding summer. However, in Sweden, a country with a colder climate, this increase is only around 15\%.



\begin{figure}[tb]
    \single
    \centering
    \includegraphics[scale=0.68]{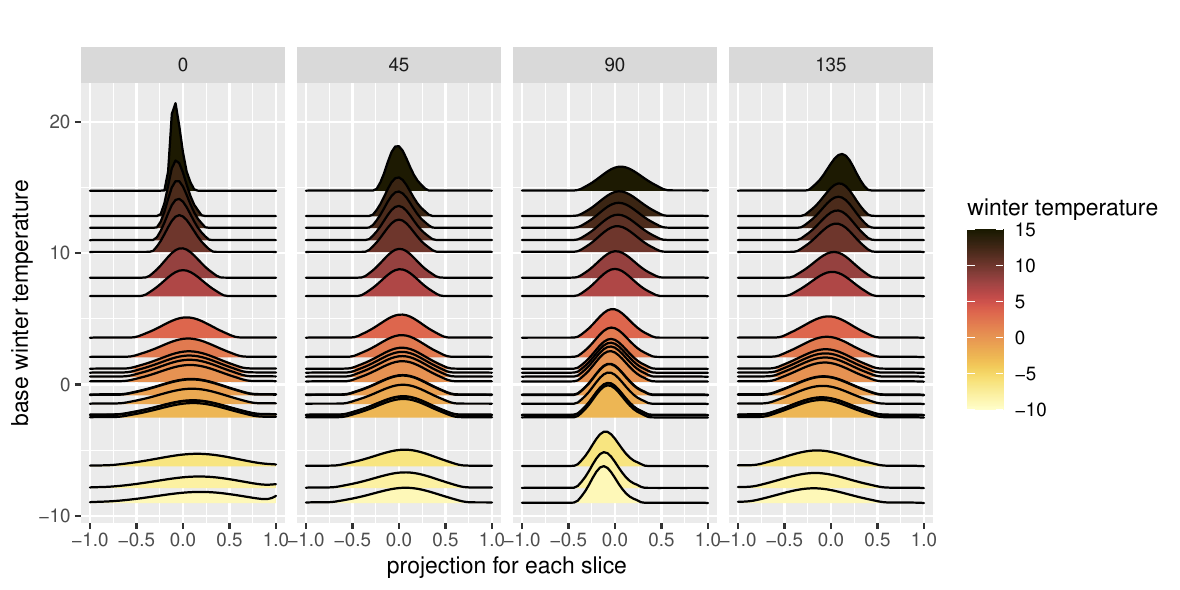}
    \caption{Fr\'{e}chet regression between the base winter temperature (predictor, on the $y$-axis) and the slicing distributions (responses, on the $x$-axis) for various projections. The number at the top of each panel indicates the angle of the respective projection with the $x$-axis in Figure \ref{fig:EUDeath} (a).}
    \label{fig:EUDeathSliceWise}
\end{figure}

Figure \ref{fig:EUDeathSliceWise} illustrates the Fr\'{e}chet regression steps for one-dimensional distributions for some representative projections. As the base temperature increases, the variance of the temperature anomalies decreases and the excess winter death rates shift to a higher level. The slice corresponding to the projection angle at 135 degrees illustrates the main effect of the regression, which is a shift to the right as winter base temperature goes up, leading to higher winter mortality, and effect that is most pronounced for Spain. Most likely, populations used to a relatively warm winter contain a higher fraction of individuals who are more susceptible to cold temperatures than populations accustomed to a cold winter.

\subsection{S\&P500 and VIX Index Joint Modeling}
\begin{figure}[]
    \single
  \centering
  \begin{subfigure}{\textwidth}
    \centering
    \includegraphics[width=\linewidth]{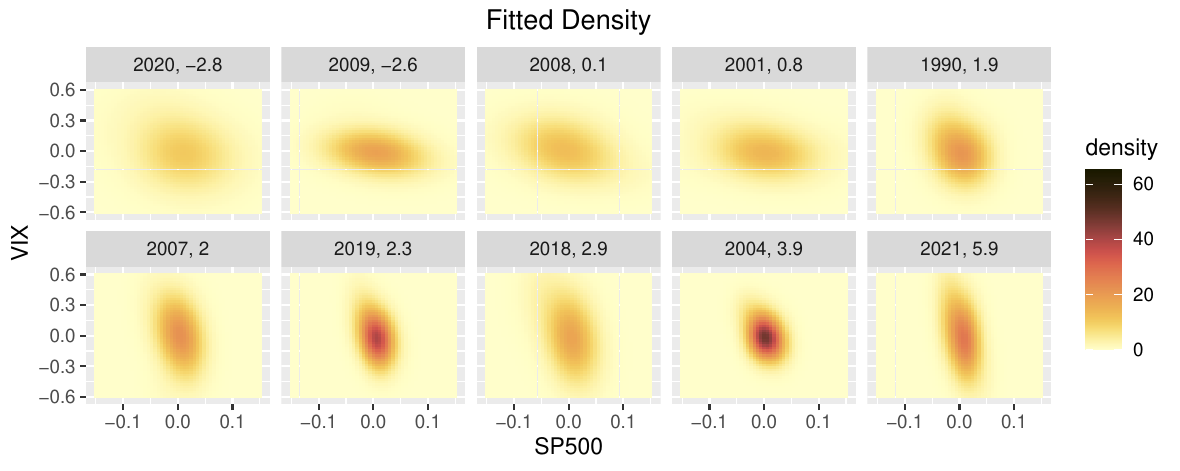}
    \caption{}
  \end{subfigure}\hfill
  \begin{subfigure}{\textwidth}
    \centering
    \includegraphics[width=\linewidth]{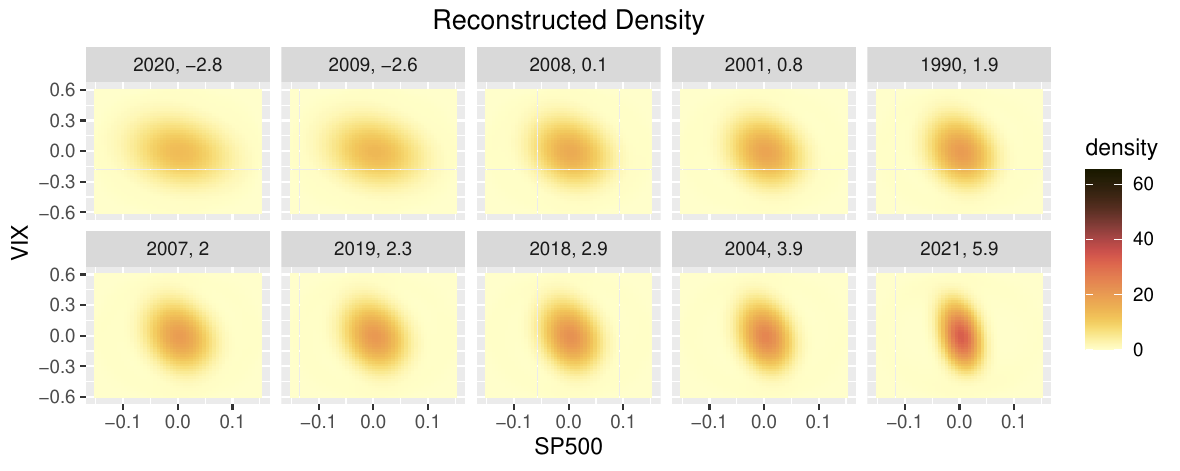}
    \caption{}
  \end{subfigure}
  \caption{(a) Observed smoothed and (b) fitted densities from global slice-wise Wasserstein regression (GSWW) of the joint distributions of weekly returns of the VIX index, which reflects expected market volatility ($y$-axis) and S\&P500 index ($x$-axis) for selected years, arranged in order of the associated yearly GDP growth rate. Each panel is labeled by year; the corresponding yearly GDP growth rate (as a percentage) is also indicated.}
  \label{fig:VIX}
\end{figure}

In the realm of financial markets, the application of distributional data modeling techniques holds promise, particularly concerning assets like stocks, exchange-traded funds or derivatives \citep{officer1972distribution,madan2020multivariate}. One central challenge for modeling of financial market modeling stems from complex market dynamics featuring non-normality, low signal-to-noise ratio, and sudden, unpredictable changes in volatility \citep{madan2017asymmetries}. Traditional modeling of multivariate distribution in finance has centered on copula-based approaches that link marginal univariate distributions such as bilateral gamma marginals through a pre-specified copula function \citep{cherubini2004copula,kuchler2008bilateral, madan2020multivariate}. The promise of sliced Wasserstein regression is that it eliminates the need to decompose the bivariate distribution into marginal distributions and to specify a copula function by directly targeting the bivariate distributions and their relation with predictors. In this context, there is growing interest in exploring the joint distribution between weekly returns of two crucial financial indices, the S\&P 500 index and the Volatility Index (VIX), where the latter tracks expected volatility in the stock market over the subsequent 30 days \citep{madan2020multivariate}. The objective is to gain a deeper understanding of this joint distribution and its connection with the Gross Domestic Product (GDP) of the United States, which is an annual monetary measure of final goods or services. The S\&P 500 and VIX index data are sourced from Yahoo Finance (\url{https://finance.yahoo.com/}). 
The observed smoothed bivariate distributions are in Figure \ref{fig:VIX} (a) for a few selected years, ordered by the associated GDP growth rate.

\begin{figure}[tb]
    \single
    \centering
    \includegraphics[width=\linewidth]{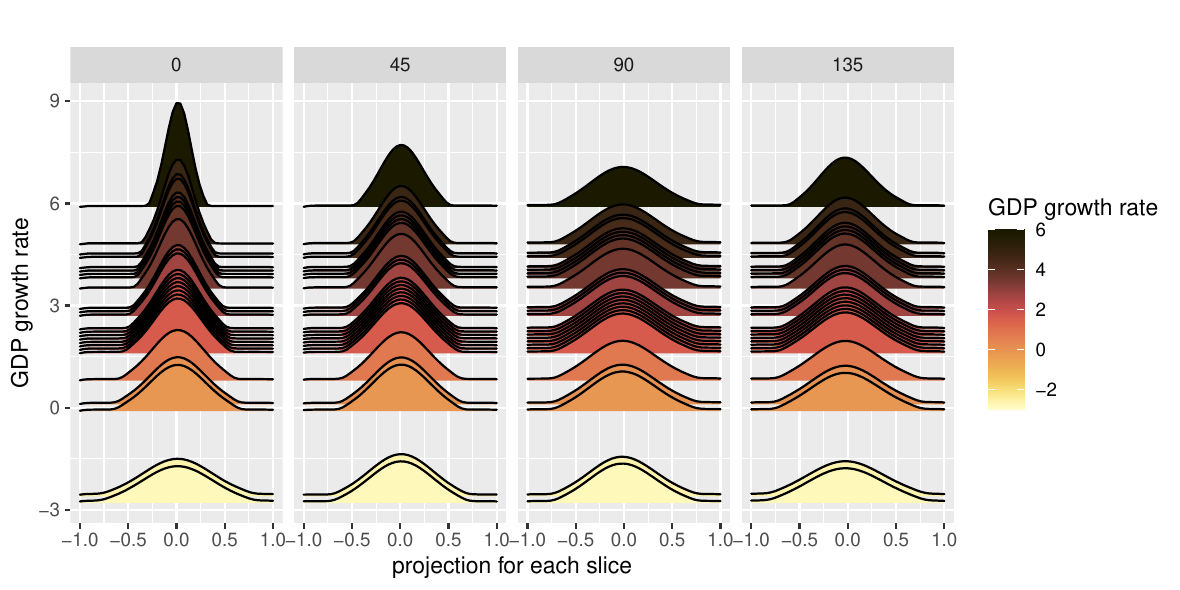}
    \caption{Fr\'{e}chet regression between the GDP growth rate (predictor, on the $y$-axis) and the slicing distributions (responses, on the $x$-axis)  
    for various projections. The number at the top of each panel indicates the angle of the respective projection with the $x$-axis in Figure \ref{fig:VIX} (a).}
    \label{fig:VIXslice}
\end{figure}

Consistent with the findings in the prior study by \citet{madan2020multivariate}, the VIX returns are seen to exhibit more variation than the S\&P 500 returns, along with an evident negative correlation between these two returns. This inverse relationship occurs because elevated VIX levels, indicating market turbulence, typically coincide with lower stock prices, reflecting investor anxiety. In contrast, periods of market stability or upward trends in stock prices are associated with lower VIX values. The fitted densities obtained from the GSWW model are shown in Figure \ref{fig:VIX} (b). They are seen to generally track the observed trends but not the sharpness of the density peaks; the Wasserstein fraction of variance explained is low at 0.15. Details of how the sliced regression works can be seen in Figure \ref{fig:VIXslice}, which illustrates the (global) Fr\'echet regression step for the univariate distributions corresponding to a few representative slices.

Further results for GSAW (instead of GSWW) and the local models (LSWW and LSAW)  can be found in Section \ref{sec:adx:figures} in the Appendix. The slice-wise fits elucidate the association between the GDP growth rate and the bivariate distributions, showcasing the profound influence of GDP growth on the marginal distribution of the S\&P 500 index's weekly return and its relationship with the VIX index. In Figure \ref{fig:VIXslice}, an increase in the GDP growth rate coincides with a substantial decrease in the variance of the S\&P 500 index return, contrasting with the relatively stable variance of the VIX index. Moreover, as depicted in Figure \ref{fig:VIX} (b), a higher GDP growth rate intensifies the responsiveness of VIX index upticks to downticks in the S\&P 500 index. In essence, during times of robust economic health (indicated by higher GDP growth rates), declines in the S\&P 500 index are more likely to be associated with market turbulence, resulting in an increase in the VIX index.

For instance, during challenging periods such as  1990 (amidst the Gulf War), 2001 (following the September 11 terrorist attacks), 2008-2009 (amidst the financial crisis and substantial recession), and 2020 (amidst the COVID-19 pandemic), when the US economy faced significant hardships, a notable increase in the variance of the S\&P 500 index was expected. In contrast, during more favorable economic phases like 2004 (highlighted by solid economic performance and low inflation rates), 2018-2019 (marked by sustained economic expansion and robust GDP growth), and 2021 (notably aided by fiscal stimulus to rebound from the pandemic's impact), when the US economy was thriving, downturns in the S\&P 500 index were more prone to coincide with surges in the VIX index. This association potentially stems from fears regarding an impending economic downturn.

\subsection{Exchange Traded Funds}
\label{etf} 
An additional data illustration is presented in the Appendix Section \ref{sec:adx:data} to highlight the application of the local versions LSWW and LSAW of the sliced Wasserstein regression models. 

\section{Discussion and Outlook}\label{sec:disc}

The proposed SAW and SWW regression approaches are new tools for the statistical analysis of multivariate distributional data that come with theoretical guarantees and with the flexibility to employ both global and local Fr\'{e}chet regression methods. 
The SAW regression model employs Fr\'{e}chet regression over the multivariate distribution space equipped with the established slice-averaged Wasserstein distance, where the pointwise rates of convergence are optimal for both global and local SAW regression models under certain regularity conditions. The SWW approach offers a new perspective on how slicing transforms can be utilized for the analysis of data that consist of multivariate distributions. SWW regression is based on a regression step for each slice, thus allowing for parallel computation and avoiding the entanglement of the slices that is a characteristic of SAW regression. SWW  is coupled with a regularized inverse transform from the sliced space to the original distribution space and is motivated by the idea that executing a regression step for each slice minimizes the slice-wise prediction error and when integrating this error it is smaller than the SAW error, as the latter seeks to minimize the aggregated error across all slices. 

A downside of SWW regression is that it is associated with a slower rate of convergence. This is due to the fact that the slice-wise minimizers are not necessarily in the image space {$\widetilde{\psi}(\mathscr{F})$}, necessitating to employ a generalized version of the inverse slicing transform {$\widetilde{\psi}_\tau^{-1}$} when transforming back to $\mathscr{F}$, which requires a regularization parameter {$\tau$} and slows down the convergence of the final distribution estimates in the distribution space  $\mathscr{F}$. Additionally, SWW requires a higher order of smoothness for the densities of the underlying distributions to achieve comparable convergence rates to SAW, since the convergence rates of SWW depend crucially on the convergence of the Fourier transform, where the undesirable assumption $k\geq p+1$ is required to ensure sufficiently fast convergence.  However, these disadvantages are mitigated by the improved performance of SWW regression compared to SAW regression in both simulation settings and real data applications. This superior performance likely is due to smaller constants in the error rates for the slice-wise optimization in SWW as compared to the slice-averaged optimization in SAW, where the better rates of convergence of SAW cannot overcome this advantage for moderate sample sizes as typically encountered in statistical data analysis. 

We investigated two cases, the theoretically ideal case where the underlying densities are all known, an assumption that has been routinely made in previous research on sliced Wasserstein methods, and a more realistic case where the densities are not known and must be estimated from data that they generate. When densities need to be estimated, several additional issues arise. One of these is that uniform convergence of the estimators over the entire domain $D$ is required,  while the resulting estimates also need to be bona fide densities on $D$. While we offer a valid approach by truncating distributions to suitable subdomains, in order to keep the focus of the paper on the sliced Wasserstein approaches, a full resolution of this issue is left for future research. Commonly available estimators such as kernel estimators and their many variants do not achieve both properties over the entire domain; see \citet{mull:16:1} for an in-depth discussion of the analogous issue for the one-dimensional case. 

An ideal solution would lead to a bona fide density estimator that takes advantage of the additional order of smoothness $k$ in the SWW case. However, our practical experience with kernel density estimation has shown that faster rates of convergence that are theoretically attainable for higher order smoothness of the underlying density functions coupled with higher order kernels rarely materialize for moderate sample sizes typically encountered in statistical data analysis. The reason for this is the same as mentioned above for the superiority of SWW over SAW in practical applications: The improvement in the rate of convergence is not strong enough to mitigate the effect of the larger constants associated with higher order kernels and one also incurs the problem of negative-valued density estimates (since higher order kernels cannot be non-negative). 

In view of these considerations, we implemented a simple approach: We assume all distributions have nice smooth densities $f$ on domains $D_f \supset D_\epsilon \supset D$, where $D$ is a convex compact and reasonably nice set, densities are estimated with positive kernels (of order 2) on $D_\epsilon$ and the truncated versions of these estimates on domain $D$ are used as estimates of the densities on $D$, which are assumed to be the targets. This allows for boundary-effect-free estimation, yielding bona fide densities on $D$ that satisfy the requisite assumptions. The individual domains $D_f$ may be known or unknown, and more often than not they will be unknown. In such cases subject-matter knowledge and common statistical sense will guide the choice of $D$. 

Concurrent with our work,  geometric properties associated with the so-called disintegrated distributional space have been recently studied \citep{kitagawa2023two} and it was shown that the SAW representation and metric associated with the slice-averaged Wasserstein distance does not lead to a geodesic space \citep{park2023geometry}. In contrast, the SWW representation and associated distribution-slicing Wasserstein metric permits geodesics and thus is a geodesic space; this is a straightforward consequence of the fact that the Wasserstein space for univariate distributions is a geodesic space with the McCann interpolants serving as geodesics \citep{gang:96}. A geodesic space is a prerequisite for adopting promising intrinsic approaches for principal component analysis \citep{bigo:17}, regression \citep{zhu2023geodesic} and extrapolation \citep{fan2021conditional}. The geodesic structure of the novel slice-wise methodology that we introduced in this paper is thus a major advantage of this approach and exploiting this structure is a promising avenue for future research.

When the response distributions are not fully observed and must be reconstructed from random samples generated by the underlying distributions—an aspect overlooked in prior research on applying the Radon transform to density functions, typically assumed to be fully observed—the domain assumption requires a slight adjustment discussed in Section \ref{sec:adx:dens} in the Appendix. This adjustment makes it possible to avoid an unproductive discussion of boundary effects when estimating densities from available data they generate while conforming with statistical practice, where invariably a user will choose a reasonable domain of interest. To guarantee that the density estimators are non-negative bona fide densities, we utilize the second-order smoothness of the multivariate density function. For those looking to take full advantage of the assumed smoothness of the multivariate density function to achieve a faster convergence rate, the option of employing a higher-order kernel $\kappa$ of order $k$ is available. However, this comes at the cost of the estimated density potentially losing non-negativity property, making it less practical for our setting where the density estimates must be bona fide densities. See Section \ref{sec:adx:dens} in the Appendix for a detailed discussion.

\single
\bibliographystyle{rss}
\bibliography{swr.bib}
\double

\clearpage
\appendix
\section{Abbreviations, Notations and Summary of Models}
\label{sec:adx:notation}

\begin{table}[h]
    \single
\centering
\caption{List of notations (1). The bracketed explanation corresponds to the second half of the notation.}
\begin{tabular}{c|c} 
\toprule
Notation & Explanation\\
 \midrule
 $D$ & known support of multivariate distribution  \\ 
 $|D|$ & Lebesgue measure of the support $D$ \\
 $p$ & dimension of the multivariate distribution   \\ 
 $q$ & dimension of the predictor $X$  \\ 
 $f(z)$, $\mathcal{F}$ & multivariate density functions (space) \\
 $g(u)$, $\mathcal{G}$ & univariate density functions (space) \\ 
 $I(g)$ & support of univariate distribution,  $g\in\mathcal{G}$\\
 $q(s)$, $\mathcal{Q}$ & quantile functions (space) \\
 $\mu,\mathscr{F}$ & multivariate distributions (space)  \\
 $\nu, \mathscr{G}$ & univariate distributions (space) \\
 $\theta, \Theta$ & slicing parameter (set) \\
 $\lambda$, $\Lambda_{\Theta}$ & density slicing functions (space) from $\Theta$ to $\mathcal{G}$ \\
 $\gamma$, $\Gamma_{\Theta}$ & quantile slicing functions (space) from $\Theta$ to $\mathscr{G}$ \\
 $\psi$, $\psi_{\tau}$ & slicing (regularized slicing) transform from $\mathcal{F}$ to $\Lambda_{\Theta}$\\
 $\mathcal{R}$, $\mathcal{R}_{\tau}$ & Radon (regularized Radon) transform from $\mathcal{F}$ to $\Lambda_{\Theta}$ \\
 $\phi_\tau(r)$ & a regularizing function on Fourier domain\\
 $\Delta_{\tau}$ & reconstruction error \\ 
 $\mathcal{J}_1, \mathcal{J}_p$ & univariate (multivariate) Fourier transform \\
 $\varphi$ & map from $\mathscr{F}$ to $\mathcal{F}$ or from $\mathscr{G}$ to $\mathcal{G}$  \\
 $\varrho$ &  map from $\Gamma_{\Theta}$ to $\Lambda_{\Theta}$\\
 $G$ & map from $\mathscr{G}$ to $\mathcal{Q}$ \\
 $\widetilde{\psi}, \widetilde{\psi}_{\tau}$ & induced slicing (regularized slicing) transform from $\mathscr{F}$  to $\Gamma_{\Theta}$\\
 \bottomrule
\end{tabular}
\end{table}

\begin{table}[]
    \single
\centering
\caption{List of notations (2). The bracketed explanation corresponds to the second half of the notation.}
\scalebox{1}{\begin{tabular}{c|c} 
\toprule
Notation & Explanation\\
 \midrule
 $d$ &  metric on $\mathscr{F}$\\
 $d_W$ & Wasserstein distance on $\mathscr{F}$ or $\mathscr{G}$ \\
 $d_{DW}$ & distribution-slicing Wasserstein metric on $\Gamma_{\Theta}$\\
 $d_{SW}$ & slice-averaged Wasserstein distance on  $\mathscr{F}$\\
 $d_2$ & $L^2$ norm\\
 $d_\infty$ &  $ L^{\infty}$ norm \\
 $Z_1, Z_2$ & random variables on $\mathbb{R}^p$ \\
 $\mathcal{P}(\mu_1,\mu_2)$ & probability measure with marginal distributions $\mu_1,\mu_2$ \\
 $\mu,X$ & random pair with $\mu\in\mathscr{F}, X\in\mathbb{R}^q$\\
 $F$& joint distribution of $X$ and $\mu$\\
 $F_X, F_\mu$ & marginal distribution of $X$ (of $\mu$)\\
 $m(x)$ & Fr\'{e}chet minimizer\\
 $M(\cdot, x)$ &  conditional Fr\'{e}chet function\\
 $s_G(X,x), s_{i,G}(x)$ & (sample) weight function of global Fr\'{e}chet regression\\
 SAW, GSAW & (global) slice-averaged Wasserstein \\
 SWW, GSWW & (global) slice-wise Wasserstein \\
 $m_G^{SAW}(x), m_{G,\tau}^{SWW}$ & GSAW (GSWW) regression minimizer\\
 $M_G^{SAW}(\cdot,x), M_{G}^{SWW}(\cdot, x)$ & GSAW (GSWW) conditional Fr\'{e}chet function\\
 $\gamma_{G,x}$ & minimizer of $M_{G}^{SWW}(\cdot, x)$\\
 $\{(X_i, \mu_i)\}_{i=1}^n$ &  a sample of random pairs of predictors and measures \\
 $\hat{\mu}_i, \hat{f}(x)$ & estimated distribution (density function) \\
 $\bar{X}, \hat{\Sigma}$ & sample mean (variance) of $\{X_i\}_{i=1}^n$\\
 $R_{\oplus}^{2}$ & slice-averaged Wasserstein $R^2$ coefficient\\
 $\mathbf{W} = \{\mathbf{W}_j\}_{j=1}^N$ & random observations \\
 $\mathbf{W}^{(i)} = \{\mathbf{W}_{ij}\}_{j=1}^{n_i}$ & random observations from $\mu_i$ \\ 
 $\mathbf{W}(\theta), \mathbf{W}^{(i)}(\theta)$ &  sliced observation (observations from $\mu_i$) \\
 $\mathcal{M}_G(\mathbf{W},x)$ & GSAW objective function \\
 $\langle \cdot, \theta\rangle$ & Radon slicing operation on random observation in $\mathbb{R}^p$\\
 $\Pi_A$ & permutation operator on $A\in\mathbb{R}^N$\\
 $\eta$ & step parameter in the gradient descent \\
 \bottomrule
\end{tabular}}
\end{table}


\begin{table}[tb]
    \single
\centering
\caption{List of notations for the Appendix. The bracketed explanation corresponds to the second half of the notation.}
\begin{tabular}{c|c} 
\toprule
Notation & Explanation\\
 \midrule
 $D_\epsilon$ & neighborhood surrounding the common support $D$ \\ 
 $D_f$ & support of the density function $f$ \\
 $D_i$ &  support of the $i$-th distribution   \\
 $f_{i, D_i}, \mu_{i,D_i}$ & the $i$-th sample density (distribution) on support $D_i$\\
 $f_{i, D_i|D_\epsilon}$ & the $i$-th sample density on the subdomain $D_\epsilon$\\
 $N_i, \tilde{N}_i$ & number of observations on $D_i$ ($D_\epsilon$)\\
\ $\kappa, b$ & kernel function (bandwidth) for density estimation \\
 $\mathcal{E}_{X|\mu}, \mathcal{E}_{X}$ & conditional (marginal) density of $X$ \\
 $K, h$ & kernel function (bandwidth) for local regression \\
 $\mathcal{T}$ & domain of the predictor $X$ when $q=1$ \\
 $s_{L}(X,x,h), s_{iL,h}(x)$ & (sample) weight function of local Fr\'{e}chet regression\\
 LSAW & local slice-averaged Wasserstein \\
 LSWW & local slice-wise Wasserstein \\
 $m_{L,h}^{SAW}(x), m_{L,h,\tau}^{SWW}$ & LSAW (LSWW) regression minimizer\\
 $M_{L,h}^{SAW}(\cdot,x), M_{L,h}^{SWW}(\cdot, x)$ & LSAW (LSWW) conditional Fr\'{e}chet function\\
 $\gamma_{L,h,x}$ & minimizer of $M_{L,h}^{SWW}(\cdot, x)$\\  $\mathcal{M}_{L,h}(\mathbf{W},x)$ & GSAW objective function \\
 ${\mathcal{D}}^{\mathbf{k}}$ & differentiation operator on $\mathcal{F}$\\
 $\mathcal{K}, a$ & kernel function (bandwidth) used for a proof \\
 $N(\epsilon, \mathscr{F}, d)$ & covering number using balls of size $\epsilon$ \\
 $B_\delta[m_G^{SAW}(x)]$ & $\delta$-ball in $\mathscr{F} $centered at $m_G^{SAW}(x)$\\
 $B_G(x)$ & best approximation \\
  $\Psi$ & map from quantile functions to distribution functions \\
 $\eta_0, \eta_1,\eta_2, \beta_0, \beta_1, \beta_2$ & constants  \\
 $A_1, A_2, A_3, B_0, B_1 $ & constants \\
 $\tau_0,\tau_1,\tau_2, C_0, C_1,C_2,C, \zeta$ & constants  \\
   \bottomrule
\end{tabular}
\end{table}

\begin{table}[tb]
    \single
    \small
    \centering
    \caption{Summary of the  global and local regression models.}
    \label{tab:summaryTable}
    \begin{tabular}{cccc}
    \toprule
         Type & Regression & Object & Result \\
    \midrule
    \multirow{8}{*}{Global} & & weight function & $s_G(X,x) = 1+(X-E(X))^\intercal\Var(X)^{-1}(x-E(X))$ \\\cmidrule(lr){2-4}
          & \multirow{3}{*}{SAW} & \multirow{2}{*}{target function} & $m_G^{SAW}(x)=\argmin_{\omega\in\mathcal{F}}M_G^{SAW}(\omega, x)$\\
         &  &  & $M_G^{SAW}(\cdot,x) = E[s_G(X,x)d_{SW}^2(\mu, \cdot)]$ \\ \cmidrule(lr){3-4}
         &  & convergence & $d_{SW}(m_G^{SAW}(x), \hat{m}_G^{SAW}(x)) = O_p(n^{-1/2})$\\ \cmidrule(lr){2-4}
         & \multirow{4}{*}{SWW} & \multirow{2}{*}{target function} & $m_{G,\tau}^{SWW}(x) = \widetilde{\psi}_{\tau}^{-1}\left[ \argmin_{\gamma\in\Gamma_{\Theta}}M_G^{SWW}(\gamma,x) \right]$  \\
         & &  &  $M_G^{SWW}(\cdot, x) = E\left[s_G(X,x)d_{DW}^2(\widetilde{\psi}(\mu),\cdot) \right]$\\\cmidrule(lr){3-4}
         & & convergence & $d_{\infty}\left( m_{G}^{SWW}(x), \hat{m}_{G,\tau}^{SWW}(x)\right) = O_p\left({C_1(\tau) + C_2(\tau)}{n^{-2/7}} \right)$  \\
    \midrule
    \multirow{12}{*}{Local} & & weight function & $s_L(X, x,h) = K_h(X-x)[\upsilon_2 - \upsilon_1(X-x)] / \sigma^2_0$ \\ \cmidrule(lr){2-4}
    & \multirow{5}{*}{SAW} & \multirow{2}{*}{target function} & $m^{SAW}(x) = \argmin_{\omega\in\mathscr{F}}M^{SAW}(\omega, x)$\\
    & & & $M^{SAW}(\cdot,x) = E\left[d_{SW}^2(\mu,\cdot)|X=x\right]$\\\cmidrule(lr){3-4}
    & & \multirow{2}{*}{local target} & $m_{L,h}^{SAW}(x) = \argmin_{\omega\in\mathcal{F}}M_{L,h}^{SAW}(\omega, x)$  \\
    &  &  & $M_{L,h}^{SAW}(\cdot, x) = E[s_{L}(X,x,h)d_{SW}^2(\mu, \cdot)]$ \\\cmidrule(lr){3-4}
    &  & convergence & $ d_{SW}(m^{SAW}(x), \hat{m}^{SAW}_{L,h}(x)) = O_p(n^{-2/5})$\\\cmidrule(lr){2-4}
    & \multirow{6}{*}{SWW} & \multirow{2}{*}{target function} & $m^{SWW} = \widetilde{\psi}^{-1}\left[\argmin_{\gamma\in\Gamma_{\Theta}}
    M^{SWW}(\gamma, x)\right]$\\
    & & & $M^{SWW}(\cdot,x) = E\left[d_{DW}^2(\widetilde{\psi}(\mu),\cdot)|X=x\right]$ \\\cmidrule(lr){3-4}
    & & \multirow{2}{*}{local target} & $m_{L,h,\tau}^{SWW}(x) = \widetilde{\psi}^{-1}_{\tau}\left[ \argmin_{\gamma\in\Gamma_{\Theta}}M_{L,h}^{SWW}(\gamma,x) \right]$\\
    & & & $M_{L,h}^{SWW}(\cdot, x) = E\left[s_L(X, x, h)d_{DW}^2(\widetilde{\psi}(\mu),\cdot) \right]$ \\\cmidrule(lr){3-4}
    & & convergence & $d_{\infty}( m^{SWW}(x), \hat{m}^{SWW}_{L,h,\tau}(x)) = O_p\left(C_1(\tau) + C_2(\tau)n^{-8/35}\right)$ \\
    \bottomrule
    \end{tabular}
\end{table}

\section{Additional Assumptions}
\label{sec:adx:assumptions}

\begin{enumerate}[label=(A\arabic*), leftmargin=1cm]
\setcounter{enumi}{2}
\item $\argmin_{\gamma\in\Gamma_{\Theta}}M_{L,h}^{SWW}(\gamma,x) \in \widetilde{\psi}(\mathscr{F})$ as per \eqref{formula:adx:transformLocReg}.  
\end{enumerate}

\begin{enumerate}[label=(L\arabic*), leftmargin=1cm]
\item The kernel $K$ used in the definition of  the local sliced Wasserstein regression in Section \ref{sec:adx:local} is a probability density function and is symmetric around zero. Furthermore, defining $K_{kj} = \int_{\mathbb{R}}K^k(u)u^jdu$, $|K_{14}|$ and $|K_{26}|$ are both finite. 

\item The marginal density of $X$ denoted as $\mathcal{E}_X(x)$ and the conditional density of $X$ given $\mu=\omega$, $\mathcal{E}_{X|\mu}(\cdot, \omega)$, exist and are twice continuously differentiable, the latter for all $\omega\in \mathscr{F}$, with $\sup_{x,\omega}|
\left(\partial^2\mathcal{E}_{X|\mu}/\partial x^2\right)(x,\omega)
|<\infty$. Additionally, for any open $U\subset \mathscr{F}$, $\int_{U}dF_{\mu|X}(x,\omega)$ is continuous as a function of $x$.

\item  
The derivative $K^\prime$ exists and is bounded on the support of $K$, i.e., $\sup_{K(u)>0}|K^\prime(u)|<\infty$; additionally, $\int_{\mathbb{R}}u^2|K^\prime(u)|(|u\log|u||)^{1/2}du<\infty$. 

\item Let $\mathcal{T}$ be a closed interval in $\mathbb{R}$ and $\mathcal{T}^o$ be its interior. Denote $\mathcal{E}_X(s)$ and $\mathcal{E}_{X|\mu}(\cdot,\omega)$ as per (L2), which exist and are twice continuously differentiable on $\mathcal{T}^o$, the latter for all $\omega \in \mathscr{F}$. The marginal density $\mathcal{E}_X(x)$ is bounded away from zero on $\mathcal{T}$, $\inf_{x\in\mathcal{T}}\mathcal{E}_X(x)>0$. The second-order derivative $\mathcal{E}_X^{\prime\prime}(x)$ is bounded, $\sup_{x\in\mathcal{T}^o}|\mathcal{E}_X^{\prime\prime}(x)| < \infty$. The second-order partial derivatives $(\partial^2\mathcal{E}_{X|\mu}(x,\omega) / \partial x^2 )(x, \omega)$ are uniformly bounded, i.e., \\ $\sup_{x\in\mathcal{T}^o, \omega\in \mathscr{F}} |(\partial^2\mathcal{E}_{X|\mu} / \partial x^2)(x,\omega)| < \infty$. Additionally, for any open set $U\subset \mathscr{F}$, $\int_{U}dF_{\mu|X}(x,\omega)$ is continuous as a function of $x$; for any $x\in \mathcal{T}$, $M(\cdot, x)$ is equicontinuous, i.e.,
\begin{align*}
    \limsup_{y\rightarrow x} \sup_{\omega \in \mathscr{F}}|M(\omega,y) - M(\omega,x)| = 0. 
\end{align*}
\end{enumerate}

\section{Additional Preliminary Density Estimation Step} 
\label{sec:adx:dens}

We assume for our main results that the multivariate distributions under consideration are well defined and possess density functions on a common domain $D$, satisfying (D1) and (F1).  In practice, the common domain  $D$ will either be pre-specified  in statistical applications or will be chosen by the analyst based on subject-matter or practical considerations. This is exemplified in our data examples.  Once  $D$ has been selected,  distributions truncated on $D$ are the targets of the analysis and serve as responses for the SAW or SWW regression models.  The actual distributions may have domains of which $D$ is a subset and these domains may vary from distribution to distribution and could be unbounded.

For the practically important case where the distributions $\mu_i$  are not known but must be inferred from observations, 
we make some robust assumptions that will suffice to side-step the issue of boundary effects in density estimation without delving into complex technical details. A first assumption is  that there exists $\epsilon>0$ so that $D_\epsilon = \bigcup_{z\in D} B(z,\epsilon)$ is a subset of the domain of each underlying distribution, where $B(z,\epsilon)$ is a ball with radius $\epsilon$ centered at $z$. This assumption makes it possible to avoid a detailed discussion of boundary effects when estimating densities from available data they generate.  
We furthermore assume that 
the continuous differentiability assumption  (F1) holds on  $D_\epsilon$. If a  random distribution in the sample with density $f_{D_f}$ has the (random) domain $D_{f}$, we require that $D_f \supset  D_\epsilon$ and that on  $D_\epsilon$  the density $f_{D_f}$  is uniformly bounded from below and above, continuously differentiable of order $k$ and has uniformly bounded partial derivatives, where the uniformity requirement extends across all $f_{D_f}$. Formally,

\begin{enumerate}[label=(F\arabic*$^\prime$), leftmargin=1cm]
\item There exists a constant $M_0>0$ and an integer $k\geq 2$ such that for all $f\in\mathcal{F}$, the density function $f$ is the density of a truncated  distribution on the domain $D$, where the original distribution has a domain $D_f$ with $D_f \supset D_\epsilon$ and a density $f_{D_f}$ that is defined on $D_f$.  It holds that $\max\{\|f_{D_f}\|_\infty, \|1/f_{D_f}\|_\infty\}\leq {M_0}$ on $D_\epsilon$ and that $f_{D_f}$ is continuously differentiable of order $k$ on $D_\epsilon$, with uniformly bounded partial derivatives. The target distribution  with density $f$ is the  truncated version of the $f_{D_f}$ on $D$, so that for all $z\in D$ one has $f(z)=f_{D_f}(z)/\int_{D}f_{D_f}(u)du.$
\end{enumerate}

Assume we have a sample of distributions $\mu_{1,D_1},\dots,\mu_{n,D_n}$ with  densities $f_{1, D_1},\dots,f_{n,D_n}$ and  domains $D_{i} \supset D_\epsilon$ and $N_i$  i.i.d. observations generated by  $f_{i,D_i}$ by a random mechanism that is independent of the random mechanism that generates the $\mu_{i,D_i}$. 
Let $\tilde{N}_i$ be the number of observations made for $f_{i,D_i}$ that fall inside the domain $D_\epsilon$. 
We impose the following assumption (P1) on the  $N_i$ to ensure that the $L^2$ convergence rate of the density estimates  
is faster than or at least as fast as the parametric rate $n^{-1/2}$. 

\begin{enumerate}[label=(P\arabic*), leftmargin=1cm]
\item $N(n) = \min_{1\leq i \leq n}N_i \gtrsim n^{(p+4)/4}$,  where $N_i$ is the number of random observations for the $i$-th distribution $\mu_{i,D_i}$.
\end{enumerate}
The following proposition shows that $\tilde{N}_i$ follows the same rate as (P1) with probability approaching one. 
\begin{proposition}
\label{prop:adx:Ni}
    Assume (P1) and ($\text{F1}^\prime)$. 
    Let $\tilde{N}=\min_{1\leq i \leq n}\tilde{N}_i$.
    There exists a constant $\tilde{c}$ such that
    \begin{align*}
        P\left(\tilde{N}\geq \tilde{c}n^{(p+4)/4}\right) \rightarrow 1, \quad \text{as }n \rightarrow\infty. 
    \end{align*}
\end{proposition}




To obtain the consistency of estimates of SAW regression and SWW regression when estimated densities are used,   one needs to  quantify the discrepancy between the  true densities of the distributions $\{\mu_i\}_{i=1}^n$ and their estimates. This preliminary density estimation step can be implemented with standard density estimation methods, which have been  extensively studied in the literature for both univariate and multivariate scenarios \citep{hall:96, haze:09}.

Let $\mu_{D_f}$ be a random probability distribution with density function $f_{D_f}$ that satisfies ($\text{F1}^\prime$), 
from which random observations ${Z}_1,...,{Z}_N$ are independently sampled.
Then a standard kernel estimator $\check{f}$  for $f_{D_f}$  
and its truncated version $\hat{f}$ for the density $f$ 
truncated to the domain $D$ is 
\begin{align}
\label{formula:adx:kerndens}
\check{f}(z) = \frac{1}{Nb^p} \sum_{j=1}^{N} \kappa\left(\frac{z-Z_j}{b}\right), \quad 
\hat{f}(z) = \left. \check{f}(z) \middle/ \int_D \check{f}(u) du\right. , \quad z\in D\subset \mathbb{R}^p.
\end{align}
Here, $\kappa$ is a kernel function and $b$ a positive bandwidth (tuning parameter) satisfying Assumptions (K1) and (K2) listed below.
\begin{enumerate}[label=(K\arabic*), leftmargin=1cm]
\item The kernel function $\kappa$ as per \eqref{formula:adx:kerndens} is a probability density function that has compact support and  is symmetric,  bounded and  $k$ times continuously differentiable (without loss of generality, the  support is assumed to be contained in the unit cube of $\mathbb{R}^p$).
\item For some $A>0$ and $\omega>0$, the class of functions $\mathcal{I}_b = \{\kappa\left(\frac{z-\cdot}{b}\right), z\in \mathbb{R}^p, b>0 \}$ satisfies 
\begin{align*}
\sup_{\mathcal{P}}M\left(\mathcal{I}_b, L_2(\mathcal{P}), \varepsilon\|F \|_{L_2(\mathcal{P})}\right)\leq \left(\frac{A}{\varepsilon}\right)^\omega,
\end{align*}
where $M(T,d,\varepsilon)$ denotes the $\varepsilon$-covering number of the metric space $(T,d)$, $F$ is the envelope function of $\mathcal{I}_b$ and the supremum is taken over the set of all probability measures on $\mathbb{R}^p$. 
\end{enumerate}

\begin{proposition}
\label{prop:adx:multidens}
Assume (D1), ($\text{F1}^\prime$) and (K1).  Choosing $b \sim N^{-\frac{1}{p+4}}$,  the kernel density estimator $\hat{f}$ in (\ref{formula:adx:kerndens}) satisfies that 
\begin{gather}
\label{formula:adx:densEst}
\sup_{f\in\mathcal{F}}E\left[d_2\left(\hat{f}, f\right)^2 \middle\vert f\right] = O(N^{-4/(4+p)}).
\end{gather}
\end{proposition}
We note that by construction, 
  $ \hat{f}\geq 0, \int_D \hat{f}(z)dz = 1$ and since $b \rightarrow 0$ as $N \rightarrow \infty$, we only need to consider the scenario where $b<\epsilon$, when due to the compactness of the support of the kernel 
  $\kappa$ there are no  boundary effects when estimating $f$, as for any $x \in D$ one has $B(x,\epsilon) \subset D_\epsilon$.
   The following proposition formalizes this in terms of a uniform convergence rate.
\begin{proposition}
\label{prop:adx:multidensUnif}
Assume (D1), ($\text{F1}^\prime$) and (K1)--(K2) and $b \sim N^{-\frac{1}{p+4}}$. The  kernel density estimator $\hat{f}$ in (\ref{formula:adx:kerndens}) satisfies 
\begin{gather}
\label{formula:adx:densEst2}
\sup_{f\in\mathcal{F}} d_\infty\left(\hat{f}, f \right)  = O_p(\sqrt{\log N} N^{-\frac{2}{p+4}}).
\end{gather}
\end{proposition}

We have assumed symmetric non-negative kernel functions, thus their first non-zero moments are of order two (alternatively one can also use products of kernels that are one-dimensional symmetric densities). If one wants to take full advantage of the higher order smoothness of the density function $f$ in view of  Assumptions (F2) or (F3), one can use higher order kernels $\kappa$ of order $k$, where the order of the first non-zero moments is $k$. We refer to 
\citet{mull:99} for a more detailed account on multivariate kernels and boundary effects in multivariate density estimation. 
It is well known that such higher order kernel functions cannot be non-negative valued and thus  it is not guaranteed anymore that the resulting kernel density estimates on $D$ satisfy 
$\hat{f} \ge 0$, however one can still show that, since boundary effects can be ignored, 
\begin{align*}
    \sup_{f\in\mathcal{F}}E\left[d_2\left(\hat{f}, f\right)^2 \middle\vert f\right] = O(N^{-2k/(2k+p)}). 
\end{align*} 
For example, under the condition $k\geq p+1$ in (F3) one has 
\begin{align*}
    \sup_{f\in\mathcal{F}}E\left[d_2\left(\hat{f}, f\right)^2 \middle\vert f\right] = O(N^{-2/3}). 
\end{align*} 
Under  (F3), the condition (P1) can then be replaced by a dimension-independent rate, if one adopts the following condition  $(\text{P1}^\prime)$.
\begin{enumerate}[label=(P\arabic*$^\prime$), leftmargin=1cm]
\item $N(n) = \min_{1\leq i \leq n}N_i \gtrsim n^{3/2}$,  where $N_i$ is the number of random observations for the $i$-th distribution $\mu_i$ when densities need to be estimated.
\end{enumerate}

The fact that $\hat{f} \ge 0$ is not guaranteed for these faster converging estimates makes this option less practical for our setting where the density estimates must be bona fide densities. The gains for higher order kernels are a consequence of their bias-reducing property,  while the variance typically suffers a substantial increase through 
larger constants that are associated with these kernels. One can artificially enforce bona fide density estimates by modifying the estimator \citep{gaje:86} but this leads to new problems. If one nevertheless follows through with these adjustments, it may be possible to relax Condition (P1) to ($\text{P1}^\prime$).

\section{Local Sliced Wasserstein Regression}
\label{sec:adx:local}

\subsection{Regression Model}
\label{subsec:adx:localReg}
For local Fr\'{e}chet regression \citep{mull:19:3} we consider the case of a scalar predictor $X\in\mathbb{R}$ while the extension to ${X\in\mathbb{R}^q}$ with $q>1$ is easily possible but tedious and for $q>3$ usually subject to the curse of dimensionality, just like ordinary nonparametric regression approaches.  For a smoothing kernel $K(\cdot)$ corresponding to a probability density and $K_h = h^{-1}K_h(\cdot/h)$, where $h$ is a bandwidth, local Fr\'{e}chet regression at $x$ is defined as 
\begin{align}
\label{formula:adx:loc}
    {m_{L,h}(x)} = \argmin_{\omega\in\mathscr{F}}M_{L,h}(\omega,  x), \quad M_{L,h}(\cdot,x) = E\left[s_L(X,x,h)d^2\left(\mu,\cdot\right)\right],
\end{align}
where $s_L(X, x,h) = K_h(X-x)[\upsilon_2 - \upsilon_1(X-x)] / \sigma^2_0$ with $\upsilon_j = E[K_h(X-x)(X-x)^j]$ for $j=0, 1, 2$ and $\sigma^2_0 = \upsilon_0\upsilon_2 - \upsilon_1^2$. 
The proposed local slice-averaged Wasserstein (LSAW) regression at $x$ is 
\begin{align}
\label{formula:adx:swLoc}
    m_{L,h}^{SAW}(x) = \argmin_{\omega\in\mathscr{F}}M^{SAW}_{L,h}(\omega,  x), \quad M^{SAW}_{L,h}(\cdot,x) = E\left[s_L(X,x,h)d^2_{SW}\left(\mu,\cdot\right)\right].
\end{align}

The proposed local slice-wise Wasserstein (LSWW) regression at $x$ is defined as 
\begin{gather}
    m_{L,h,\tau}^{SWW}(x) = \widetilde{\psi}^{-1}_{\tau}\left[ \argmin_{\gamma\in\Gamma_{\Theta}}M_{L,h}^{SWW}(\gamma,x) \right], \quad M_{L,h}^{SWW}(\cdot, x) = E\left[s_L(X, x, h)d_{DW}^2(\widetilde{\psi}(\mu),\cdot) \right].      \label{formula:adx:transformLocReg} 
\end{gather}
In analogy to Proposition \ref{prop:transformFrechet} for GSWW, the following result shows that the LSWW regression is equivalent to applying the Fr\'{e}chet regression along each slice $\theta$ followed by an inverse transform. 
\begin{proposition}
\label{prop:adx:transformFrechet}
    The minmizing argument  $\gamma_{L,h,x} = \argmin_{\gamma\in\Gamma_{\Theta}}M_{L,h}^{SWW}(\gamma, x)$, see \eqref{formula:adx:transformLocReg}, is  characterized as 
    \begin{align*}
        \gamma_{L,h,x}(\theta) = \argmin_{\nu\in\mathscr{G}}E \left[s_L(X,x, h)d_W^2\left(G^{-1}\left(\widetilde{\psi}(\mu)(\theta)\right), \nu\right)  \right], \quad {\text{for almost all }} \theta\in \Theta.
    \end{align*}
\end{proposition}

\subsection{Estimation}
\label{subsec:adx:estimation}
Suppose we have a sample of independent random pairs $\{(X_i, \mu_i)\}_{i=1}^n\sim F$, then the sample mean and variance are
\begin{align*}
    \bar{X} = n^{-1}\sum_{i=1}^nX_i,\quad \hat{\Sigma}=n^{-1}\sum_{i=1}^n(X_i-\bar{X})(X_i-\bar{X})^\intercal.
\end{align*}
If random distributions $\mu_i$ are fully observed, sample estimators of LSAW and LSWW are obtained as 
\begin{gather}
    \check{m}_{L,h}^{SAW}(x) = \argmin_{\omega\in\mathscr{F}}\check{M}_{L,h}^{SAW}(\omega,  x), \quad \check{M}_{L,h}^{SAW}(\cdot,x) = n^{-1}\sum_{i=1}^n\left[s_{iL, h}(x)d^2_{SW}\left(\mu_i,\cdot\right)\right],     \label{formula:adx:sample0:swLoc}\\
    \check{m}_{L,h,\tau}^{SWW}(x) = \widetilde{\psi}_{\tau}^{-1}\left[ \argmin_{\gamma\in\Gamma_{\Theta}}\check{M}_{L,h}^{SWW}(\gamma,x) \right], \, \check{M}_{L,h}^{SWW}(\cdot, x) = n^{-1}\sum_{i=1}^n\left[s_{iL, h}(x)d_{DW}^2(\widetilde{\psi}(\mu_i),\cdot) \right],  \label{formula:adx:sample0:transformLoc}
\end{gather}
where $s_{iL, h}(x)=K_h(X_i-x)[\hat{\upsilon}_2 - \hat{\upsilon}_1(X_i-x)] / \hat{\sigma}_0^2$ with $\hat{\upsilon}_j = n^{-1}\sum_{i=1}^nK_h(X_i-x)(X_i-x)^j$, $j=0,1,2$ and $\hat{\sigma}_0^2=\hat{\upsilon}_0\hat{\upsilon}_2-\hat{\upsilon}_1^2$. If the random distributions $\mu_i$ are estimated by $\hat{\mu}_i$ as per \eqref{formula:adx:kerndens}, the corresponding estimators are
\begin{gather}
    \hat{m}_{L,h}^{SAW}(x) = \argmin_{\omega\in\mathscr{F}}\hat{M}_{L,h}^{SAW}(\omega,  x), \quad \hat{M}_{L,h}^{SAW}(\cdot,x) = n^{-1}\sum_{i=1}^n\left[s_{iL, h}(x)d^2_{SW}\left(\hat{\mu}_i,\cdot\right)\right];     \label{formula:adx:sample:swLoc} \\
    \hat{m}_{L,h,\tau}^{SWW}(x) = \widetilde{\psi}_{\tau}^{-1}\left[ \argmin_{\gamma\in\Gamma_{\Theta}}\hat{M}_{L,h}^{SWW}(\gamma,x) \right], \, \hat{M}_{L,h}^{SWW}(\cdot, x) = n^{-1}\sum_{i=1}^n\left[s_{iL, h}(x)d_{DW}^2(\widetilde{\psi}(\hat{\mu}_i),\cdot) \right].      \label{formula:adx:sample:transformLocReg} 
\end{gather}

A practical data-driven approach to select the tuning parameter $\tau$ and bandwidth $h$ when an i.i.d. sample of random pairs $\{(X_i, \mu_i)\}_{i=1}^n$ is available can be obtained through leave-one-out cross-validation. 
Specifically, we aim to minimize the discrepancy between predicted and observed distributions, given by
\begin{align*}
    \hat{\tau}, \hat{h} = \argmin_{\tau, h}\sum_{i=1}^nd_{SW}^2\left({\mu_i}, \hat{m}_{L,h,\tau,-i}^{SWW}(X_i)\right),
\end{align*}
where $\hat{m}_{L,h,\tau,-i}^{SWW}(X_i)$ is the prediction at $X_i$ from the LSWW regression of the $i$th-left-out sample $\{(X_{i^{'}}, \mu_{i^{'}})\}_{i^{'}\neq i}$. When the sample size $n$ exceeds $30$, we substitute leave-one-out cross-validation with 5-fold cross-validation to strike a balance between computational efficiency and the accuracy of the tuning parameter selection. 

\subsection{Asymptotic Convergence}
\label{subsec:adx:convergence}
Some additional assumptions listed in Section \ref{sec:adx:assumptions} are required  to derive asymptotic convergence result for the local models. 
Assumption (A3) ensures the underlying minimizer of the LSWW regression belongs to the image space of the slicing transform. 
Additional kernel and distributional assumptions (L1)--(L4) are standard for local regression estimation. 
We provide the following convergence result for LSAW. Here Theorem \ref{thm:adx:swLoc0}  provides rates of convergence for the case where the densities in the random sample are known and Theorem \ref{thm:adx:swLoc} for the case where they are unknown and must be estimated from the data. 

\begin{theorem}
\label{thm:adx:swLoc0}
   (LSAW for known distributions).  Assume (D1), (F1)--(F2), (A1), (T0)--(T2) and (L1)--(L2). When  adopting the slice-averaged Wasserstein distance, for a fixed $x\in \mathbb{R}$ and $m^{SAW}(x), m^{SAW}_{L,h}(x)$, and $\check{m}^{SAW}_{L,h}(x)$ as per \eqref{formula:swFrechet}, \eqref{formula:adx:swLoc} and \eqref{formula:adx:sample0:swLoc}, 
    \begin{gather*}
        d_{SW}(m^{SAW}(x), m^{SAW}_{L,h}(x)) = O\left(h^2\right), \\
        d_{SW}(m_{L,h}^{SAW}(x), \check{m}_{L,h}^{SAW}(x)) = O_p\left(\left(nh\right)^{-1/2}\right),
    \end{gather*}
    and when taking $h\sim n^{-1/5}$, 
    \begin{align*}
        d_{SW}(m^{SAW}(x), \check{m}^{SAW}_{L,h}(x)) = O_p(n^{-2/5}).
    \end{align*}
    Furthermore, assuming (L3)--(L4) for a closed interval $\mathcal{T}\subset\mathbb{R}$, if $h\rightarrow 0 $, $nh^2(-\log h)^{-1}\rightarrow \infty$ as $n\rightarrow \infty$, for any $\epsilon > 0$,
    \begin{gather*}
        \sup_{x\in\mathcal{T}}d_{SW}(m^{SAW}(x), m^{SAW}_{L,h}(x)) = O\left(h^2\right), \\
        \sup_{x\in\mathcal{T}}d_{SW}(m^{SAW}_{L,h}(x), \check{m}^{SAW}_{L,h}(x)  ) = O_p(\max \{ \left(nh^2\right)^{-1/(2+\epsilon)}, [nh^2(-\log h)^{-1}]^{-1/2} \}), 
    \end{gather*}
    and when taking $h\sim n^{-1/(6+2\epsilon)}$, 
    \begin{align*}
        \sup_{x\in\mathcal{T}}d_{SW}(m^{SAW}(x), \check{m}^{SAW}_{L,h}(x)) = O_p\left(n^{-1/(3+\epsilon)}\right). 
    \end{align*}
\end{theorem}

\begin{theorem}
\label{thm:adx:swLoc}
     (LSAW for estimated distributions). Assume (D1), (F1)--(F2), (A1), (T0)--(T2), (L1)--(L2) and assumptions based on kernel density estimation (P1), ($\text{F1}^\prime$), (K1)--(K2). When  adopting the slice-averaged Wasserstein distance, for a fixed $x\in \mathbb{R}$ and $m^{SAW}(x)$, $m^{SAW}_{L,h}(x)$, and $\hat{m}^{SAW}_{L,h}(x)$ as per \eqref{formula:swFrechet}, \eqref{formula:adx:swLoc} and \eqref{formula:adx:sample:swLoc}, 
    \begin{gather*}
        d_{SW}(m^{SAW}(x), m^{SAW}_{L,h}(x)) = O\left(h^2\right), \\
        d_{SW}(m_{L,h}^{SAW}(x), \hat{m}_{L,h}^{SAW}(x)) = O_p\left(\left(nh\right)^{-1/2}\right),
    \end{gather*}
    and when taking $h\sim n^{-1/5}$, 
    \begin{align*}
        d_{SW}(m^{SAW}(x), \hat{m}^{SAW}_{L,h}(x)) = O_p(n^{-2/5}).
    \end{align*}
    Furthermore, assuming (L3)--(L4) for a closed interval $\mathcal{T}\subset\mathbb{R}$, if $h\rightarrow 0 $, $nh^2(-\log h)^{-1}\rightarrow \infty$ as $n\rightarrow \infty$, for any $\epsilon > 0$,
    \begin{gather*}
        \sup_{x\in\mathcal{T}}d_{SW}(m^{SAW}(x), m^{SAW}_{L,h}(x)) = O\left(h^2\right), \\
        \sup_{x\in\mathcal{T}}d_{SW}(m^{SAW}_{L,h}(x), \hat{m}^{SAW}_{L,h}(x)  ) = O_p(\max \{ \left(nh^2\right)^{-1/(2+\epsilon)}, [nh^2(-\log h)^{-1}]^{-1/2} \}), 
    \end{gather*}
    and when taking $h\sim n^{-1/(6+2\epsilon)}$, 
    \begin{align*}
        \sup_{x\in\mathcal{T}}d_{SW}(m^{SAW}(x), \hat{m}^{SAW}_{L,h}(x)) = O_p\left(n^{-1/(3+\epsilon)}\right). 
    \end{align*}
\end{theorem}
The pointwise convergence rate of the LSAW estimator achieves the optimal rates established for local linear estimators for the special case of real-valued responses. 

Turning to the convergence of  LSWW, under Assumption (A3), we define the population-level targets $m^{SWW}$ as 
\begin{gather}
    m^{SWW} = \widetilde{\psi}^{-1}\left[\argmin_{\gamma\in\Gamma_{\Theta}}
    M^{SWW}(\gamma, x)\right]. \label{formula:adx:transformFrechet}
\end{gather}
 For LSWW we again obtain results for two scenarios, when the densities are known in Theorem \ref{thm:adx:transformLoc0}
and when they are unknown and must be estimated in Theorem  \ref{thm:adx:transformLoc}.

\begin{theorem}
\label{thm:adx:transformLoc0}
    (LSWW for known distributions).  Assume (D1), (F1), (F3), (A3), (T0)--(T4) and (L1)--(L2). When  adopting the slice-averaged Wasserstein distance, for a fixed $x\in \mathbb{R}$ and $ m^{SWW}(x)$,  $m^{SWW}_{L,h,\tau}(x)$ and $\check{m}^{SWW}_{L,h,\tau}(x)$ as per \eqref{formula:adx:transformFrechet}, \eqref{formula:adx:transformLocReg} and \eqref{formula:adx:sample0:transformLoc}, with $C_1(\tau)$ and $C_2(\tau)$  from (T3),
    \begin{gather*}
        d_{\infty}( m^{SWW}(x), m^{SWW}_{L,h,\tau}(x)) = O\left(C_1(\tau) + C_2(\tau)h^{8/7}\right), \\
        d_{\infty}(m^{SWW}_{L,h,\tau}(x), \check{m}_{L,h,\tau}^{SWW}(x)) = O_p\left(C_2(\tau)\left(nh\right)^{-2/7}\right),
    \end{gather*}
    and when taking $h\sim n^{-1/5}$, 
    \begin{align*}
        d_{\infty}( m^{SWW}(x), \check{m}^{SWW}_{L,h,\tau}(x)) = O_p\left(C_1(\tau) + C_2(\tau)n^{-8/35}\right).
    \end{align*}
    {Furthermore, assume (L3)--(L4) for a closed interval $\mathcal{T}\subset\mathbb{R}$, if $h\rightarrow 0 $, $nh^2(-\log h)^{-1}\rightarrow \infty$ as $n\rightarrow \infty$, then for any $\epsilon > 0$,
    \begin{gather*}
        \sup_{x\in\mathcal{T}}d_{\infty}( m^{SWW}(x), m^{SWW}_{L,h,\tau}(x)) = O\left(C_1(\tau) + C_2(\tau)h^{8/7}\right), \\
        \sup_{x\in\mathcal{T}}d_{\infty}({m}^{SWW}_{L,h,\tau}(x), \check{m}^{SWW}_{L,h,\tau}(x)  ) = O_p\left(C_2(\tau)\max \left\{ \left(nh^2\right)^{-2/(7+\epsilon)}, \left[nh^2(-\log h)^{-1}\right]^{-2/7} \right\}\right), 
    \end{gather*}
   and  when taking $h\sim n^{-1/(6+\epsilon)}$, 
    \begin{align*}
        \sup_{x\in\mathcal{T}}d_{\infty}\left( m^{SWW}(x), \check{m}^{SWW}_{L,h,\tau}(x)\right) = O_p\left(C_1(\tau) + C_2(\tau)n^{-4/(21+\epsilon)}\right). 
    \end{align*}}
\end{theorem}

\begin{theorem}
\label{thm:adx:transformLoc}
 (LSWW for estimated  distributions).    Assume (D1), (F1), (F3), (A3), (T0)--(T4), (L1)--(L2) and assumptions based on kernel density estimation (P1), ($\text{F1}^\prime$), (K1)--(K2). When  adopting the slice-averaged Wasserstein distance, for a fixed $x\in \mathbb{R}$ and $ m^{SWW}(x)$,  $m^{SWW}_{L,h,\tau}(x)$ and $\hat{m}^{SWW}_{L,h,\tau}(x)$ as per \eqref{formula:adx:transformFrechet}, \eqref{formula:adx:transformLocReg} and \eqref{formula:adx:sample:transformLocReg}, with $C_1(\tau)$ and $C_2(\tau)$  from (T3),
    \begin{gather*}
        d_{\infty}( m^{SWW}(x), m^{SWW}_{L,h,\tau}(x)) = O\left(C_1(\tau) + C_2(\tau)h^{8/7}\right), \\
        d_{\infty}(m^{SWW}_{L,h,\tau}(x), \hat{m}_{L,h,\tau}^{SWW}(x)) = O_p\left(C_2(\tau)\left(nh\right)^{-2/7}\right),
    \end{gather*}
    and when taking $h\sim n^{-1/5}$, 
    \begin{align*}
        d_{\infty}( m^{SWW}(x), \hat{m}^{SWW}_{L,h,\tau}(x)) = O_p\left(C_1(\tau) + C_2(\tau)n^{-8/35}\right).
    \end{align*}
    {Furthermore, assume (L3)--(L4) for a closed interval $\mathcal{T}\subset\mathbb{R}$, if $h\rightarrow 0 $, $nh^2(-\log h)^{-1}\rightarrow \infty$ as $n\rightarrow \infty$, then for any $\epsilon > 0$,
    \begin{gather*}
        \sup_{x\in\mathcal{T}}d_{\infty}( m^{SWW}(x), m^{SWW}_{L,h,\tau}(x)) = O\left(C_1(\tau) + C_2(\tau)h^{8/7}\right), \\
        \sup_{x\in\mathcal{T}}d_{\infty}({m}^{SWW}_{L,h,\tau}(x), \hat{m}^{SWW}_{L,h,\tau}(x)  ) = O_p\left(C_2(\tau)\max \left\{ \left(nh^2\right)^{-2/(7+\epsilon)}, \left[nh^2(-\log h)^{-1}\right]^{-2/7} \right\}\right), 
    \end{gather*}
   and  when taking $h\sim n^{-1/(6+\epsilon)}$, 
    \begin{align*}
        \sup_{x\in\mathcal{T}}d_{\infty}\left( m^{SWW}(x), \hat{m}^{SWW}_{L,h,\tau}(x)\right) = O_p\left(C_1(\tau) + C_2(\tau)n^{-4/(21+\epsilon)}\right). 
    \end{align*}}
\end{theorem}

This result demonstrates the convergence rate of  LSWW regression and provides a decomposition of the reconstruction error into two components, similar to the situation for GSWW. In the special case of a Radon transform, Corollary \ref{cor:adx:RadonLoc} below shows that the curse of dimensionality is manifested in both $C_1(\tau)$ and $C_2(\tau)$, as a higher order of smoothness is required for a higher dimensional distribution to achieve the same convergence rate. Here we only give an explicit result for the  more intricate scenario where densities are not fully observed and must be estimated;  however, analogous results are available for scenarios with fully observed densities.

\begin{corollary}
\label{cor:adx:RadonLoc}
    {When taking the Radon transform $\mathcal{R}$ and the corresponding regularized inverse $\mathcal{R}^{-1}_{\tau}$ as per \eqref{formula:radon} and \eqref{formula:regback}, under the assumptions of Theorem \ref{thm:adx:transformLoc}, 
    \begin{gather*}
        d_{\infty}\left( m^{SWW}(x), \hat{m}_{L,h,\tau}^{SWW}(x)\right) = O_p\left(\tau^{-(k-p)} + \tau^p{n^{-8/35}} \right), \\
        \sup_{x\in\mathcal{T}}d_{\infty}\left( m^{SWW}(x), \hat{m}_{L,h,\tau}^{SWW}(x)\right) = O_p\left(\tau^{-(k-p)} + \tau^pn^{-4/(21+\epsilon)} \right),
    \end{gather*}
    and with $\tau \sim n^{8/(35k)}$,
    \begin{gather*}
        d_{\infty}\left( m^{SWW}(x), \hat{m}_{L,h,\tau}^{SWW}(x)\right) = O_p\left(n^{-8(k-p) / 35k}\right).
    \end{gather*}
    Furthermore, for  $\tau\sim n^{4/(21k)}$, 
    \begin{gather*}
        \sup_{x\in\mathcal{T}}d_{\infty}\left( m^{SWW}(x), \hat{m}_{L,h,\tau}^{SWW}(x)\right) = O_p\left(n^{-4(k-p)/(21k+\epsilon)}\right).
    \end{gather*}}
\end{corollary}

\subsection{Numerical Algorithm}
\label{subsec:adx:alg}
Following the notations in Section \ref{sec:numerical}, the LSAW regression of \eqref{formula:adx:sample:swLoc} given $X=x$ is defined as, 
\begin{align}
\label{formula:adx:SAW_Loc_Numeric}
    \argmin_{\mathbf{W}\in\mathbb{R}^{p\times N}}\mathcal{M}_{L,h}(\mathbf{W},x) = n^{-1}\sum_{i=1}^n\left[ s_{iL,h}(x)d_{SW}^2(\mu_{\mathbf{W}^{(i)}}, \mu_{\mathbf{W}})\right]. 
\end{align}
The following proposition states that the target function of LSAW is smooth for the case of  the Radon transform. 
   
\begin{proposition}[Theorem 1 \citep{bonneel2015sliced}]
    For each fixed $x$ and $N_i \equiv N$, $\mathcal{M}_{L,h}(\mathbf{W},x):\mathbb{R}^{p\times N} \rightarrow \mathbb{R}$ is a $L^1$ function with a uniformly $\rho_L$-Lipschitz gradient for some $\rho_L > 0$ given by 
     \begin{align*}
        \nabla \mathcal{M}_{L,h}(\mathbf{W}, x) = n^{-1}\sum_{i=1}^n\left[ s_{iL,h}(x)  
        \int_{\Theta} \theta
        \left(\mathbf{W}(\theta) - 
        \Pi_{\mathbf{W}(\theta)}^{-1}\circ \Pi_{\mathbf{W}^{(i)}(\theta)}\circ
        \mathbf{W}^{(i)}(\theta)\right)^\intercal d\theta 
        \right].
    \end{align*}
\end{proposition}

In analogy to Algorithm \ref{alg:SAW}, we use the following gradient descent algorithm to find a stationary point.

\begin{algorithm}[h]
\caption{LSAW Algorithm when using the Radon Transform} 
\begin{algorithmic}[1]
    \State Initialize a grid $(\theta_1,\theta_2,...,\theta_L)$ along $\Theta$
    \State Set $N = \min_{i=1,...,n}N_i$, convergence threshold $\varepsilon$ and learning rate $\eta$
    \State For each $\mu_{\mathbf{W}^{(i)}}$, downsample $\mathbf{W}^{(i)}$ such that  $\mathbf{W}^{(i)} \in \mathbb{R}^{p\times N}$
    \State Initialize $\mathbf{W}^{[0]}\in\mathbb{R}^{p\times N}$ arbitrarily and fix the output predictor $X=x$
    \Repeat 
        \State Calculate $\nabla 
        \mathcal{M}_{L,h}(\mathbf{W}^{[k]}, x)$ through
        \begin{align*}
            \nabla 
        \mathcal{M}_{L,h}(\mathbf{W}^{[k]}, x) = (nL)^{-1} \sum_{i=1}^n 
        \sum_{l=1}^L
        s_{iL,h}(x)  \theta_l\left(\mathbf{W}(\theta_l) - \Pi_{\mathbf{W}(\theta_l)}^{-1}\circ \Pi_{\mathbf{W}^{(i)}(\theta_l)}\circ \mathbf{W}^{(i)}(\theta_l)\right)^\intercal
        \end{align*}
        \State {$\mathbf{W}^{[k+1]} = \mathbf{W}^{[k]} - \eta \nabla 
        \mathcal{M}_{L,h}(\mathbf{W}^{[k]}, x) $} 
    \Until Algorithm converges with $\|\mathbf{W}^{[k+1]} - \mathbf{W}^{[k]} \|_2 / \| \mathbf{W}^{[k]}\|_2 < \varepsilon$ to $\mathbf{W}^{[\infty]}$
    \State Consider each column of $\mathbf{W}^{[\infty]}$ as a sample from $\hat{m}_{L,h}^{SAW}(x)$ and apply the kernel density estimator \eqref{formula:adx:kerndens} to derive the density estimator $\hat{f}$
	\end{algorithmic}
\label{alg:LSAW}
\end{algorithm}

\section{Additional Data Application}
\label{sec:adx:data}
\subsection{Exchange Traded Funds Modeling}
We provide an additional data application to further illustrate the LSAW and LSWW regression models, aiming to demonstrate that these models can be used for the smoothing (nonparametric regression) of multivariate distributions against a one-dimensional covariate, for which we use calendar year in this application. Exchange Traded Funds (ETFs) are investment vehicle that invests assets to track a benchmark, such as a general index, sector, bonds, fixed income, etc. Sector ETFs that track a particular industry have become popular among investors and are widely used for hedging and statistical arbitrage. Historical data for sector ETFs can be obtained from Yahoo finance \url{https://finance.yahoo.com/}.

\begin{figure}[]
    \single
  \centering
  \begin{subfigure}{\textwidth}
    \centering
    \includegraphics[width=0.82\linewidth]{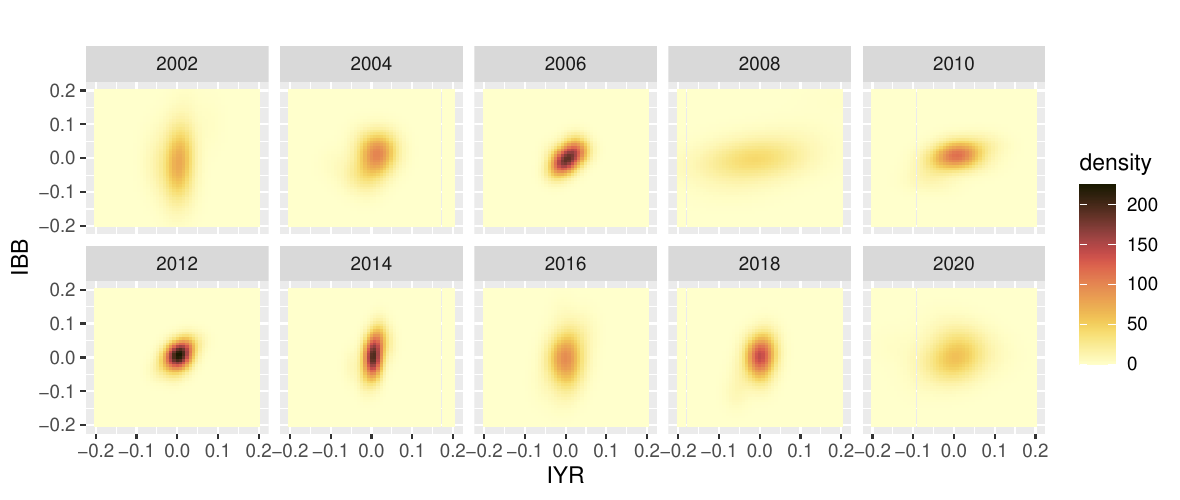}
    \caption{}
  \end{subfigure}\hfill
  \begin{subfigure}{\textwidth}
    \centering
    \includegraphics[width=0.82\linewidth]{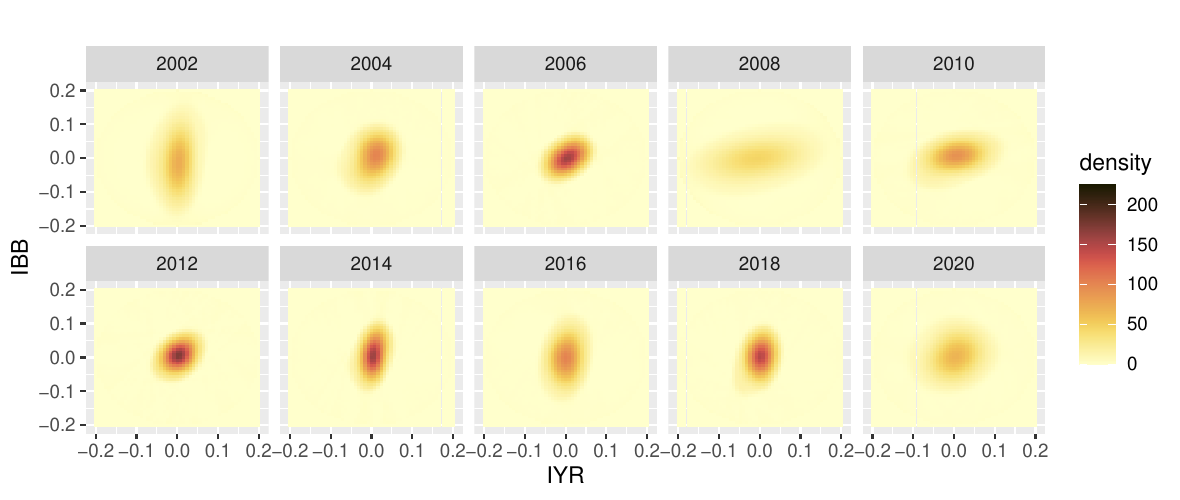}
    \caption{}
  \end{subfigure}\hfill
  \begin{subfigure}{\textwidth}
    \centering
    \includegraphics[width=0.82\linewidth]{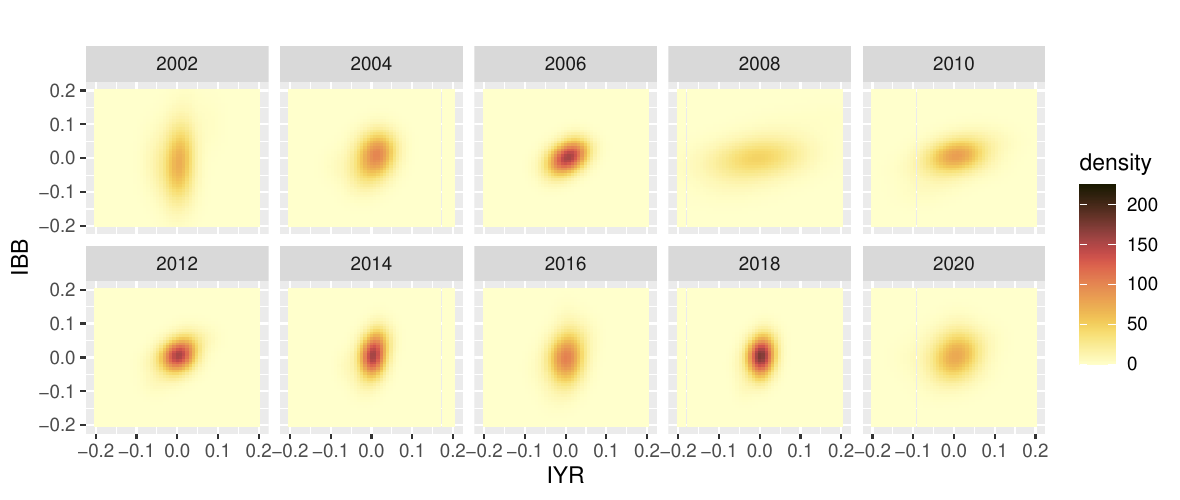}
    \caption{}
  \end{subfigure}
  \caption{(a) Observed smoothed densities and fitted densities from (b) local slice-wise Wasserstein regression (LSWW) and (c) local sliced-average Wasserstein regression (LSAW) of the joint distributions of Biotechnology (IBB) ($y$-axis) and Real Estate (IYR) ETFs ($x$-axis). Each panel is labeled by year. The sliced Wasserstein fractions of variance explained are 0.95 for LSWW and 0.74 for LSAW.}
  \label{fig:ETF}
\end{figure}

\begin{figure}[tb]
    \single
    \centering
    \includegraphics[width=\linewidth]{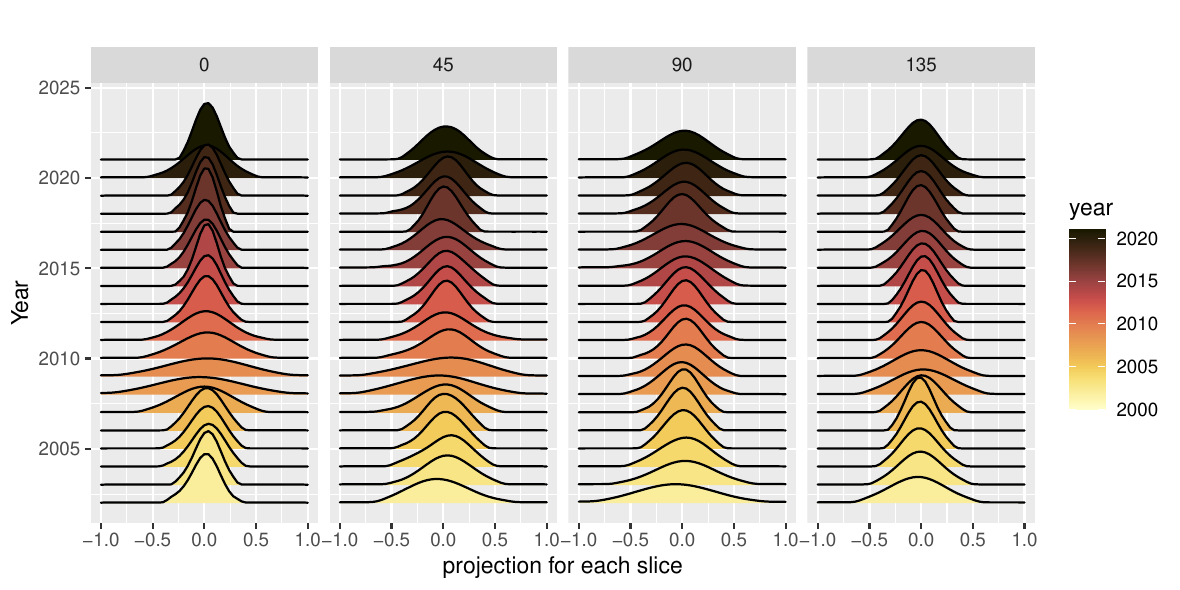}
    \caption{Fr\'{e}chet regressions for LSWW between year (predictor, on the $y$-axis) and fitted slicing distributions (response, on the $x$-axis) for various projections. The number at the top of each panel indicates the angle of the respective projection with the $x$-axis (IYR) in Figure \ref{fig:ETF} (a).}
    \label{fig:ETFSliceWise}
\end{figure}

For each year, we model the bivariate distribution between the weekly return of iShares Biotechnology ETF (IBB) and the weekly return of iShares U.S. Real Estate ETF (IYR). Kernel density estimates of the data are in Figure \ref{fig:ETF} (a). Figure \ref{fig:ETFSliceWise} displays the local Fr\'{e}chet regressions for representative angles for LSWW. The variance of weekly returns for the two included sectors increased during the financial crisis of 2008-2009 and the COVID-19 pandemic in 2020, indicating increased market uncertainty during these periods. The real estate market was impacted more significantly than biotechnology, likely due to its sensitivity to economic and financial conditions.

Reconstructed density surfaces obtained from LSWW are shown in Figure \ref{fig:ETF} (b) while those obtained from  LSAW are shown in Figure \ref{fig:ETF} (c). The sliced Wasserstein fraction of variance explained for LSWW and LSAW models, as per \eqref{formula:SWFVE}, is 0.95 and 0.74. During periods of economic expansion and positive investor sentiment, the biotech and real estate sectors both benefit and become positively correlated, as was the case during the housing bubble from 2002 to 2008 and the post-crisis period from 2010 to 2018. Conversely, during times of economic uncertainty or market volatility, correlations between sectors tend to decrease as investors shift toward safe-haven assets. This was evident during the financial crisis of 2008-2009 and the COVID-19 pandemic in 2020. Both LSWW and LSAW models reveal the general trends of the joint distribution and achieve satisfactory values of the sliced Wasserstein fraction of variance explained.




\section{Slicing Transforms}
\label{sec:adx:slicingTransform}
Various slicing transforms besides the Radon transform may be of interest. One of these is the circular Radon transform $\mathcal{CR}$ \citep{kuchment2006generalized}. This transform  is the integral of a function $f$ over a sphere centered at $\theta \in \Theta$, where 
\begin{align*}
    \mathcal{CR}(f)(\theta,u) = \int_{|z-\theta| =u}f(z)d\sigma(z), \quad \theta \in \Theta.
\end{align*}
The injectivity property of the circular Radon transform has been studied previously, especially for the case when $\Theta$ is a unit sphere \citep{agranovsky1996injectivity, ehrenpreis2003universality}. However, the analytic inverse formulas for even dimensions are still unknown \citep{finch2004determining}.

Another transform of interest is the generalized Radon transform, which extends the classic Radon transform \citep{beylkin1984inversion, ehrenpreis2003universality}.
The generalized Radon transform $\mathcal{GR}$ is defined as an integral  over $I_{u,\theta} = \{z\in D | \chi(z, \theta) = u\}$,
\begin{align*}
    \mathcal{GR}\left(f\right)(\theta,u) = \int_{\chi(z,\theta)=u}f(z)d\sigma(z),
\end{align*}
where $d\sigma(z)$ integrates the surface area  on $I_{u,\theta}$ and $\chi$ is a so-called defining function if it satisfies some regularity conditions \citep{beylkin1984inversion}. The sliced Wasserstein distance has been extended to the generalized Radon transform \citep{kolouri2019generalized}.

\section{Auxiliary Lemmas and Propositions}
\label{sec:adx:lemmas}
\begin{lemma}
\label{lemma:adx:RadonAssumption}
Assume (D1) and (F1). For any vector $\mathbf{k} = (k_1,...,k_p)$ of $p$ non-negative integers with $\sum_{l=1}^pk_l= k$ let $\mathcal{D}^\mathbf{k} =  \frac{\partial^{k}}{\partial z_{k_1}\cdots \partial z_{k_p}}$. It holds that  $\mathcal{R}(f)(\theta,\cdot), f\in\mathcal{F},$ is $k-$times differentiable, and 
\begin{align}
\label{formula:adx:RadonDiff}
    \mathcal{R}\left[
    \mathcal{D}^{\mathbf{k}}f
    \right](\theta,u) = (-1)^k\left(\prod_{l=1}^p\theta_l^{k_l}\right) \frac{\partial^k\left(\mathcal{R}(f)\right)}{\partial u^k}
    (\theta,u), 
\end{align}
for each $\theta=(\theta_1,...,\theta_p)\in \Theta$. 
Furthermore, 
\begin{align*}
    \left| \frac{\partial^k(\mathcal{R}(f))  }{ \partial u^k}(\theta,u) \right|\leq B_1p^{k/2}<\infty,
\end{align*}
for each $\theta\in \Theta$ where the constant $B_1$ does not depend on $f$ or $\theta$.
\end{lemma}

\begin{proof}
Assumptions (D1) and (F1) imply $f$ has bounded support and continuous partial derivatives of order $k$, and  formula \eqref{formula:adx:RadonDiff} follows from Proposition 6.1.3 in \citet{epstein2007introduction}. For each $\theta = (\theta_1,...,\theta_p) \in \Theta$, $\sum_{j=1}^p\theta_j^2=1$, 
$$
\max_{j=1,..,p}|\theta_j| \geq p^{-1/2}.
$$
Let $j(\theta) = \argmax_{j=1,\cdots,p}|\theta_j|$. Taking $\mathbf{k}(\theta) = (0,...,0,k,0, ...,0)$ with all components equal to $0$ except the $j(\theta)$-th as $k$, so that 
\begin{align*}
    \left|\left(\prod_{l=1}^p\theta_l^{k_l}\right)\right| = \left(\prod_{l=1}^p|\theta_l|^{k_l}\right) \geq p^{-k/2}. 
\end{align*}
Since $\norm{\mathcal{D}^{\mathbf{k}(\theta)}f}_\infty \leq B_1$ for a constant $B_1$ from (F1), it follows from the definition of the Radon transform that $\norm{\mathcal{R}\left[\mathcal{D}^{\mathbf{k}(\theta)}f\right](\theta,u)}_\infty \leq B_1$. Hence, 
\begin{align*}
    \abs{\frac{\partial^k\left(\mathcal{R}(f)\right)}{\partial u^k}
    (\theta,u)} \leq B_1p^{k/2}. 
\end{align*}
\end{proof}

\begin{lemma}
\label{lemma:adx:derivative}
    Suppose $h_1, h_2$ are three times continuously differentiable  on [0,1] with \newline  $\max \{  \|h_1^{(3)}(s)\|_\infty,  \|h_2^{(3)}(s)\|_\infty\}\leq B_0$ where $B_0$ is a constant, then 
    \begin{align*}
        |h_1^\prime(s) - h_2^\prime(s)| \leq C(B_0)d_2^{4/7}(h_1, h_2), \quad \forall z\in[0,1].
    \end{align*}
\end{lemma}

\begin{proof}
        Consider a kernel function $\mathcal{K}(\cdot)$ that is a symmetric probability density function with compact support on [-1,1] and a bounded derivative $\mathcal{K}^\prime$. Furthermore, assume $\mathcal{K}$ satisfies that $\sigma^2(\mathcal{K}) = \int u^2\mathcal{K}(u)du<\infty$ and $\sigma^3(\mathcal{K}) = \int |u|^3\mathcal{K}(u)du<\infty$.
        From Lemma 1 in \citet{chen2020rank}, arbitrarily fix $s\in(0,1)$, and assume for the  bandwidth $a$ that $a \leq \min\{s, 1-s\}$. As $a\rightarrow 0$, 
    \begin{gather}
         \int_0^1a^{-1}h_1(u)\mathcal{K}\left(\frac{s-u}{a}\right)du = h_1(s) + \frac{1}{2}a^2\sigma^2(\mathcal{K})h_1^{(2)}(s) + R_{11}(a), \label{formula:adx:kernel1} \\  
    \int_0^1a^{-2}h_1(u)\mathcal{K}^\prime\left(\frac{s-u}{a}\right)du = h_1^\prime(s) + \frac{1}{2}a^2\sigma^2(\mathcal{K})h_1^{(3)}(s) + R_{12}(a). \label{formula:adx:kernel2}  
    \end{gather}

    Expanding  $h_1(u)$, 
    \begin{align*}
        h_1(u) = h_1(s) + h_1^\prime(s)(u-s) + h_1^{(2)}(s)(u-s)^2/2 + \int_s^u\frac{h_1^{(3)}(t)(u-t)^2}{3!}dt. 
    \end{align*}
    Combining   the expansion of $h_1(u)$  with  \eqref{formula:adx:kernel1} and \eqref{formula:adx:kernel2}, the 
     remainder terms $R_{11}(a), R_{12}(a)$ can be bounded as follows 
    \begin{align*}
        \left|R_{11}(a)\right| = \left|\int_0^1a^{-1}\mathcal{K}\left(\frac{s-u}{a}\right)\int_s^u\frac{h_1^{(3)}(t)(u-t)^2}{3!}dtdu\right| 
        \lesssim B_0a^3,
    \end{align*}
    \begin{align*}
        R_{12}(a)&= \int_0^1a^{-2}\mathcal{K}^\prime\left(\frac{s-u}{a}\right)\int_s^u\frac{h_1^{(3)}(t)(u-t)^2}{3!}dtdu \\
        & \quad -\int_0^1a^{-1}\mathcal{K}\left(\frac{s-u}{a}\right)\frac{h_1^{(3)}(s)(u-s)^2}{3!} du  \lesssim B_0a^2. 
    \end{align*}
    Similarly for $h_2(u)$, we have
    \begin{align*}
        \int_0^1a^{-2}h_2(u)\mathcal{K}^\prime\left(\frac{s-u}{a}\right)du = h_2^\prime(s) + \frac{1}{2}a^2\sigma^2(\mathcal{K})h_2^{(3)}(s) + R_{22}(a).
    \end{align*}
    Then 
    \begin{align*}
        |h_1^{\prime}(s) - h_2^{\prime}(s)| &\leq a^{-2}\int_0^1|h_1(u)-h_2(u)|\mathcal{K}^\prime\left(\frac{s-u}{a}\right)du + \frac{1}{2}a^2\sigma^2(\mathcal{K})|h_1^{(3)}(u) - h_2^{(3)}(u)| \\
        &\quad + \left|R_{12}(a)-R_{22}(a) \right| \\
        &\leq a^{-3/2}d_2(h_1, h_2)\left(\int \left(\mathcal{K}^\prime(u)\right)^2du\right)^{1/2} + 
        B_0a^2\sigma^2(\mathcal{K}) + \left|R_{12}(a)+R_{22}(a) \right|\\
        &\lesssim C_1(B_0)\left( a^{-3/2}d_2(h_1, h_2) \right)  +C_2(B_0)a^2,
    \end{align*}
    for constants $C_1(B_0), C_2(B_0)$ which only depend on $B_0$. 
    By choosing $a \sim d_2^{2/7}(h_1, h_2)$, 
     \begin{align*}
        |h_1^\prime(s) - h_2^\prime(s)| \leq C(B_0)d_2^{4/7}(h_1, h_2).
    \end{align*}
\end{proof}

\begin{lemma}
\label{lemma:adx:distRelation}
    If $D$ is compact, the univariate Wasserstein distance is bounded by the $L^2$ distance of the corresponding densities 
    \begin{align}
        d_{W}(\nu_1, \nu_2) \lesssim d_2\left(\varphi(\nu_1), \varphi(\nu_2)\right), \quad \nu_1, \nu_2 \in \mathscr{G}.
    \end{align}
\end{lemma}

\begin{proof}
    From Theorem 4 in \citet{gibbs2002choosing}, 
    \begin{align*}
        d_W(\nu_1, \nu_2) \leq \text{diam}(D)d_{TV}(\nu_1, \nu_2), \quad \nu_1, \nu_2 \in \mathscr{G},
    \end{align*}
    where $\text{diam}(D) = \sup\{ d(z_1, z_2): z_1, z_2\in D \}$ and 
    $d_{TV}(\nu_1, \nu_2)$ represents the total variation distance $d_{TV} = \sup_{S\subset D}|\nu_1(S)- \nu_2(S)|$. Note that when $\nu_1$ and $\nu_2$ have densities $\varphi(\nu_1)$ and $\varphi(\nu_2)$, it follows that 
    \begin{align*}
        d_{TV}(\nu_1, \nu_2) = \frac{1}{2}\int_{D}|\varphi(\nu_1)(u)-\varphi(\nu_2)(u)|du \leq \frac{1}{2}\sqrt{\text{diam}(D)}d_2(\varphi(\nu_1), \varphi(\nu_2)).
    \end{align*}
    If $D$ is a compact set, $\text{diam}(D)$ is bounded. We conclude that 
    \begin{align}
        d_{W}(\nu_1, \nu_2) \lesssim d_2(\varphi(\nu_1), \varphi(\nu_2)).
    \end{align}
\end{proof}

\begin{proposition}
\label{prop:adx:Connection}
{Assume (T0) and (T2), 
for $m_G^{SAW}$ as per \eqref{formula:swGlob} and $M_G^{SWW}$ as per \eqref{formula:transformGlobReg}, 
    \begin{align*}
        m_G^{SAW}(x) = \widetilde{\psi}^{-1}\left[ \argmin_{\gamma \in \Gamma_{\Theta}\cap \widetilde{\psi}(\mathscr{F})}M_G^{SWW}(\gamma,x) \right].
    \end{align*}
Similarly, for $m_{L,h}^{SAW}$ as per \eqref{formula:adx:swLoc} and $M_{L,h}^{SWW}$ as per \eqref{formula:adx:transformLocReg}, 
    \begin{align*}
        m_{L,h}^{SAW}(x) = \widetilde{\psi}^{-1}\left[ \argmin_{\gamma \in \Gamma_{\Theta}\cap \widetilde{\psi}(\mathscr{F})}M_{L,h}^{SWW}(\gamma,x) \right].
    \end{align*}}
\end{proposition}

\begin{proof}
 The change of variable ${\gamma} = \widetilde{\psi}(\omega)\in \widetilde{\psi}(\mathscr{F})$, $\omega\in \mathscr{F}$  is a bijection because of the injectivity of the transform $\widetilde{\psi}$. Hence, 
    \begin{align*}
        m_G^{SAW}(x) &= \argmin_{\omega\in\mathscr{F}} E\left[s_G(X,x)d_{SW}^2(\mu, \omega)\right]  \\
        & = \argmin_{\omega\in\mathscr{F}} E\left[s_G(X,x)d_{DW}^2(\widetilde{\psi}(\mu), \widetilde{\psi}(\omega))\right]  \\
        & = \widetilde{\psi}^{-1}\left(\argmin_{{\gamma}\in\Gamma_{\Theta}\bigcap \widetilde{\psi}(\mathscr{F})}
         E\left[s_G(X,x)d_{DW}^2(\widetilde{\psi}(\mu), \gamma)\right] \right) \\
        & = \widetilde{\psi}^{-1}\left[ \argmin_{{\gamma}\in\Gamma_{\Theta}\bigcap \widetilde{\psi}(\mathscr{F})}M_{G}^{SWW}(\omega,x) \right].
    \end{align*}
Similarly,  
\begin{align*}
    m_{L,h}^{SAW}(x) = \widetilde{\psi}^{-1}\left[ \argmin_{\gamma \in \Gamma_{\Theta}\cap \widetilde{\psi}(\mathscr{F})}M_{L,h}^{SWW}(\gamma,x) \right].
\end{align*}
\end{proof}

\section{Proofs}
\label{sec:adx:proof}

\subsection{Proof of Proposition \ref{prop:adx:Ni}}
\begin{proof}
    Denoting by 
 $\tilde{N}_i$ the number of observations made for $f_{i, D_i|D_\epsilon}$ that fall  within the domain $D_\epsilon$, it is clear that each $\tilde{N}_i$ follows a binomial distribution $\mathscr{B}(p_i,N_i)$, where $p_i=\int_{D_\epsilon} f_{i, D_i|D_\epsilon}(z)dz \ge M_0^{-1} |D_\epsilon|$ according to ($\text{F1}^\prime$).
From (P1), there exists a constant $c>0$, such that $\min_{1\leq i \leq n}N_i\geq c n^{1+p/4}$. 
Let $\tilde{c} = c\cdot |D_\epsilon|/(2M_0)$. 
Hoeffding's inequality \citep{hoeffding1994probability} implies that for any $\rho \le N_ip_i$, 
\begin{align*}
    P\left(\tilde{N}_i\leq \rho\right) \leq \exp\left(-2N_i\left(p_i-\frac{\rho}{N_i} \right)^2\right).
\end{align*}
Whence, 
\begin{align*}
    P\left(\min_{1\leq i\leq n}\tilde{N_i}> \tilde{c}n^{1+p/4} \right) & = \prod_{i=1}^n P\left(\tilde{N}_i >  \tilde{c}n^{1+p/4} \right) \\
    &= \prod_{i=1}^n \left(1 - P\left(\tilde{N}_i \leq   \tilde{c}n^{1+p/4} \right) \right) \\
    &\geq \prod_{i=1}^n \left( 1 - e^{-2N_i\left(p_i - |D_\epsilon| / (2M_0) \right)^2} \right).
\end{align*}
Denote $\tilde{m}_i = p_i - |D_\epsilon| / (2M_0) \geq |D_\epsilon| / (2M_0) > 0$, then 
\begin{align*}
    P\left(\min_{1\leq i\leq n}\tilde{N_i}> \tilde{c}n^{1+p/4} \right) & \geq 
    \left( 1 - e^{-2N_i\tilde{m}_i^2} \right)^n \\
    &\geq \left( 1 - e^{-2c\tilde{m}_i^2 n^{1+p/4}} \right)^n \\
    & = e^{n\log\left( 1 - e^{-2c\tilde{m}_i^2 n^{1+p/4}} \right)} \\
    &\rightarrow 1 \text{  as }n\rightarrow \infty. 
\end{align*}
\end{proof}

\subsection{Proof of Proposition \ref{prop:radonT0T1}}
\begin{proof}
First, we show that if Assumptions (D1), (F1) are satisfied, $\mathcal{R}(\mathcal{F})$ satisfies Assumptions (D2) and (G1).
It is clear that $\mathcal{R}(f) \geq 0$ and $\int_u\mathcal{R}(f)(\theta, u)du = 1$, where we denote $\mathcal{R}(f)(\theta, u)$ as $\mathcal{R}(f)(\theta)(u)$ for simplicity.
The validity of (D2) follows from (D1) and the definition of the Radon transform.
Furthermore, (G1) holds based on (D1), (F1) and the implications of Lemma \ref{lemma:adx:RadonAssumption}.
Next, we prove that (T0) and (T1) are satisfied.  The injectivity of the Radon transform is evident from the definition. Hence, (T0) is satisfied. 
Since $D$ is compact, there exists a constant $C$ such that 
$\norm{D}_\infty \leq C$. Computing the $L^2$-distance between $\mathcal{R}(f_1)$ and $\mathcal{R}(f_2)$ for fixed $\theta\in\Theta$,
\begin{align*}
    &{\int_{\mathbb{R}}\abs{\mathcal{R}(f_1)(\theta,u) - \mathcal{R}(f_2)(\theta,u)}^2du }\\
    & = \int_{-C}^C \left(\int_{-C}^C\cdots \int_{-C}^C 
    \left(\left(f_1-f_2\right)\left(u\theta+\sum_{j=1}^{p-1}s_je_j\right)
    \right)ds_1\cdots ds_p\right)^2du \\
    & \leq (2C)^{p-1} \int_{-C}^C\cdots \int_{-C}^C \left(\left(f_1-f_2\right)\left(u\theta+\sum_{j=1}^{p-1}s_je_j\right)\right)^2ds_1\cdots ds_pdu \\
    &\leq (2C)^{p-1} d_2^2(f_1,f_2). 
\end{align*}
The first inequality follows from the Cauchy-Schwarz inequality. Thus, (T1) holds as well.
Assumptions (T2) and (T3) naturally follow as a corollary from Theorem \ref{thm:reginv}.
\end{proof}

\subsection{Proof of Theorem \ref{thm:reginv}}
\begin{proof}
    Set ${\widebar{\mathcal{R}(f)}(\theta,r)}  = \mathcal{J}_1(\mathcal{R}(f)(\theta))(r)$, $r\in\mathbb{R}$ and $\check{\Delta}_{\tau,1} = f - \check{\mathcal{R}}_\tau(\mathcal{R}(f))$. Then we have
        \begin{align*}
            \check{\Delta}_{\tau,1} &= \frac{1}{2(2\pi)^p} \int_{\Theta}\int_{|r|>\tau} \widebar{\mathcal{R}(f)}(\theta,r)e^{ir\langle \theta, z\rangle}|r|^{p-1}drd\theta.
        \end{align*} 
    Since $\mathcal{R}(f)(\theta)(\cdot)$ has uniform bounded $k$-derivative for each $\theta\in\Theta$ from (G1), it follows from Proposition 4.2.1 
     in \citet{epstein2007introduction} that 
    \begin{align*}
        |\widebar{\mathcal{R}(f)}(\theta,r)| \leq M_3|r|^{-k}, 
    \end{align*}
    where $M_3$ is a constant such that $\|\mathcal{R}(f)(\theta)(\cdot)\|_\infty\leq M_3$. Therefore, we get
    \begin{align*}
        \norm{\check{\Delta}_{\tau,1}}_\infty &\leq 
        \frac{1}{2(2\pi)^p} \int_{\Theta}\int_{|r|>\tau} \abs{\widebar{\mathcal{R}(f)}(\theta,r)}|r|^{p-1}drd\theta \\
        &\leq \frac{M_3}{2(2\pi)^p}  \int_{\Theta}d\theta \int_{|r|>\tau}|r|^{p-k-1}dr \\
        &=  O\left(\tau^{-(k-p)}\right),
    \end{align*}
    where we note that $k\geq p+1$ from {(F3)}. 
    Since $\int_{D} f(z)dz=1>0$ and $\|f\|_\infty\leq M_0$ from (F1),  we obtain $\|\Delta_{\tau,1}\|_\infty = O\left(\tau^{-(k-p)}\right)$.
    
    Next, set ${\lambda_{f} = \mathcal{R}(f)}$, $\lambda_{f}^{*} = \widetilde{\mathcal{R}(f)}\in \Lambda_{\Theta}$ and $\check{\Delta}_{\tau,2} = \check{\mathcal{R}}^{-1}_\tau(\lambda_{f}) - \check{\mathcal{R}}^{-1}_\tau(\lambda_{f}^*)$,. This yields 
    \begin{align*}
        \norm{\check{\Delta}_{\tau,2} }_\infty &= 
        \frac{1}{2(2\pi)^p}  \int_{\Theta}\int_{|r|\leq \tau} (\widebar{\lambda_{f}}(\theta,r) - \widebar{\lambda_{f}^*}(\theta,r) )|r|^{p-1}e^{ir\langle \theta, z\rangle}drd\theta \\
        &\leq \frac{\tau^{p-1}}{2(2\pi)^p}  \int_{\Theta}\int_{|r|\leq \tau}\abs{\widebar{\lambda_{f}}(\theta,r) -\widebar{\lambda_{f}^*}(\theta,r) }drd\theta \\
        &\leq \frac{\tau^{p-1}}{2(2\pi)^p} \int_{\Theta} \int_{|r|\leq \tau}\int_{\mathbb{R}} |\lambda_{f}(\theta,u) - \lambda_{f}^*(\theta,u)|dudrd\theta \\
        &\lesssim \frac{ \tau^{p}}{2(2\pi)^p} d_2(\lambda_{f}, \lambda_{f}^*).
    \end{align*}
    The last inequality follows from the bounded support of $\Lambda_{\Theta}$ and $\mathcal{F}$. Note that  $\|\check{\mathcal{R}}_{\tau}(\mathcal{R}(f))- f\|_\infty= O\left(\tau^{-(k-p)}\right)$ as $\tau \rightarrow \infty$. 
    The boundedness of $\Delta_{\tau,2}$ follows from the fact  $\int_D f(z)dz=1$ and the condition that $d_2(\lambda_{f}^*, \lambda_{f})\rightarrow 0$, whence
    \begin{align*}
        \norm{\Delta_{\tau,2} }_\infty = O\left(\tau^pd_2(\lambda_{f},\lambda_{f}^*)\right) = O\left(\tau^pd_2\left({\mathcal{R}(f), \widetilde{\mathcal{R}(f)}}\right)\right).
    \end{align*}
\end{proof}

\subsection{Proof of Proposition \ref{prop:well-defined}}
This proof follows a similar approach as  Proposition 1 of \citet{kolouri2019generalized}, but we present it for a more generalized version of the slicing transform as follows. 
If $\mu_1 = \mu_2, \,\mu_1, \mu_2\in\mathscr{F}$,  the slice-averaged Wasserstein distance satisfies $d_{SW}(\mu_1, \mu_2) =0$. 
The non-negativity and symmetry properties of the slice-averaged Wasserstein distance are direct consequences of the Wasserstein distance being a metric. We next consider the triangle inequality. For any $\mu_1,\mu_2,\mu_3\in\mathscr{F}$,
\begin{align*}
    d_{SW}(\mu_1, \mu_3) &= \left(\int_{\Theta}d_W^2(G^{-1}\left(\widetilde{\psi}(\mu_1)(\theta)\right), G^{-1}\left(\widetilde{\psi}(\mu_3)(\theta)\right)) \right)^{1/2}\\
    &\leq \left(\int_{\Theta}d_W^2(G^{-1}\left(\widetilde{\psi}(\mu_1)(\theta)\right), G^{-1}\left(\widetilde{\psi}(\mu_2)(\theta)\right)) \right)^{1/2} \\
    &\quad +\left(\int_{\Theta}d_W^2(G^{-1}\left(\widetilde{\psi}(\mu_2)(\theta)\right), G^{-1}\left(\widetilde{\psi}(\mu_3)(\theta)\right)) \right)^{1/2}\\
    &= d_{SW}(\mu_1, \mu_2) + d_{SW}(\mu_2, \mu_3)
\end{align*}
The last inequality is obtained using the Minkowski inequality. We have thus established that the slice-averaged Wasserstein distance satisfies non-negativity, symmetry and the triangle inequality, hence it is a pseudo-metric. If $d_{SW}(\mu_1, \mu_2) = 0$,  
\begin{align*}
    d_W\left(G^{-1}\left(\widetilde{\psi}(\mu_1)(\theta)\right) , G^{-1}\left(\widetilde{\psi}(\mu_2)(\theta)\right)\right) = 0, \quad \text{for almost all }\theta \in \Theta.
\end{align*}
Equivalently, considering that the Wasserstein distance is a metric, we have
    $\widetilde{\psi}(\mu_1) = \widetilde{\psi}(\mu_2)$.
Therefore, the slice-averaged Wasserstein distance is a distance if and only if $\widetilde{\psi}(\mu_1) = \widetilde{\psi}(\mu_2)$ implies $\mu_1=\mu_2$, which is equivalent to the injectivity of $\widetilde{\psi}$. Note that $\widetilde{\psi}$ is induced from $\psi$, and since $\varphi$ and $\varrho$ are bijective, the injectivity of $\widetilde{\psi}$ is equivalent to  (T0).

\subsection{Proof of Proposition \ref{prop:transformFrechet} and Proposition \ref{prop:adx:transformFrechet}}
\begin{proof}
    For any $\gamma_1,\gamma_2\in\Gamma_{\Theta}$, 
    \begin{align*}
        d_{DW}^2(\gamma_1, \gamma_2 ) &= \int_{\Theta} d_{W}^2(G^{-1}\left(\gamma_1(\theta)\right), G^{-1}\left(\gamma_2(\theta)\right))d\theta \\
        &\lesssim \int_{\Theta}d_2^2(\varrho(\gamma_1)(\theta), \varrho(\gamma_2)(\theta))d\theta,\\
        &\lesssim \int_{\Theta}d_\infty^2(\varrho(\gamma_1)(\theta), \varrho(\gamma_2)(\theta))d\theta.
    \end{align*}
    The first inequality comes from the distance relationship discussed in Lemma \ref{lemma:adx:distRelation}. 
    Following the boundedness condition in (D1) and (G1), 
    $ d_{DW}(\gamma_1, \gamma_2 ) $ is bounded from above. Therefore,  
    \begin{align*}
    &\quad E\left[s_G(X,x)d_{DW}^2(\widetilde{\psi}(\mu),\gamma) \right] \\
    &= \int_{X, \mu}s_G(X,x)\int_{\Theta} d_W^2\left(G^{-1}\left(\widetilde{\psi}(\mu)(\theta) \right), G^{-1}\left(\gamma(\theta)\right)\right)d\theta d_F(X,\mu)\\
    &= \int_{\Theta}\int_{X, \mu}s_G(X,x) d_W^2\left(G^{-1}\left(\widetilde{\psi}(\mu)(\theta) \right), G^{-1}\left(\gamma(\theta)\right)\right)d_F(X,\mu)d\theta\\
    &= \int_{\Theta}  E \left[s_G(X,x)d_W^2\left(G^{-1}\left(\widetilde{\psi}(\mu)(\theta) \right), G^{-1}\left(\gamma(\theta)\right) \right) \right]  d\theta.
    \end{align*}
   Since  $s_G(X,x)>0$, the second equation follows from the Fubini theorem. 
    The last equation indicates that finding the minimum in $E\left[s_G(X,x)d_{DW}^2(\widetilde{\psi}(\mu),\gamma) \right]$  
    is equivalent to finding the minimum in $E \left[s_G(X,x)d_W^2\left(G^{-1}\left(\widetilde{\psi}(\mu)(\theta) \right), G^{-1}\left(\gamma(\theta)\right) \right) \right]$ for almost all $\theta \in \Theta$.
    The result for the local slice-wise Wasserstein regression can be derived analogously.
\end{proof}

\subsection{Proof of Proposition \ref{prop:adx:multidens} and Proposition \ref{prop:adx:multidensUnif}}
\begin{proof}
    Set $z = (z_1,...,z_p)^\intercal\in D\subset D_\epsilon$, $t=(t_1,...,t_p)^\intercal\in D_\epsilon$ and the density $f_{D_f}$ on domain $D_\epsilon$ as $f_{D_\epsilon} = f_{D_f}\mathbf{1}_{D_\epsilon}$. We can express the expected value as follows
    \begin{align}
        E\left[\check{f}(z)\middle\vert f\right] &= \frac{1}{b^p}\int_{D_\epsilon}\kappa\left(\frac{z-t}{b}\right)f_{D_\epsilon}(t)dt, \quad z\in D. 
    \label{formula:adx:kernelDensExp}
    \end{align}
    Expanding the multivariate density function $f_{D_\epsilon}(t)$ yields
    \begin{align*}
        f_{D_\epsilon}(t) = f_{D_\epsilon}(z) + \sum_{s=1}^p\frac{\partial f_{D_\epsilon} (z)}{\partial z_s}(t_s-z_s) 
        + \sum_{s_1=1}^p\sum_{s_2=1}^p\frac{\partial f_{D_\epsilon}^2(z^*)}{\partial z_{s_1}z_{s_2}}(t_{s_1}-z_{s_1})(t_{s_2}-z_{s_2})/2, \quad t\in D_\epsilon, 
    \end{align*}
    where $z^* = z+\alpha(t-z) \in D_\epsilon$ for a $\alpha\in (0,1)$. With \eqref{formula:adx:kernelDensExp},
    \begin{align*}
        E\left[\check{f}(z)\middle\vert f\right] = f_{D_\epsilon}(z) + \frac{1}{2b^p}\sum_{s=1}^p\int_{D_\epsilon} 
        \frac{\partial^2 f_{D_\epsilon}(z^*)}{\partial z_s^2}
        \kappa\left( \frac{z-t}{b}\right)(z_s-t_s)^2dt, \quad z\in D.
    \end{align*}
    Here we use the symmetry property (K1) of the kernel function. From  ($\text{F1}^\prime$), there exists a constant  $M_3$ that does not depend on the function $f_{D_\epsilon}$ such that the partial derivative $|\partial^2 f_{D_\epsilon}(z^*) / \partial z_s^2|$ is uniformly bounded by $M_3$ for $s=1,...,p$. Then 
    \begin{align}
    \label{formula:adx:densBias}
        \sup_{z\in D}\sup_{f\in\mathcal{F}}\left| E\left[\check{f}(z)\middle\vert f\right] - f_{D_\epsilon}(z)\mathbf{1}_{D} \right| = O(b^2).
    \end{align} 
    Next, we establish the boundedness of the variance of $\check{f}(z)$, 
    \begin{align*}
        \Var \left(\check{f}(z) \middle\vert f \right) &= \frac{1}{N}\Var\left(\kappa\left(\frac{z-Z_{j}}{b}\right)b^{-p} \middle\vert f \right) \\ 
        &\leq \frac{1}{N}\int_{D_\epsilon}\kappa\left(\frac{z-t}{b}\right)^2b^{-2p}f_{D_\epsilon}(t)dt\\
        &\leq \frac{M_0}{N} \int_{D_\epsilon}\kappa\left(\frac{z-t}{b}\right)^2b^{-2p}dt,
    \end{align*}
    where $M_0$ from Assumption (F1) is a constant that does not depend on the density function $f_{D_\epsilon}$. From (K2), 
    \begin{align*}
        \sup_{z\in D}\sup_{f\in\mathcal{F}}\Var \left(\check{f}(z) \middle\vert f \right) = O\left(\frac{1}{Nb^p}\right).
    \end{align*}
    Combining the above results, we obtain
    \begin{align*}
        \sup_{f\in\mathcal{F}} E\left(d_2\left(\check{f}, f_{D_\epsilon}\mathbf{1}_D \right)^2 \middle\vert f\right) = O\left(\frac{1}{Nb^p} + b^4\right).
    \end{align*}
    Choosing $b\sim N^{-\frac{1}{p+4}}$ leads to 
    \begin{align*}
        \sup_{f\in\mathcal{F}} E\left(d_2\left(\check{f}, f_{D_\epsilon}\mathbf{1}_D\right)^2\middle\vert f \right) = O\left(N^{-4/(4+p)}\right).
    \end{align*}
For a fixed $f\in\mathcal{F}$ and the corresponding $f_{D_\epsilon}$, by \eqref{formula:adx:densBias} and Proposition 9 in \citet{rina:10},
\begin{align*}
    \sup _{f \in \mathcal{F}} d_\infty\left(\check{f}, f_{D_\epsilon}\mathbf{1}_D\right) = O(b^2) + O_p\left(\sqrt{\frac{\log N}{Nb^p}}\right) = O_p(\sqrt{\log N} N^{-\frac{2}{p+4}}).
\end{align*}
The error of uniform bound can be decomposed into a deterministic term $E[\check{f}(z)]-f_{D_\epsilon}(z)$ and a probabilistic term $\check{f} - E[\check{f}(z)]$. The first part depends on smoothness properties of $f_{D_\epsilon}$ only while the second part is characterized via empirical process techniques \citep{rina:10, jiang2017uniform, chen2017tutorial}.

Note that the truncated density on domain $D$ is presented as $f(z) = {f_{D_\epsilon}(z)}/{\int_{D}}f_{D_\epsilon}(u)du, z\in D$. 
We can then provide the convergence result for the truncated density $\hat{f}$. Note that when $N$ is large enough we have, 
\begin{gather*}
    \int_D {f_{D_\epsilon}}(u) du \geq \frac{|D|}{M_0}, \quad \int_D \check{f}(u)du \geq \frac{|D|}{2M_0}.
\end{gather*}
Whence,
\begin{align*}
    d_2\left(\hat{f}, f\right) &\leq d_2\left(\frac{\check{f}(z)}{\int_D \check{f}(u)du}, \frac{f_{D_\epsilon}(z)\mathbf{1}_{D}}{\int_D \check{f}(u)du} \right) + d_2\left(\frac{{f_{D_\epsilon}}(z)\mathbf{1}_{D}}{\int_D \check{f}(u)du}, \frac{f_{D_\epsilon}(z)\mathbf{1}_{D}}{\int_D {f_{D_\epsilon}}(u)du} \right) \\
    &\leq \frac{2M_0}{|D|}d_2(\check{f}, f_{D_\epsilon}\mathbf{1}_D) + \frac{2M_0^3}{|D|^{3/2}}d_{2}(\check{f}, f_{D_\epsilon}\mathbf{1}_D). 
\end{align*}
Similarly, we have 
\begin{align*}
    d_\infty\left(\hat{f}, f \right) 
    &\leq \frac{2M_0}{|D|}d_\infty(\check{f}, f_{D_\epsilon}\mathbf{1}_D) + \frac{2M_0^3}{|D|}d_{\infty}(\check{f}, f_{D_\epsilon}\mathbf{1}_D). 
\end{align*}
It follows that when choosing $b\sim N^{-\frac{1}{p+4}}$
\begin{gather*}
    \sup _{f \in \mathcal{F}} E\left[d_2\left(\hat{f}, f\right)^2 \right]=O\left(N^{-4 /(4+p)}\right), \\
    \sup _{f \in \mathcal{F}} 
    d_\infty\left(\hat{f}, f\right) =  O_p(\sqrt{\log N} N^{-\frac{2}{p+4}}).
\end{gather*}

\end{proof}

\subsection{Proof of Theorem \ref{thm:swGlob}}
\begin{proof} The proof proceeds in two steps. 

    {\bf Step 1.}  We first prove $d_{SW}\left(m_G^{SAW}(x), \check{m}_G^{SAW}(x)\right) = O_p\left(n^{-1/2}\right)$ and  establish the following three properties for the SAW estimator $\check{m}_G^{SAW}(x)$. 
\begin{enumerate}[label=(R\arabic*), leftmargin=1cm]
\item The objects $m_G^{SAW}(x)$ and $\check{m}_G^{SAW}(x)$ exist and are unique, the latter almost surely, and, for any $\epsilon>0$, 
    \begin{align*}
        \inf _{d_{SW}\left(m_G^{SAW}(x), \omega\right)>\epsilon} M_G^{SAW}(\omega, x)>M_G^{SAW}\left(m_G^{SAW}(x), x\right).
    \end{align*}
\item Let $B_{\delta}\left[m_G^{SAW}(x)\right]$ be the $\delta$-ball in $\mathscr{F}$ centered at $m_G^{SAW}(x)$ and $N(\epsilon, \mathscr{F}, d_{SW})$  its covering number using balls of size $\epsilon$. Then
    \begin{align*}
        \int_0^1\left(1+\log N\left\{\delta \epsilon, B_\delta\left[m_G^{SAW}(x)\right], d_{SW}\right\}\right)^{1 / 2} d \epsilon=O(1) \quad \text { as } \delta \rightarrow 0.
    \end{align*}
\item There exists $\eta_0>0, \beta_0>0,$ possibly depending on $x$, such that 
    \begin{align*}
        \inf _{d_{SW}\left(m_G^{SAW}(x), \omega\right)<\eta_0}\left\{M_G^{SAW}(\omega, x)-M_G^{SAW}\left(m_G^{SAW}(x), x\right)-\beta_0 d_{SW}\left(m_G^{SAW}(x), \omega\right)^{2}\right\} \geq 0.
    \end{align*}
\end{enumerate}

\paragraph{Proof of (R2).} We note that $d_W^2(\nu_1, \nu_2) = O(d_2(\varphi(\nu_1), \varphi(\nu_2)) )$ from Lemma \ref{lemma:adx:distRelation}, whence 
    \begin{align}
        d_{SW}(\mu_1,\mu_2) &= \left(\int_{\Theta}d_W^2(G^{-1}\left(\widetilde{\psi}(\mu_1)(\theta)\right), G^{-1}\left(\widetilde{\psi}(\mu_2)(\theta)\right))d\theta \right)^{1/2}\nonumber \\
       &\lesssim\left(\int_{\Theta}d_{2}^2(
       \psi\circ \varphi (\mu_1)(\theta), \psi\circ \varphi(\mu_2)(\theta))d\theta \right)^{1/2}\nonumber \\
        &\lesssim d_{2}(\varphi(\mu_1), \varphi(\mu_2)), \label{formula:SWRelation}
    \end{align}
    where the last inequality follows from (T1) and the compactness of $\Theta$. 
    Using Theorem 2.7.1 of \citet{vaart1996weak}, 
    there exists a constant $A_1$ depending only on $k$ and $p$ such that 
    \begin{align*}
      \log N(\epsilon,\mathcal{F}  , \| \|_\infty) \leq A_1{\epsilon}^{-p/k},
    \end{align*}
    for every $\epsilon > 0 $ and $k>p/2$ from (F2). 
    Note that $d_{SW}(\mu_1, \mu_2) = O(d_{2}(\varphi(\mu_1), \varphi(\mu_2))) = O(d_{\infty}(\varphi(\mu_1), \varphi(\mu_2))$ and  from the boundedness of the support $D$, we have $B_{A_2\epsilon}(\mu, \|\|_\infty)\subset B_{\epsilon}(\mu, d_{SW})$ for some constant $A_2$. Thus, we have $N(\epsilon, \mathscr{F}, d_{SW})\leq A_3 N(\epsilon, \mathcal{F},\|\|_\infty)$. 
    For $\tilde{k}=p/k<2$
    it follows that 
    \begin{align}
    \label{formula:adx:entropy}
        \int_0^1\sqrt{1 + \log N\{\delta \epsilon, B_{\delta}[m_G^{SAW}(x)], d_{SW}\}}d\epsilon  
        &< \int_0^1\sqrt{1 + \log N\{\delta \epsilon, \mathscr{F}, d_{SW}\}}d\epsilon \nonumber \\
        &\leq \int_0^1\sqrt{1+A_3A_1\epsilon^{-p/k}}d\epsilon \\
        &\lesssim \int_0^1 \epsilon^{-\tilde{k}/2} < \infty. 
    \end{align}

\paragraph{Proof of (R1) and (R3).}
    We define the Hilbert space $\mathcal{H}$ as the set of all functions 
    \begin{align*}
        \mathcal{H}:=\{\varsigma\in\mathcal{H}: \Theta \times [0,1] \rightarrow \mathbb{R}, \int_{\Theta}\int_{[0,1]}\varsigma^2(\theta, s)ds d\theta<\infty\}
    \end{align*}
    with the inner product 
    \begin{align*}
        \langle \varsigma_1, \varsigma_2 \rangle =\int_{\Theta}\int_{[0,1]} \varsigma_1(\theta,s)\varsigma_2(\theta,s)ds d\theta.
    \end{align*} 
    Here the integral is well defined due to the Cauchy-Schwarz inequality. 
    It is easy to verify that the space $\mathcal{H}$ is a vector space over the field $\mathbb{R}$. The inner product satisfies the conditions of conjugate symmetry, linearity, and positive definiteness. 
    The compactness of the Hilbert space follows from the fact that $\mathcal{H}$ consists of measurable functions that are square integrable.
    We  define the $L^2$ distance between two functions $\varsigma_1,\varsigma_2\in\mathcal{H}$ as $d_2(\varsigma_1, \varsigma_2) = \langle \varsigma_1-\varsigma_2, \varsigma_1-\varsigma_2 \rangle$. 
     Note that the quantile slicing space $\Gamma_{\Theta}$ is a subspace of $\mathcal{H}$ and the distribution slicing Wasserstein metric coincides with the $L^2$ distance, i.e., $d_{DW}(\gamma_1,\gamma_2) = d_2(\gamma_1, \gamma_2)$ for $\gamma_1, \gamma_2\in\Gamma_{\Theta}$.
    Let $B_G(x) = E[s_G(X,x)\widetilde{\psi}(\mu)]$, then for any fixed $\omega\in\mathscr{F}$,
    \begin{align*}
    M_G^{SAW}(\omega, x) 
        &=E[s_{G}(X,x)d_{SW}^2(\mu, \omega)] \\
        &=E[s_G(X,x)d_2^2(\widetilde{\psi}(\mu), \widetilde{\psi}(\omega))] \\
        & = E[s_G(X,x) \langle \widetilde{\psi}(\mu)-\widetilde{\psi}(\omega), \widetilde{\psi}(\mu)-\widetilde{\psi}(\omega) \rangle] \\
        & = E[s_G(X,x)\langle \widetilde{\psi}(\mu)-B_G(x), \widetilde{\psi}(\mu)-B_G(x) \rangle] \\
        & \qquad +E[s_G(X,x)\langle B_G(x)-\widetilde{\psi}(\omega), B_G(x)-\widetilde{\psi}(\omega) \rangle] \\
        & \qquad +2E[s_G(X,x) \langle \widetilde{\psi}(\mu)-B_G(x), B_G(x)-\widetilde{\psi}(\omega)  \rangle]\\
        &= E[s_G(X,x)d_2^2(\widetilde{\psi}(\mu), B_G(x))] + E[s_G(X,x)d_2^2(B_G(x), \widetilde{\psi}(\omega))]\\
        &= 
        E[s_G(X,x)d_2^2(\widetilde{\psi}(\mu), B_G(x))] + d_2^2(B_G(x), \widetilde{\psi}(\omega)),
    \end{align*}
     where the last equation follows from the fact that $E[s_G(X,x)] = 1$.  Thus, 
    \begin{align*}
        m_G^{SAW}(x) =  \argmin_{\omega \in \mathscr{F}}d_2^2(B_G(x), \widetilde{\psi}(\omega)).
    \end{align*}
    Using  $n^{-1}\sum_{i=1}^ns_{iG}(x)=1$, one can similarly show that 
    \begin{align}
        \label{formula:adx:SWminimizer} 
        \check{m}_G^{SAW}(x) = \argmin_{\omega \in \mathscr{F}}d_2^2(\check{B}_G(x), \widetilde{\psi}(\omega)),
    \end{align}
    where $\check{B}_G(x) = n^{-1}\sum_{i=1}^ns_{iG}(x)\widetilde{\psi}(\mu_i)$. 
    From the convexity and closedness of space $\widetilde{\psi}(\mathscr{F})$, the minimizers
    $m_G^{SAW}(x)$ and $\check{m}_G^{SAW}(x)$ exist and are unique for any $x\in\mathbb{R}^q$, so that (R1) is satisfied. 
    The best approximation $m_G^{SAW}(x)\in\mathscr{F}$ can be characterized by \cite[chap.4]{deutsch2012best}
    \begin{align}
    \label{formula:adx:characterization}
        \langle B_G(x)-\widetilde{\psi}(m_G^{SAW}(x)), \widetilde{\psi}(\omega)-\widetilde{\psi}(m_G^{SAW}(x) )\rangle \leq 0, \text{ for all }\omega \in \mathscr{F}. 
    \end{align}
    Consequently, $d_2^2(B_G(x), \widetilde{\psi}(\omega)) \geq d_2^2(B_G(x), \widetilde{\psi}(m_G^{SAW}(x))) + d_2^2(\widetilde{\psi}(m_G^{SAW}(x)), \widetilde{\psi}(\omega))$. Then, 
    \begin{align*}
        M_G^{SAW}(\omega, x) &\geq M_G^{SAW}(m_G^{SAW}(x),x) +  d_{2}^2(\widetilde{\psi}(m_G^{SAW}(x)), \widetilde{\psi}(\omega))\\
        &= M_G^{SAW}(m_G^{SAW}(x),x) +  d_{SW}^2(m_G^{SAW}(x), \omega),
    \end{align*}
    for all $\omega \in \mathscr{F}$. Hence, we may choose $\beta_0=1$. 
    
    Under Properties (R1)--(R3), it follows from \citet{mull:19:3} that 
    \begin{align}
        d_{SW}(m_G^{SAW}(x), \check{m}_G^{SAW}(x)) = O_p(n^{-1/2}).
    \label{formula:adx:globSW1}
    \end{align}

    {\bf Step 2.}  Here we show that $
        d_{SW}(m_G^{SAW}(x), \hat{m}_G^{SAW}(x)) = O_p(n^{-1/2})$.
    As before,  $\varphi(\mu_i)$ and $\varphi(\hat{\mu}_i)$ denote the  density functions of the $i$-th sample distribution $\mu_i$ and the estimator $\hat{\mu}_i$ (see \eqref{formula:adx:kerndens}). From Proposition \ref{prop:adx:multidens}, 
    \begin{align*}
        \max_{i=1,...,n} d_2(\varphi(\mu_i), \varphi(\hat{\mu}_i)) = O_p\left(N^{-2/ (4+p)}\right),
    \end{align*}
    and the derivation of \eqref{formula:SWRelation} implies 
    \begin{align*}
        d_2(\widetilde{\psi}(\mu_i), \widetilde{\psi}(\hat{\mu}_i)) = d_{SW}(\mu_i, \hat{\mu}_i) \lesssim d_2\left(\varphi(\mu_i),\varphi(\hat{\mu}_i)\right).
    \end{align*}
    Consequently,
    \begin{align*}
        \max_{i=1,...,n}d_2(\widetilde{\psi}(\mu_i), \widetilde{\psi}(\hat{\mu}_i)) = O_p\left(N^{-2/ (4+p)}\right). 
    \end{align*}
    Setting  $\hat{B}_G(x) = n^{-1}\sum_{i=1}^ns_{iG}(x)\widetilde{\psi}(\hat{\mu}_i)$ and  noting  that $n^{-1}\sum_{i=1}^ns_{iG}(x) = 1$, 
    \begin{align*}
        d_2(\check{B}_G(x), \hat{B}_G(x)) = O_p\left(N^{-2/ (4+p)}\right). 
    \end{align*}
    Similarly from the derivation of \eqref{formula:adx:SWminimizer}, we have 
    \begin{align*}
        \hat{m}_G^{SAW}(x) = \argmin_{\omega \in \mathscr{F}}d_2(\hat{B}_G(x), \widetilde{\psi}(\omega)).
    \end{align*}
     By Theorem 5.3 of \citet{deutsch2012best}, considering the closeness and convexity of $\widetilde{\psi}(\mathcal{F})$, we obtain
     \begin{align*}
         d_{2}(\widetilde{\psi}(\check{m}_G^{SAW}(x)), \widetilde{\psi}(\hat{m}_G^{SAW}(x)) ) = O_p\left(N^{-2/ (4+p)}\right),
     \end{align*}
     whence
     \begin{align}
     \label{formula:adx:SWminimizerEst}
         d_{SW}\left(\check{m}_G^{SAW}(x), \hat{m}_G^{SAW}(x)\right) = O_p\left(N^{-2/ (4+p)}\right). 
     \end{align}
     From \eqref{formula:adx:globSW1} and Assumption (P1), we conclude that 
      \begin{align*}
         d_{SW}(m_G^{SAW}(x), \hat{m}_G^{SAW}(x)) = O_p(n^{-1/2}). 
     \end{align*}

     Uniform convergence results require stronger versions of these properties provided by (R4)--(R6) as stated below.  Let $\|\cdot\|_E$ be the Euclidean norm on $\mathbb{R}^q$ and $B>0$ a constant. 
\begin{enumerate}[label=(R\arabic*), leftmargin=1cm]
\setcounter{enumi}{3}
\item Almost surely, for all $\|x \|_E\leq B$, the objects $m_G^{SAW}(x)$ and $\check{m}_G^{SAW}(x)$ exist and are unique. Additionally, for any $\epsilon> 0$,
     \begin{align*}
         \inf _{\|x\|_E \leq B} \inf_{d_{SW}\left(m_G^{SAW}(x), \omega\right)>\epsilon}\left\{M_G^{SAW}(\omega, x)-M_G^{SAW}\left[m_G^{SAW}(x), x\right]\right\}>0
     \end{align*}
        and there exists $\zeta = \zeta(\epsilon)>0$ such that
        \begin{align*}
            \operatorname{pr}\left(\inf_{\|x\|_E \leq B} \inf_{d_{SW}\left[\check{m}_G^{SAW}(x), \omega\right]>\epsilon}\left\{\check{M}_G^{SAW}(\omega, x)-\check{M}_G^{SAW}\left[\check{m}_G^{SAW}(x), x\right]\right\} \geq \zeta\right) \rightarrow 1.
        \end{align*}
\item With $B_{\delta}[m_G^{SAW}(x)]$ and $N\left\{\epsilon, B_\delta\left[m_G^{SAW}(x)\right], d_{SW}\right\}$ as in Condition (R2), 
     \begin{align*}
         \int_0^1 \sup _{\|x\|_E \leq B}\left(1+\log N\left\{\delta \epsilon, B_\delta\left[m_G^{SAW}(x)\right], d_{SW}\right\}\right)^{1 / 2} d \epsilon=O(1) \quad \text { as } \delta \rightarrow 0.
     \end{align*}
\item There exist $\tau_0>0$, $C_0>0$, possibly depending on $B$, such that 
     \begin{align*}
         \inf_{\substack{\|x\|_E \leq B, \\ 
         d_{SW}\left(m_G^{SAW}(x), \omega\right)<\tau_0 }}\left\{M_G^{SAW}(\omega, x)-M_G\left[m_G^{SAW}(x), x\right]-C_0 d_{SW}\left[m_G^{SAW}(x), \omega\right]^{2}\right\} \geq 0 .
     \end{align*}
\end{enumerate}

The derivation of (R1) and (R3) lead to (R4) and (R6) with $C_0=1$. From equation \eqref{formula:adx:entropy}, we find that (R5) is satisfied. From Theorem 2 of \citet{mull:19:3}, we conclude that 
\begin{align*}
 \sup_{\|x\|_E\leq B}d_{SW}\left(m_G^{SAW}(x), \check{m}_G^{SAW}(x)\right) = O_p\left(n^{-1/(2+\epsilon)}\right).
\end{align*}
Note that formula \eqref{formula:adx:SWminimizerEst} is uniform for $\|x\|\leq B$, leading to 
\begin{align*}
 \sup_{\|x\|_E\leq B}d_{SW}\left(m_G^{SAW}(x), \hat{m}_G^{SAW}(x)\right) = O_p\left(n^{-1/(2+\epsilon)}\right).
\end{align*}

\end{proof}

\subsection{Proof of Theorem \ref{thm:adx:swLoc}} 
\begin{proof}
We first show that 
\begin{gather}
    d_{SW}\left(m^{SAW}(x), m_{L,h}^{SAW}(x)\right) = O\left(h^2\right) \label{formula:adx:swLoc1},\\  d_{SW}\left(m_{L,h}^{SAW}(x), \check{m}_{L,h}^{SAW}(x)\right) = O_p\left(\left(nh\right)^{-1/2}\right). \label{formula:adx:swLoc2}
\end{gather}
We establish the following three properties for the SAW estimator $\check{m}_{L,h}^{SAW}(x)$.
\begin{enumerate}[label=(U\arabic*), leftmargin=1cm]
\item The minimizers $m^{SAW}(x), m^{SAW}_{L,h}(x)$ and $\check{m}_{L,h}^{SAW}(x)$ exist and are unique, the last almost surely. Additionally, for any $\epsilon>0$,
\begin{gather*}
\inf _{d_{SW}\left(m^{SAW}(x), \omega\right)>\epsilon}\{M^{SAW}(\omega, x)-M^{SAW}[m^{SAW}(x), x]\}>0, \\
\liminf _{h \rightarrow 0} \inf _{d_{SW}\left(m^{SAW}_{L,h}(x), \omega\right)>\epsilon}\left\{M^{SAW}_{L, h}(\omega, x)-M^{SAW}_{L, h}\left[m^{SAW}_{L,h}(x), x\right]\right\}>0 .
\end{gather*}

\item Let $B_\delta[m^{SAW}(x)]$ be the ball of radius $\delta$ centered at $m^{SAW}(x)$ with covering number $N\left\{\delta \epsilon, B_\delta[m^{SAW}(x)], d_{SW}\right\}$. Then 
\begin{align*}
    \int_0^1\left(1+\log N\left\{\delta \epsilon, B_\delta[m^{SAW}(x)], d_{SW}\right\}\right)^{1 / 2} d \epsilon=O(1) \quad \text { as } \delta \rightarrow 0.
\end{align*}

\item There exists $\eta_1,\eta_2>0$, $\beta_1, \beta_2>0$ such that
\begin{multline*}
\inf _{d_{SW}\left(m^{SAW}(x), \omega\right)<\eta_1}\left\{M^{SAW}(\omega, x)-M^{SAW}[m^{SAW}(x), x]-\beta_1 d_{SW}\left(m^{SAW}(x), \omega\right)^{2}\right\} \geq 0, \\
\liminf _{h \rightarrow 0} \inf _{d_{SW}\left(m^{SAW}_{L,h}(x), \omega\right)<\eta_2}\left\{M^{SAW}_{L,h}(\omega, x)-M^{SAW}_{L,h}\left[m^{SAW}_{L,h}(x), x\right] \right.\\
\left. -\beta_2 d_{SW}\left(m^{SAW}_{L,h}(x), \omega\right)^{2}\right\} \geq 0.
\end{multline*}
\end{enumerate}

\paragraph{Proof of (U2).} Similar to the derivation of \eqref{formula:adx:entropy}, we have
\begin{align*}
        \int_0^1\sqrt{1 + \log N\{\delta \epsilon, B_{\delta}[m^{SAW}(x)], d_{SW}\}}d\epsilon  
        &< \int_0^1\sqrt{1 + \log N\{\delta \epsilon, \mathscr{F}, d_{SW}\}}d\epsilon \nonumber \\
        &\leq \int_0^1\sqrt{1+A_3A_1\epsilon^{-p/k}} d\epsilon < \infty. 
    \end{align*}

\paragraph{Proof of (U1) and (U3).} For any distribution $\mu\in\mathscr{F}$, let $\widetilde{\psi}(\mu)\in\widetilde{\psi}(\mathscr{F})$  and $B(x) = E[\widetilde{\psi}(\mu)|X=x]$. Then 
\begin{align*}
    M^{SAW}(\omega, x) &= E[d_{SW}^2(\mu, \omega)|X=x] \\
                   &= E[d_2^2(\widetilde{\psi}(\mu), \widetilde{\psi}(\omega))|X=x] \\ 
                   &= E[d_2^2(\widetilde{\psi}(\mu), B(x))|X=x] + E[d_2^2(B(x), \widetilde{\psi}(\omega))|X=x] \\
                   &= M^{SAW}\left[B(x), x\right] + d_2^2\left(B(x), \widetilde{\psi}(\omega)\right)
\end{align*} 
for all $\omega \in \mathscr{F}$,  whence
\begin{align*}
    m^{SAW}(x) = \argmin_{\omega\in\mathscr{F}}d_2^2\left(B(x), \widetilde{\psi}(\omega)\right).
\end{align*}
Set $B_{L,h}(x) = E\left[s_L(X, x,h)\widetilde{\psi}(\mu)\right], \check{B}_{L,h}(x) = n^{-1}\sum_{i=1}^ns_{iL,h}(x)\widetilde{\psi}(\mu_i)$. Considering $E[s_L(x,h)]=1$ and $n^{-1}\sum_{i=1}^ns_{iL,h}(x)=1$, one finds 
\begin{gather*}
    m^{SAW}_{L,h}(x) = \argmin_{\omega\in\mathscr{F}}d_2^2\left(B_{L,h}(x), \widetilde{\psi}(\omega)\right), \\
    \check{m}^{SAW}_{L,h}(x) = \argmin_{\omega\in\mathscr{F}}d_2^2\left(\check{B}_{L,h}(x), \widetilde{\psi}(\omega)\right).
\end{gather*}

From the convexity and closedness of space $\widetilde{\psi}(\mathscr{F})$, the minimizers $m^{SAW}(x)$, $m_{L,h}^{SAW}(x)$ and $\check{m}_{L,h}^{SAW}(x)$ exist  uniquely for any $x\in\mathbb{R}$, so that (U1) is satisfied. Following the characterization of the best approximation as per \eqref{formula:adx:characterization}, we have 
\begin{align*}
    M^{SAW}(\omega, x) \geq M^{SAW}\left(m^{SAW}(x), x\right) + d_{SW}^2\left(m^{SAW}(x), \omega\right).
\end{align*}
Similarly, 
\begin{align*}
    M^{SAW}_{L,h}(\omega, x) \geq M^{SAW}_{L,h}\left(m^{SAW}_{L,h}(x), x\right) + d_{SW}^2\left(m^{SAW}_{L,h}(x), \omega\right).
\end{align*}
Thus, (U3) is satisfied with $\beta_1=\beta_2=1$ and $\eta_1, \eta_2$ chosen arbitrarily. 
Under Conditions (L1)--(L2) and (U1)--(U3), it follows from Theorem 3 and Theorem 4 in \citet{mull:19:3} that 
\begin{gather*}
    d_{SW}\left(m^{SAW}(x), m_{L,h}^{SAW}(x)\right) = O\left(h^{2 }\right), \\ d_{SW}\left(m_{L,h}^{SAW}(x), \check{m}_{L,h}^{SAW}(x)\right) = O_p\left(\left(nh\right)^{-1/2}\right).
\end{gather*}
Hence, we conclude that \eqref{formula:adx:swLoc1} and \eqref{formula:adx:swLoc2} are satisfied. Similarly to the derivation of \eqref{formula:adx:SWminimizerEst}, it follows that 
\begin{align*}
    d_{SW}\left(m^{SAW}_{L,h}(x), \hat{m}^{SAW}_{L,h}(x)\right) = O_p\left(\left(nh\right)^{-1/2} + N^{-4/(4+p)}\right). 
\end{align*}
From Assumption (P1), we conclude that 
\begin{align*}
    d_{SW}\left(m^{SAW}_{L,h}(x), \hat{m}^{SAW}_{L,h}(x)\right) = O_p\left(\left(nh\right)^{-1/2}\right). 
\end{align*}

Next, we provide stronger versions of the assumptions and then show that the previous results can be extended to hold uniformly over a closed interval $\mathcal{T}\subset \mathbb{R}$. 
\begin{enumerate}[label=(U\arabic*), leftmargin=1cm]
\setcounter{enumi}{3}
\item For all $x\in\mathcal{T}$, the minimizers $m^{SAW}(x), m^{SAW}_{L,h}(x)$ and $\check{m}^{SAW}_{L,h}(x)$ exist and are unique, the latter  almost surely. Additionally, for any $\epsilon>0$, 
\begin{gather*}
\inf _{x \in \mathcal{T}} \inf _{d_{SW}\left(m(x), \omega\right)>\epsilon}\left\{M^{SAW}(\omega, x)-M^{SAW}[m^{SAW}(x), x]\right\}>0, \\
\liminf _{h \rightarrow 0} \inf _{x \in \mathcal{T}} \inf _{d_{SW}\left(m^{SAW}_{L, h}(x), \omega\right)>\epsilon}\left\{M^{SAW}_{L, h}(\omega, x)-M^{SAW}_{L, h}\left[m^{SAW}_{L, h}(x), x\right]\right\}>0,
\end{gather*}
and there exists $\zeta = \zeta(\epsilon)>0$ such that
\begin{align*}
    \operatorname{pr}\left(\inf _{x \in \mathcal{T}} \inf _{d_{SW}\left(\check{m}^{SAW}_{L, n}(x), \omega\right)>\epsilon}\left\{\check{M}^{SAW}_{L, n}(\omega, x)-\check{M}^{SAW}_{L, n}\left[\check{m}^{SAW}_{L, n}(x), x\right]\right\} \geq \zeta\right) \rightarrow 1.
\end{align*}
\item With $B_{\delta}\left[m^{SAW}(x)\right]$ and $N\left\{\delta \epsilon, B_\delta[m^{SAW}(x)], d_{SW}\right\}$ as in Condition (R5),
\begin{align*}
    \int_0^1 \sup _{x \in \mathcal{T}}\left(1+\log N\left\{\delta \epsilon, B_\delta[m^{SAW}(x)], d_{SW}\right\}\right)^{1 / 2} d \epsilon=O(1) \quad \text { as } \delta \rightarrow 0.
\end{align*}
\item There exists $\tau_1, \tau_2>0$ and $C_1, C_2>0$, such that
\begin{multline*}
\inf _{x \in \mathcal{T}} \inf _{d_{SW}\left(m^{SAW}(x), \omega\right)<\tau_1}\left\{M^{SAW}(\omega, x)-M^{SAW}[m^{SAW}(x), x]-C_1 d_{SW}(m^{SAW}(x), \omega)^{2}\right\} \geq 0, \\
\liminf _{h \rightarrow 0} \inf _{x \in \mathcal{T}} \inf _{d_{SW}\left(m^{SAW}_{L, h}(x), \omega\right)<\tau_2}\left\{M^{SAW}_{L, h}(\omega, x)-M^{SAW}_{L, h}\left[m^{SAW}_{L, h}(x), x\right] \right. \\
\left. -C_2 d_{SW}\left(m^{SAW}_{L, h}(x), \omega\right)^{2}\right\} \geq 0.
\end{multline*}
\end{enumerate}

The arguments provided for  (U1) and (U3) lead to (U4) and (U6) with $C_1=C_2=1$. Using  \eqref{formula:adx:entropy}, one finds that (U5) is satisfied. With Assumptions (L1)--(L2) from Theorem 1 of \citet{chen2022uniform}, we may  conclude that 
\begin{gather*}
    \sup_{x\in\mathcal{T}} d_{SW}\left(m^{SAW}(x), m_{L,h}^{SAW}(x)\right) = O\left(h^{2}\right),\\
    \sup_{x\in\mathcal{T}}d_{SW}\left(m^{SAW}_{L,h}(x), \check{m}^{SAW}_{L,h}(x)\right) = O_p\left(\max \left\{ \left(nh^2\right)^{-1/(2+\epsilon)}, [nh^2(-\log h)^{-1}]^{-1/2} \right\}\right). 
\end{gather*}
From  \eqref{formula:adx:SWminimizerEst} which holds uniformly for $x\in\mathcal{T}$ and Assumption (P1), 
\begin{align*}
    \sup_{x\in\mathcal{T}}d_{SW}\left(m^{SAW}_{L,h}(x), \hat{m}^{SAW}_{L,h}(x)\right) = O_p\left(\max \left\{ \left(nh^2\right)^{-1/(2+\epsilon)}, \left[nh^2(-\log h)^{-1}\right]^{-1/2} \right\}\right). 
\end{align*}
\end{proof}

\subsection{Proof of Theorem \ref{thm:transformGlob}}
\begin{proof}
    Let $\gamma_{G} = \argmin_{\gamma \in \Gamma_{\Theta}}M_G^{SWW}(\gamma, x)$ and $\hat{\gamma}_{G} =$ $\argmin_{\gamma \in \Gamma_{\Theta}}$ $\hat{M}_G^{SWW}(\gamma, x)$. 
    Note that the space {$\Gamma_{\Theta}$} is convex and bounded. Then applying  the arguments used for Theorem \ref{thm:swGlob} to the metric space $(\Gamma_{\Theta}, d_{DW})$, it follows that 
    \begin{align}
    \label{formula:gammaL2}
        d_{DW}^2\left(\gamma_{G}(x), \hat{\gamma}_{G}(x)\right) = \int_{\Theta}\int_{[0,1]}\left(\gamma_{G}(\theta)(s) - \hat{\gamma}_{G}(\theta)(s)\right)^2dsd\theta= O_p\left(n^{-1}\right). 
    \end{align}
    Recall that $G$ maps a univariate distribution to the corresponding quantile function while $G^{-1}$ maps a quantile function to the corresponding distribution. Let $\Psi$ map the quantile function to the cumulative distribution function. The arguments given in the  proof of Proposition 2 in \citet{mull:16:1} lead to 
    \begin{align*}
 &\quad \int_{\Theta}\int_{\mathbb{R}}\left(\Psi\left(\gamma_{G}(\theta)\right)(u) - \Psi\left(\hat{\gamma}_{G}(\theta)\right)(u)  \right)^2dud\theta \\
 &\lesssim \int_{\Theta}\int_{[0,1]}\left(\gamma_{G}(\theta)(s) - \hat{\gamma}_{G}(\theta)(s)\right)^2dsd\theta,
    \end{align*}
where we observe that by  Assumption (G1)  the derivatives of $\Psi\left(\gamma_G(\theta)\right)$ and $\Psi\left(\hat{\gamma}_G(\theta)\right)$ are bounded above and below. 
    From Lemma \ref{lemma:adx:derivative} and the fact that $\varphi(G^{-1}\left(\gamma_{G}(\theta))\right) = \partial \Psi\left(\gamma_{G}(\theta)\right)(u) / \partial u$ we obtain
    \begin{align*}
        &\quad \int_{\Theta}
        \|\varphi(G^{-1}(\gamma_{G}(\theta) ) )(u) - \varphi(G^{-1}(\hat{\gamma}_{G}(\theta)))(u)  \|_\infty^2 d\theta \\
        &\leq \int_{\Theta}\left(\int_{\mathbb{R}} \left(\Psi(\gamma_G(\theta))(u)-\Psi(\hat{\gamma}_G(\theta)(u)) \right)^2du\right)^{4/7}d\theta
        \\
        &\lesssim
        \left(\int_{\Theta}\int_{\mathbb{R}} \left(\Psi(\gamma_G(\theta))(u)-\Psi(\hat{\gamma}_G(\theta)(u)) \right)^2dud\theta\right)^{4/7}
        =O_p(n^{-4/7}).
    \end{align*}
    Here the first inequality follows from Lemma \ref{lemma:adx:derivative} and the second inequality  from H\"{o}lder's inequality and the compactness of $\Theta$. By Assumption (D2) the support of elements in $\mathcal{G}$ is uniformly bounded, whence 
    \begin{align*}
       \int_{\Theta}\int_\mathbb{R}\left(\varphi(G^{-1}(\gamma_{G}(\theta) ) )(u) - \varphi(G^{-1}(\hat{\gamma}_{G}(\theta)))(u)  \right)^2 du d\theta = O_p(n^{-4/7}).
    \end{align*}
    Recalling  the notation $\varrho(\gamma_G)(\theta) = \varphi(G^{-1}(\gamma_G(\theta)))$ and the definition of $L^2$ distance in \eqref{formula:LambdaL2}, we conclude 
    \begin{align}
    \label{formula:adx:quantile2dens}
        d_2\left(\varrho(\gamma_G), \varrho(\hat{\gamma}_G)\right) = O_p\left(n^{-2/7}\right).
    \end{align}
    By Theorem \ref{thm:reginv}, we have
    \begin{align*}
        d_{\infty}\left( m_G^{SWW}(x), \hat{m}_{G,\tau}^{SWW} \right) = O_p\left(C_1(\tau) + C_2(\tau)n^{-2/7}\right). 
    \end{align*}
    Uniform convergence then follows similarly to the  uniform convergence in Theorem \ref{thm:swGlob},
    \begin{align*}
        \sup_{\|x\|\leq B}d_{DW}\left(\gamma_{G}(x), \hat{\gamma}_{G}(x)\right) = O_p\left(n^{-1/(2+\epsilon)}\right),
    \end{align*}
    for a given $B>0$ and any $\epsilon > 0$.  Since this holds uniformly across $x$, we have 
    \begin{align*}
        \sup_{\|x\|\leq B}d_{\infty}\left( m_G^{SWW}(x), \hat{m}_{G,\tau}^{SWW} \right) = O_p\left(C_1(\tau) + C_2(\tau)n^{-2/(7+\epsilon)}\right). 
    \end{align*}  

\end{proof}

\subsection{Proof of Theorem \ref{thm:adx:transformLoc}}
Consider the following notations
\begin{gather*}
    \gamma_x = \argmin_{\gamma\in{\Gamma_{\Theta}}}M^{SWW}(\gamma, x),\\
    \gamma_{x,L,h} = \argmin_{\gamma\in\Gamma_{\Theta}}M_{L,h}^{SWW}(\gamma, x),\\
    \hat{\gamma}_{x,L,h}= \argmin_{\gamma\in\Gamma_{\Theta}}\hat{M}_{L,h}^{SWW}(\gamma, x).
\end{gather*}
Note that the space $\Gamma_{\Theta}$ is convex and closed. Applying the arguments in the proof  of Theorem \ref{thm:adx:swLoc} to the metric space $(\Gamma_{\Theta}, d_{DW})$, it follows that 
\begin{gather*}
        d_{DW}(\gamma_{x,L,h}, {\gamma}_x) = O_p\left(h^2\right), \\
        d_{DW}(\gamma_{x,L,h}, \hat{\gamma}_{x,L,h}) = O_p\left(\left(nh\right)^{-1/2}\right). 
\end{gather*}
In analogy to  the derivation of \eqref{formula:adx:quantile2dens}, we have 
\begin{gather*}
        d_{2}\left(\varrho(\gamma_{x,L,h}), \varrho({\gamma}_x) \right) = O_p\left(h^{8/7}\right), \\
        d_{2}\left(\varrho(\gamma_{x,L,h}), \varrho(\hat{\gamma}_{x,L,h})\right)  = O_p\left(\left(nh\right)^{-2/7}\right). 
\end{gather*}
From Assumption (T3), we obtain 
\begin{gather*}
    d_{\infty}\left( m^{SWW}(x), m_{L,h,\tau}^{SWW}(x)  \right)  = O_p\left(C_1(\tau) + C_2(\tau)h^{8/7}\right),\\
    d_{\infty}\left( m_{L,h,\tau}^{SWW}(x), \hat{m}_{L,h,\tau}^{SWW}(x)  \right)  = O_p\left(C_2(\tau)\left(nh\right)^{-2/7}\right).
\end{gather*}
When choosing $h \sim n^{-1/5}$, the resulting convergence rate is 
\begin{gather*}
    d_{\infty}\left( m^{SWW}(x), \hat{m}_{L,h,\tau}^{SWW}(x)  \right)  = O_p\left(C_1(\tau) + C_2(\tau)n^{-8/35}\right).
\end{gather*}
Under the additional assumptions (L3)--(L4), the uniform convergence rate similarly is obtained as 
\begin{gather*}
        \sup_{x\in\mathcal{T}}d_{\infty}( m^{SWW}(x), m^{SWW}_{L,h,\tau}(x)) = O\left(C_1(\tau) + C_2(\tau)h^{8/7}\right), \\
        \sup_{x\in\mathcal{T}}d_{\infty}({m}^{SWW}_{L,h,\tau}(x), \hat{m}^{SWW}_{L,h,\tau}(x)  ) = O_p\left(C_2(\tau)\max \left\{ \left(nh^2\right)^{-2/(7+\epsilon)}, \left[nh^2(-\log h)^{-1}\right]^{-2/7} \right\}\right), 
\end{gather*}
and when taking $h\sim n^{-1/(6+\epsilon)}$, 
\begin{align*}
    \sup_{x\in\mathcal{T}}d_{\infty}\left( m^{SWW}(x), \hat{m}^{SWW}_{L,h,\tau}(x)\right) = O_p\left(C_1(\tau) + C_2(\tau)n^{-4/(21+\epsilon)}\right). 
\end{align*}

\section{Additional Figures}
\label{sec:adx:figures}

\begin{figure}[h]
    \single
    \centering
    \includegraphics[width=0.9\linewidth]{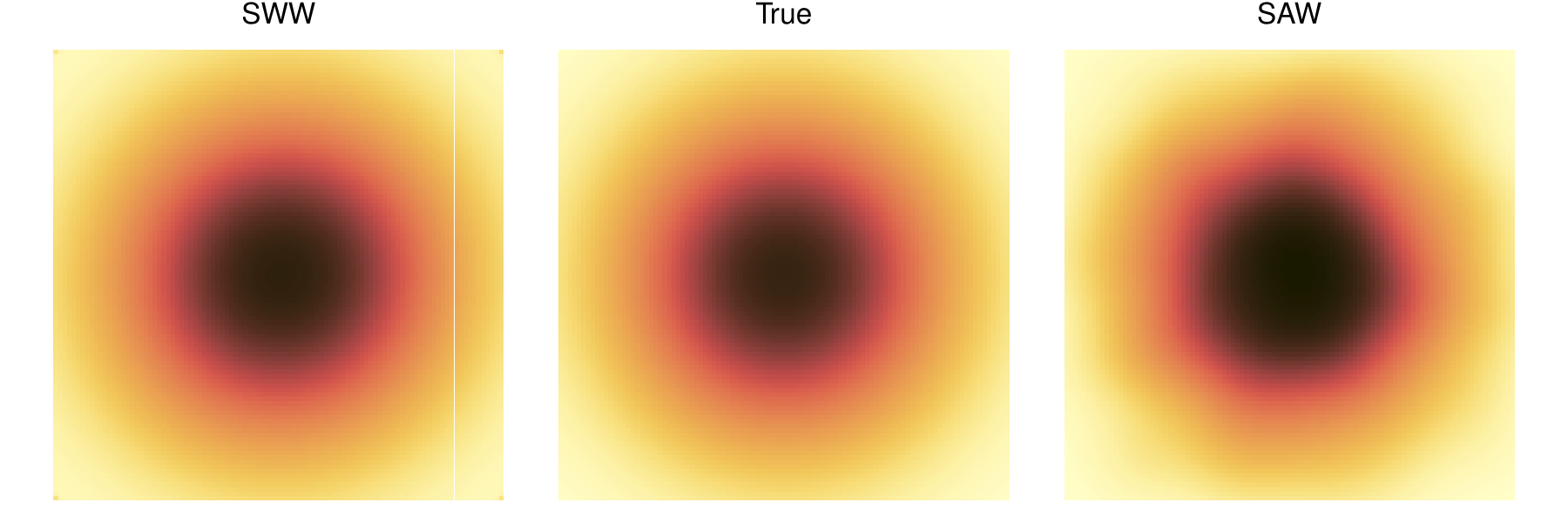}
    \caption{Comparison of the fitted bivariate Gaussian densities obtained by SWW (left) and SAW (right) with the true density (center).}
    \label{fig:simcom}
\end{figure}

\begin{figure}[h]
    \single
    \centering
    \includegraphics[width=\linewidth]{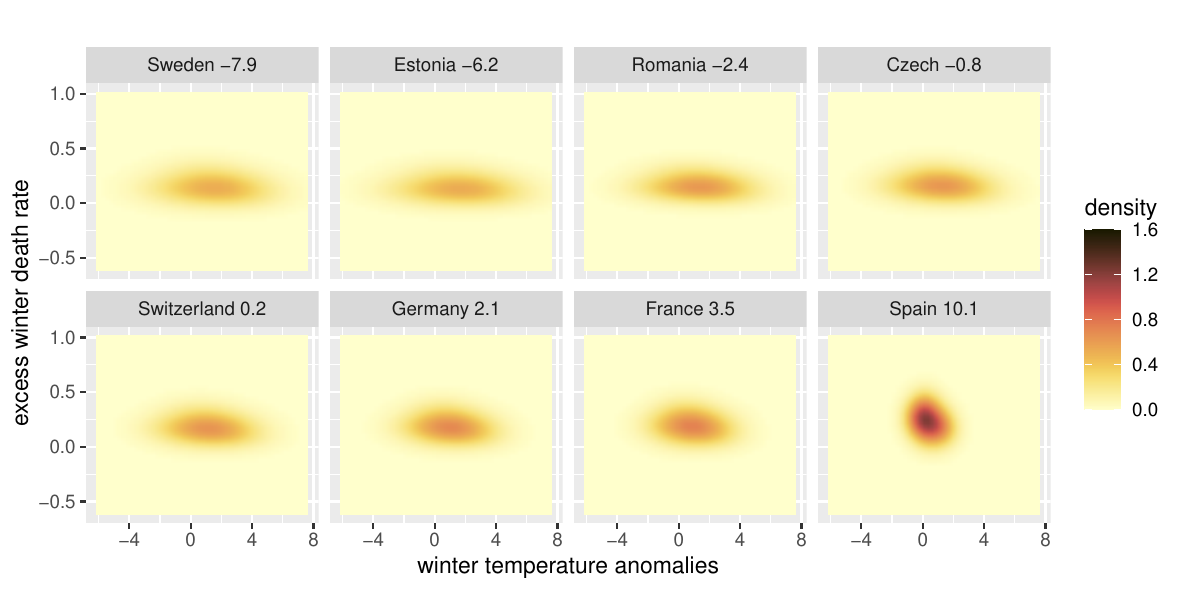}
    \caption{Excess Winter Death Rates: Fitted density surfaces for randomly selected countries obtained by the GSAW version of sliced Wasserstein regression with sliced Wasserstein fraction of variance explained at level 0.28. }
    \label{fig:EUDeathGSAW}
\end{figure}

\begin{figure}[htbp]
    \single
    \centering
    \includegraphics[width=\linewidth]{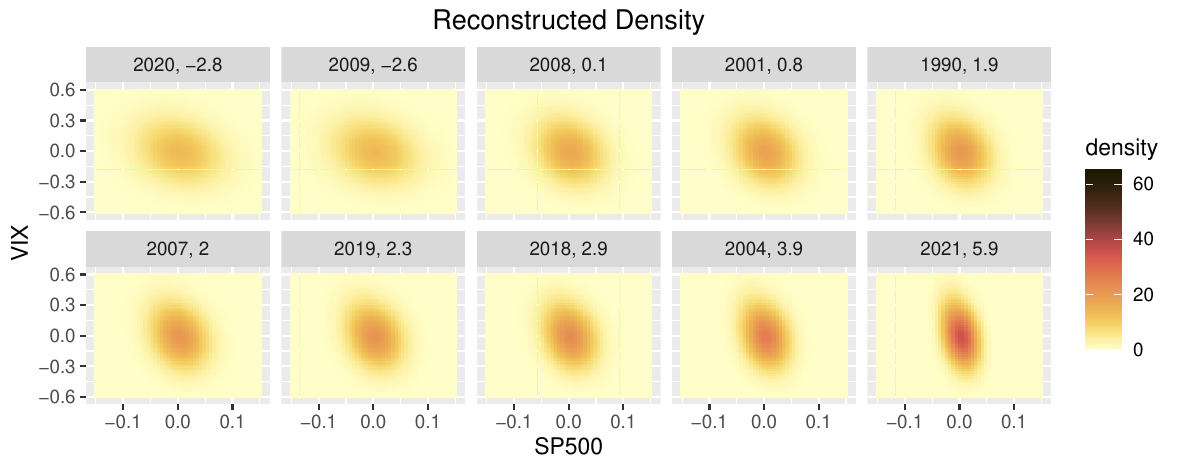}
    \caption{S\&P500 and VIX Index Data: Fitted density surfaces for randomly selected years obtained by the GSAW version of sliced Wasserstein regression with sliced Wasserstein  fraction of variance explained at level 0.13. }
    \label{fig:VIXGSAW}
\end{figure}

\begin{figure}[htbp]
    \single
    \centering
    \includegraphics[width=\linewidth]{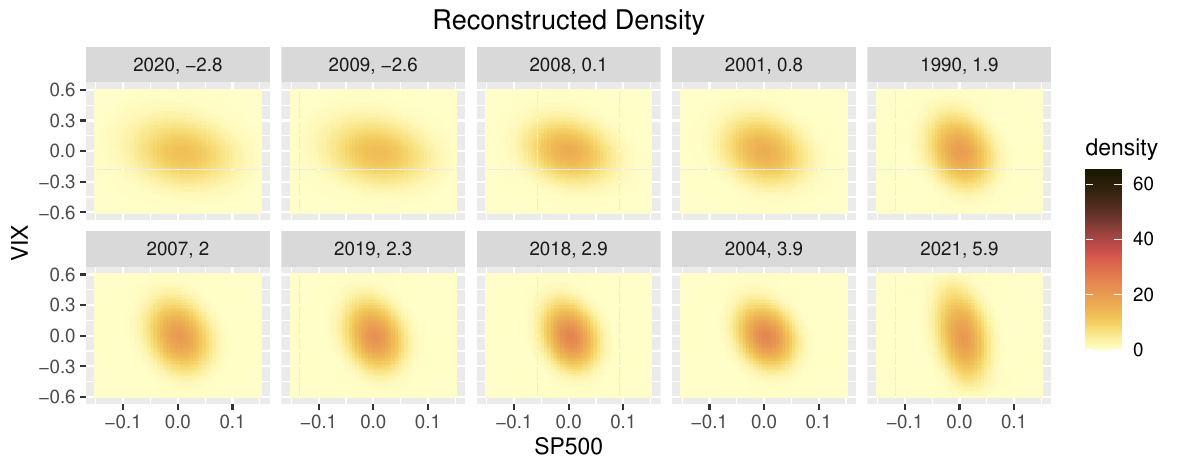}
    \caption{S\&P500 and VIX Index data: Fitted density surfaces for randomly selected years obtained by the LSWW version of sliced Wasserstein regression with sliced Wasserstein  fraction of variance explained at level 0.31. }
    \label{fig:VIXLSWW}
\end{figure}

\end{document}